\documentclass[brazil,american,portuges,english]{ws-ijtaf}

\usepackage[T1]{fontenc}
\usepackage[latin9]{inputenc}
\usepackage{float}
\usepackage{textcomp}
\usepackage{amsmath}
\usepackage{amssymb}
\usepackage{stackrel}
\usepackage{graphicx}
\usepackage{esint}
\usepackage{babel}

\makeatletter

\begin{document}

\markboth{A. Catal\~{a}o \& R. Rosenfeld}
{Analytical Path-Integral Pricing of Moving-Barrier Options under non-Gaussian Distributions}

\title{ANALYTICAL PATH-INTEGRAL PRICING OF MOVING-BARRIER OPTIONS UNDER NON-GAUSSIAN DISTRIBUTIONS}

\author{ANDR\'{E} CATAL\~{A}O}

\address{Instituto de F\'{i}sica Te\'{o}rica, Universidade Estadual Paulista - UNESP\\ R. Dr. Bento Teobaldo Ferraz, 271\\
S\~{a}o Paulo, SP, 01140-070, Brazil\\
andre.silvana74@gmail.com}

\author{ROGERIO ROSENFELD}

\address{Instituto de F\'{i}sica Te\'{o}rica, Universidade Estadual Paulista - UNESP\\
South American Institute for Fundamental Research\\ R. Dr. Bento Teobaldo Ferraz, 271\\
S\~{a}o Paulo, SP, 01140-070, Brazil\\
rosenfel@ift.unesp.br}

\maketitle

\begin{history}
\received{(Day Month Year)}
\revised{(Day Month Year)}
\end{history}
\begin{abstract}
In this work we present an analytical model, based on the path-integral
formalism of Statistical Mechanics, for pricing 
options using first-passage time problems involving both fixed
and deterministically moving absorbing barriers under possible non-gaussian
distributions of the underlying object. We adapt to our problem a model
originally proposed to describe the formation of galaxies in the universe
of \cite{MagIV}, which uses cumulant expansions in terms of the
Gaussian distribution, and we generalize it to take into acount drift and cumulants
of orders higher than three. From the probability density function,
we obtain an analytical pricing model, not only for vanilla options (thus
removing the need of volatility smile inherent to the \cite{BS73}
model), but also for fixed or deterministically moving barrier options.
Market prices of vanilla options are used to calibrate the model,
and barrier option pricing arising from the model is compared to the
price resulted from the relative entropy model.

\medskip{}

\end{abstract}
\vfill{}

\hyphenpenalty=10000\sloppy

\medskip{}

\section{Introduction\label{sec:Intgroduction}}

\noindent In Stochastic Processes, the first passage time $\tau_{f}$,
defined as the time a system takes to cross a barrier for the first
time - usually associated to survival analysis - appears in several
branches of science: from Biology, in cell transport phenomena, to
Economics, in credit default events; Sociology, in group decisions;
Physics, in Statistical Mechanics, Optics, Solid State; Chemistry,
in reactions and corrosion; and Cosmology. In this latter case, to
form a galaxy, the concentration of mass needs to reach a critical
value, which can be seen as a barrier, not necessarily fixed, and
possibly subject to a stochastic process.

The problem of finding the probability distribution of first passage
time was first studied by \cite{Sch15} in the context of a Brownian
motion in a physical medium, and it was shown to be given by the Inverse
Normal distribution. In Statistics, the distribution was first obtained
by \cite{W47} in likelihood-ratio tests. In stochastic calculus,
the problem can be studied in terms of the transition probability
distribution between states emerging from boundary conditions imposed
to the Fokker-Planck equation (\cite{Gar04,Ris89}), from which the
cumulative probability distribution that the first passage time occurs
after a given instant $T$, that is, $P(\tau_{f}>T)$, is derived.

The study of first passage time depends on the distribution of the
underlying process that is assumed. In the case of galaxy formation,
\cite{MagI} and \cite{MagII} discusses the treatment of Gaussian
distributions, while \cite{MagIII} and \cite{MagIV} treat non-Gaussian
diffusion, the latter including the case of moving barriers. Usually,
the non-Gaussian approach is developed in terms of expansions based
on a benchmark distribution, which is commonly taken as the Gaussian
one.

In Finance, the first passage time problem may arise in derivative
contracts that establish deactivation or activation conditions, upon
the passage of a time dependent variable $P_{t}$ through a barrier
$B$:

\medskip{}

\begin{equation}
\tau_{f}=min\left\{ t\;|\;P_{t}>B\right\} .
\end{equation}

\medskip{}

The most frequently used distribution in Finance is the lognormal
distribution for prices, in the context of the \cite{BS73} model
hypothesis. A single barrier knock-up-and-out european call option
(KUO european call) is then a contract that enables its holder to
buy a certain asset $S_{T}$, the underlying asset, at maturity date
$T$, paying the contract strike price $K$, as long as the underlying
asset does not cross a contractual barrier level $B$. Mathematically,
the payoff of such contract is

\medskip{}

\begin{equation}
Call_{KUO}\left(T\right)=1_{S_{t}<B,\forall t\in\left[0,T\right]}max\left(S_{T}-K,0\right),
\end{equation}

\medskip{}

\noindent and its value at any time under the \cite{BS73} hypothesis
has closed-form analytical solution, as shown in \cite{Shr04}. However,
although the lognormal distribution requires a single parameter of
volatility, evidence is that vanilla (non-barrier) options of different
strikes demand different implied volatilities, given their market
prices. This structure of volatility, dependent on the strike values,
is called volatility smile, and through time it generates the volatility
surface. An additional issue is related to barrier clauses in the
contract: should we select the volatility according to the strike,
to the barrier, or both? Furthermore, in the special case of moving
barrier options, \cite{Kun92} provide analytical solution to the
pricing problem, but in a Black-Scholes setup of one volatility.

Several approaches have been proposed in order to explain the volatility
smile and, subsequently, to allow exotic options' pricing, such as
barrier options. Among them we may cite the local volatility model
(\cite{Dup94}), the stochastic volatility model (\cite{He93}), the
jump model (\cite{Merton76}), the relative entropy model (\cite{Ave01});
and also non-Gaussian distribution models based on expansions, an
example of which is the Edgeworth expansion (\cite{Rubinstein 98}
and \cite{Balieiro e Rosenfeld}). While some of them pose difficulties
related to the need of a complete set of market prices to build the
volatility surface, others that rely on numerical or simulation implementations
demand careful attention regarding the behaviour of the process surounding
the barrier region.

In this paper we adapt the non-Gaussian model of galaxy formation,
based on cumulant expansion, developed by \cite{MagIV}, to the pricing
of both fixed and deterministically moving up-and-out barrier options.
By doing so, in the limiting case of infinite barrier values, we also
obtain a non-Gaussian vanilla pricing model. Our adaptation consists
on introducing a drift term in the expansion and also extending it
to an arbitrary number of cumulants. In addition, we derive the martingale
condition for risk-neutral pricing. The methodology employs the path
integral formalism of Statistical Mechanics (\cite{Ris89}), and results
in closed-form expressions for vanillas, fixed and deterministically
moving barrier options. The development also takes into account the
behaviour of the expansion in the neighbourhood of the barrier.

The model parameters, which are the cumulants, are calibrated with
vanilla options and are afterwards used to price barrier options.
As long as data for barrier options almost always refer to market-to-model
quotes, we compare our results to the ones delivered by the relative
entropy model (\cite{Ave01}).

With respect to the organization of the paper, we begin by describing
the theoretical framework, which encompasses the development of \cite{MagI}
to \cite{MagIV}. We first present the general formulation of cumulant
expansion in terms of path integrals and recover the result for the
Gaussian fluctuation and fixed barrier. Then we generalize to the
case of non-Gaussian fluctuation and moving barrier. Next, we describe
the calibration procedure, followed by the barrier options pricing.
Finally, we present our conclusions. We collect in the appendices some results 
used in the text.

\medskip{}

\section{Cumulant expansion and the path integral formalism\label{sec:Cumulant-expansion-and}}

\medskip{}

\hspace*{0.25in}In this section, we present the cumulant expansion,
connecting it to the path integral formalism. Let $\omega=\omega(t)=\omega_{t}$
be the stochastic variable whose distribution we wish to model. In
our case, $\omega=\frac{1}{\sigma}ln\frac{S_{t}}{S_{0}}$, where $\sigma$
is the volatility parameter, $S_{0}$ and and $S_{t}$ are the underlying
values at $t=0$ and $t$, respectively. A path begins at $t=0$,
with $\omega_{0}=0$, and evolves until the final instant of time
$t_{n}$, where $\omega_{n}=\omega(t_{n})=\omega(T)$. We assume the
time discretization $\Delta t=\epsilon$, with $t_{k}=k\epsilon$.
A price path is a collection $\left\{ \omega_{1},...,\omega_{n}\right\} $,
such that $\omega(t_{k})=\omega_{k}$. If there is no absorbing barrier,
$\omega_{t}\in\left(-\infty,\infty\right)$. The probability density
in the space of trajectories can be described by the expected value
of a product of Dirac delta functions:

\medskip{}

\begin{equation}
W_{n}=W\left(\omega_{0},\omega_{1},...,\omega_{n};t_{n}\right)\equiv\left\langle \delta\left(\omega\left(t_{1}\right)-\omega_{1}\right)...\delta\left(\omega\left(t_{n}\right)-\omega_{n}\right)\right\rangle ,\label{eq:3}
\end{equation}

\medskip{}

which follows from

\medskip{}

\[
\left\langle \delta\left(x_{1}-\bar{x}_{1}\right)\delta\left(x_{2}-\bar{x}_{2}\right)...\delta\left(x_{n}-\bar{x}_{n}\right)\right\rangle =
\]

\[
\int_{-\infty}^{\infty}...\int_{-\infty}^{\infty}p\left(x_{1},x_{2},...,x_{n}\right)\delta\left(x_{1}-\bar{x}_{1}\right)\delta\left(x_{2}-\bar{x}_{2}\right)...\delta\left(x_{n}-\bar{x}_{n}\right)dx_{1}dx_{2}...dx_{n}=
\]

\begin{equation}
p\left(\bar{x}_{1},\bar{x}_{2},...,\bar{x}_{n}\right),
\end{equation}

\medskip{}
 which is the probability density. \footnote{See \cite{Ris89}, section 2.4.}
In terms of $W$, the probability that the variable assumes the value
$\omega_{n}$ at instant $t_{n}$, from $\omega_{0}$, at $t=0$,
in trajectories that never exceed $\omega_{c}$, is given by:

\noindent \medskip{}

\begin{equation}
\Pi_{\epsilon}\left(\omega_{0},\omega_{n};t_{n}\right)=\int_{-\infty}^{\omega_{c}}d\omega_{1}...\int_{-\infty}^{\omega_{c}}d\omega_{n-1}W\left(\omega_{0},\omega_{1},...,\omega_{n};t_{n}\right).\label{eq:densProbConfinamento}
\end{equation}

\medskip{}

And the probability that the path remains in the region $\omega<\omega_{c}$,
for all instants lower than $t_{n}$, is:

\medskip{}

\begin{equation}
\Pi\left(\omega_{0};t_{n}\right)=\int_{-\infty}^{\omega_{c}}d\omega_{n}\Pi_{\epsilon}\left(\omega_{0},\omega_{n};t_{n}\right).\label{eq:ProbConfinamento}
\end{equation}

\medskip{}

This equation represents the sum over all possible paths, thus representing
the path integral that computes the probability function. We will
express it in terms of the cumulants of the distribution. The characteristic
function is the Fourier transform of the distribution \footnote{See \cite{Ris89}, section 2.3.}:

\medskip{}

\[
C_{n}\left(u_{1},...,u_{n}\right)=\left\langle e^{iu_{1}\omega_{1}+\cdots+iu_{n}\omega_{n}}\right\rangle =
\]

\begin{equation}
\int\ldots\int e^{iu_{1}\omega_{1}+\cdots+iu_{n}\omega_{n}}W\left(\omega_{0},\omega_{1},...,\omega_{n};t_{n}\right)d\omega_{1}...d\omega_{n}.\label{eq:8}
\end{equation}

\medskip{}

The joint moment function is defined by

\medskip{}
 \medskip{}

\[
M_{m_{1},...,m_{n}}=\left\langle \omega_{1}^{m_{1}}\ldots\omega_{n}^{m_{n}}\right\rangle 
\]

\begin{equation}
=\left(\frac{\partial}{\partial\left(iu_{1}\right)}\right)^{m_{1}}\ldots\left(\frac{\partial}{\partial\left(iu_{n}\right)}\right)^{m_{n}}\left.C_{n}\left(u_{1},...,u_{n}\right)\right|_{u_{1}=...=u_{n}=0}.
\end{equation}

\medskip{}

The joint moments are the coefficients of the Taylor expansion of
the characteristic function:

\medskip{}

\begin{equation}
C_{n}\left(u_{1},...,u_{n}\right)=\underset{m_{1},...,m_{n}}{\sum}M_{m_{1},...,m_{n}}\frac{\left(iu_{1}\right)^{m_{1}}}{m_{1}!}\cdots\frac{\left(iu_{n}\right)^{m_{n}}}{m_{n}!}.
\end{equation}

\medskip{}

The joint cumulants $\kappa_{m_{1},...,m_{n}}$ of a distribution
are related to the characteristic function by

\medskip{}

\begin{equation}
C_{n}\left(u_{1},...,u_{n}\right)=\exp\left(\stackrel[m_{1},...,m_{n}]{\infty}{\sum}\kappa_{m_{1},...,m_{n}}\frac{\left(iu_{1}\right)^{m_{1}}}{m_{1}!}\cdots\frac{\left(iu_{n}\right)^{m_{n}}}{m_{n}!}\right)
\end{equation}

\begin{equation}
\kappa_{m_{1},...,m_{n}}=\left(\frac{\partial}{\partial\left(iu_{1}\right)}\right)^{m_{1}}\ldots\left(\frac{\partial}{\partial\left(iu_{n}\right)}\right)^{m_{n}}\left.\ln\left[C_{n}\left(u_{1},...,u_{n}\right)\right|_{u_{1}=...=u_{n}=0}\right].
\end{equation}

\medskip{}

$\Pi_{\epsilon}\left(\omega_{0},\omega_{n};t_{n}\right)$ can be expressed
in terms of the cumulants. To see this we use the following representation
of the Dirac delta function

\medskip{}

\begin{equation}
\delta\left(\omega\right)=\int_{-\infty}^{\infty}\frac{du}{2\pi}\cdot e^{-iu\omega}.
\end{equation}

\medskip{}

Substituting in (\ref{eq:3}),

\medskip{}

\[
W\left(\omega_{0},\omega_{1},...,\omega_{n};t_{n}\right)=\left\langle \int_{-\infty}^{\infty}\frac{du_{1}}{2\pi}\cdots\frac{du_{n}}{2\pi}\cdot e^{-i\stackrel[j=1]{n}{\sum}u_{j}\left(\omega\left(t_{j}\right)-\omega_{j}\right)}\right\rangle 
\]

\begin{equation}
=\int_{-\infty}^{\infty}\frac{du_{1}}{2\pi}\cdots\frac{du_{n}}{2\pi}\cdot e^{i\stackrel[j=1]{n}{\sum}u_{j}\omega_{j}}\left\langle e^{-i\stackrel[j=1]{n}{\sum}u_{j}\omega\left(t_{j}\right)}\right\rangle .\label{eq:14}
\end{equation}

\medskip{}

We define the integration measure

\medskip{}

\begin{equation}
\int_{-\infty}^{\infty}Du\equiv\int_{-\infty}^{\infty}\frac{du_{1}}{2\pi}\cdots\frac{du_{n}}{2\pi}.
\end{equation}

\medskip{}

Therefore, using the definition (\ref{eq:8}) in (\ref{eq:14}) we
can write






\medskip{}

\[
W\left(\omega_{0},\omega_{1},...,\omega_{n};t_{n}\right)=\int_{-\infty}^{\infty}Du\cdot
\]

\begin{equation}
\cdot exp\left(i\stackrel[j=1]{n}{\sum}u_{j}\omega_{j}+\stackrel[m_{1},...,m_{n}]{\infty}{\sum}\kappa_{m_{1},...,m_{n}}\frac{\left(-iu_{1}\right)^{m_{1}}}{m_{1}!}\cdots\frac{\left(-iu_{n}\right)^{m_{n}}}{m_{n}!}\right).\label{eq:17}
\end{equation}

This is the expansion of the probability of a given path in terms
of the joint cumulants. \medskip{}

Keeping only terms with $m_{i}$ is equal to $0$ or $1$, that is,
$m_{1}=0,1;\;m_{2}=0,1;...m_{n}=0,1$ (we will justify this shortly)
results in:







\medskip{}

\[
W\left(\omega_{0},\omega_{1},...,\omega_{n};t_{n}\right)=\int_{-\infty}^{\infty}Du\cdot exp\left(i\stackrel[j=1]{n}{\sum}u_{j}\omega_{j}-i\stackrel[i=1]{n}{\sum}u_{i}\kappa_{i}-\frac{1}{2}\stackrel[i,j=1]{n}{\sum}u_{i}u_{j}\kappa_{ij}\right.
\]

\begin{equation}
\left.+\frac{\left(-i\right)^{3}}{3!}\stackrel[i,j,k=1]{n}{\sum}u_{i}u_{j}u_{k}\kappa_{ijk}+\frac{\left(-i\right)^{4}}{4!}\stackrel[i,j,k,l=1]{n}{\sum}u_{i}u_{j}u_{k}u_{l}\kappa_{ijkl}+...\right).\label{eq:19}
\end{equation}

\medskip{}

In this notation, $\kappa_{1}\equiv\kappa_{m_{1}=1,m_{2}=0...,m_{n}=0}$,
$\kappa_{2}\equiv\kappa_{m_{1}=0,m_{2}=1,m_{3}=0,...,m_{n}=0}$, $\kappa_{12}\equiv\kappa_{m_{1}=1,m_{2}=1,m_{3}=0,...,m_{n}=0}$,
$\kappa_{13}\equiv\kappa_{m_{1}=1,m_{2}=0,m_{3}=1,m_{4}=0,...,m_{n}=0}$,
and so on. (\ref{eq:19}) will be the version of equation (\ref{eq:17})
that we will use.

Had we considered other values of $m_{1}$, $m_{2}$, $etc$, different
from zero and one, we would have included generalized moments, beyond
the usual covariance between two variables. For instance, the covariance
between the fourth power of a variable $i$ and the cube of another
variable $j$, in the case of $m_{i}=2$ and $m_{j}=3$, etc, and
they contribute at higher orders. In our case, where we seek to calibrate
market data, it will be enough to consider just the usual moments
(variance, kurtosis, etc) and, thus, we will not consider covariances
and its generalizations in the combinations of the several orders
of the variables. In this notation, for example, in $\kappa_{ij}$,
when $i=j$, we have the cumulant linked to the variance; in $\kappa_{ijk}$,
when $i=j=k$, the cumulant related to asymmetry, etc.

Using this expansion in (\ref{eq:densProbConfinamento}) one obtains

\medskip{}

\[
\Pi_{\epsilon}\left(\omega_{0},\omega_{n};t_{n}\right)=\int_{-\infty}^{\omega_{c}}d\omega_{1}...\int_{-\infty}^{\omega_{c}}d\omega_{n-1}\int_{-\infty}^{\infty}Du\cdot
\]

\[
exp\left(i\stackrel[j=1]{n}{\sum}u_{j}\omega_{j}-i\stackrel[i=1]{n}{\sum}u_{i}\kappa_{i}-\frac{1}{2}\stackrel[i,j=1]{n}{\sum}u_{i}u_{j}\kappa_{ij}\right.
\]

\begin{equation}
\left.+\frac{\left(-i\right)^{3}}{3!}\stackrel[i,j,k=1]{n}{\sum}u_{i}u_{j}u_{k}\kappa_{ijk}+\frac{\left(-i\right)^{4}}{4!}\stackrel[i,j,k,l=1]{n}{\sum}u_{i}u_{j}u_{k}u_{l}\kappa_{ijkl}+...\right).\label{eq:20}
\end{equation}

\medskip{}

This equation is the path integral representation of the probability
distribution in termos of the cumulants.

\medskip{}

\section{Gaussian Fluctuations\label{sec:Gaussian-Fluctuations}}

\medskip{}

\hspace*{0.25in} In this Section we show that our formalism reproduces
the well-known formulae for the price of barrier options for Gaussian
fluctuations as a sanity check. In the case of Gaussian fluctuations,
the cumulants are zero, except those satisfying $m_{1}+m_{2}+...+m_{r}\leq2$\footnote{\cite{Ris89}, section 2.3.3.}:

\medskip{}

\begin{equation}
\left\langle e^{i\stackrel[j=1]{n}{\sum}\left(-u_{j}\right)\omega\left(t_{j}\right)}\right\rangle =exp\left(-i\stackrel[i=1]{n}{\sum}u_{i}\kappa_{i}-\frac{1}{2}\stackrel[i,j=1]{n}{\sum}u_{i}u_{j}\kappa_{ij}\right).\label{eq:FuncaoCaracteristicaGaussiana}
\end{equation}

\medskip{}

In this case equations (\ref{eq:19}) and (\ref{eq:20}) assume the
form (we put a superscritp ``g\textquotedbl{} to indicate Gaussian)

\medskip{}

\begin{equation}
W^{g}\left(\omega_{0},\omega_{1},...,\omega_{n};t_{n}\right)=\int_{-\infty}^{\infty}Du\cdot exp\left(i\stackrel[j=1]{n}{\sum}u_{j}\omega_{j}-i\stackrel[i=1]{n}{\sum}u_{i}\kappa_{i}-\frac{1}{2}\stackrel[i,j=1]{n}{\sum}u_{i}u_{j}\kappa_{ij}\right);\label{eq:Wgaussiana}
\end{equation}

\[
\Pi_{\epsilon}^{g}\left(\omega_{0},\omega_{n};t_{n}\right)=\int_{-\infty}^{\omega_{c}}d\omega_{1}...\int_{-\infty}^{\omega_{c}}d\omega_{n-1}\int_{-\infty}^{\infty}Du\cdot
\]

\begin{equation}
exp\left(i\stackrel[i=1]{n}{\sum}u_{i}\omega_{i}-i\stackrel[i=1]{n}{\sum}u_{i}\kappa_{i}-\frac{1}{2}\stackrel[i,j=1]{n}{\sum}u_{i}u_{j}\kappa_{ij}\right).\label{eq:distGaussiana}
\end{equation}

\medskip{}

Besides, $\kappa_{ij}=\sigma_{ij}$, where $\sigma_{ij}$ is the covariance
between $i$ and $j$.

Consider the case of Markovian processes, where just the previous
state of the variable influences the present state:

\medskip{}

\begin{equation}
\Pi\left(\omega\left(t_{n}\right)\leq\omega_{n}|\omega\left(t_{n-1}\right),...,\omega\left(t_{1}\right)\right)=\Pi\left(\omega\left(t_{n}\right)\leq\omega_{n}|\omega\left(t_{n-1}\right)\right).
\end{equation}

\medskip{}

We will denote \foreignlanguage{brazil}{$\Pi_{\epsilon}^{gm}$ and
$W^{gm}$ the Gaussian probability and probability density under the
Markov hypothesis, that is, when the particle executes a Markovian
Gaussian Brownian motion. In a stationary stochastic process, the
moments are constant along time, and their values only depend on the
least instante between periods. If the variable $\omega$ is standard
Gaussian, as in a Wiener process,}

\medskip{}

\begin{equation}
\sigma_{ij}=<\omega_{i}\omega_{j}>=\epsilon min(i,j)\equiv\epsilon A_{ij}.\label{eq:covarWiWj}
\end{equation}

\medskip{}

The probability density (\ref{eq:Wgaussiana}) becomes:

\medskip{}

\[
W^{gm}\left(\omega_{0},\omega_{1},...,\omega_{n};t_{n}\right)=\int_{-\infty}^{\infty}Du\cdot exp\left(i\stackrel[i=1]{n}{\sum}u_{i}\omega_{i}-i\stackrel[i=1]{n}{\sum}u_{i}\kappa_{i}-\frac{\epsilon}{2}\stackrel[i,j=1]{n}{\sum}u_{i}u_{j}A_{ij}\right)
\]

\begin{equation}
=\int_{-\infty}^{\infty}Du\cdot exp\left(i\stackrel[i=1]{n}{\sum}u_{i}\left(\omega_{i}-\kappa_{i}\right)-\frac{\epsilon}{2}\stackrel[i,j=1]{n}{\sum}u_{i}u_{j}A_{ij}\right).\label{eq:Wgaussiana1}
\end{equation}

\medskip{}

To illustrate, consider one variable $\omega_{i}=\omega$. Then,

\medskip{}

\[
W^{g}(\omega)=\frac{1}{2\pi}\int_{-\infty}^{\infty}du\cdot e^{iu\left(\omega-\kappa_{1}\right)-\frac{1}{2}u^{2}\kappa_{2}}
\]
\begin{equation}
=\frac{1}{\sqrt{2\pi\cdot\kappa_{2}}}e^{-\frac{\left(\omega-\kappa_{1}\right)^{2}}{2\kappa_{2}}},
\end{equation}

\medskip{}

\noindent where $\kappa_{2}=<\omega^{2}>=\epsilon$.














\medskip{}

In the case of $n$ Gaussian Markovian variables, with $\kappa_{2}=\epsilon$
and $\kappa_{1}=\epsilon\alpha$, where $\alpha$ is the \emph{drift}:

\medskip{}

\begin{equation}
W^{gm}\left(\omega_{0},\omega_{1},...,\omega_{n};t_{n}\right)=\frac{1}{\left(2\pi\epsilon\right)^{n/2}}e^{-\stackrel[i=0]{n-1}{\sum}\frac{\left(\omega_{i+1}-\omega_{i}+\alpha\epsilon\right)^{2}}{2\epsilon}}.\label{eq:30}
\end{equation}
\medskip{}

Therefore we can write

\medskip{}

\begin{equation}
W^{gm}\left(\omega_{0},\omega_{1},...,\omega_{n};t_{n}\right)=\Psi_{\epsilon}\left(\omega_{n}-\omega_{n-1}\right)W^{gm}\left(\omega_{0},\omega_{1},...,\omega_{n-1};t_{n-1}\right)\label{eq:separacao}
\end{equation}

\begin{equation}
\Psi_{\epsilon}\left(\Delta\omega\right)\equiv\frac{1}{\sqrt{2\pi\epsilon}}e^{-\frac{\left(\Delta\omega+\alpha\epsilon\right)^{2}}{2\epsilon}}\label{eq:PsiDeltaOmega}
\end{equation}

\begin{equation}
\Delta\omega=\omega_{n}-\omega_{n-1}.\label{eq:deltaOmega}
\end{equation}

\medskip{}

Thus,

\medskip{}

\begin{equation}
\Pi_{\epsilon}^{gm}\left(\omega_{0},\omega_{n};t_{n}\right)=\int_{-\infty}^{\omega_{c}}d\omega_{n-1}\Psi_{\epsilon}\left(\omega_{n}-\omega_{n-1}\right)\Pi_{\epsilon}^{gm}\left(\omega_{0},\omega_{n-1};t_{n-1}\right).\label{eq:pi gm}
\end{equation}

\medskip{}

\selectlanguage{brazil}%
In the presence of a fixed barrier, the probability density in the
case of and up absorbing barrier $B$, to be used in the call KUO
pricing, under the Black-Scholes assumptions, is (\cite{Shr04}):

\selectlanguage{english}%
\medskip{}

\begin{equation}
\Pi_{\epsilon\rightarrow0}^{gm}\left(\omega_{0},\omega_{n};t_{n}\right)=\frac{1}{\sqrt{2\pi t_{n}}}e^{\alpha\left(\omega_{n}-\omega_{0}\right)-\frac{1}{2}\alpha^{2}t_{n}}\left[e^{-\frac{\left(\omega_{n}-\omega_{0}\right)^{2}}{2t_{n}}}-e^{-\frac{\left(2\omega_{c}-\omega_{n}-\omega_{0}\right)^{2}}{2t_{n}}}\right].\label{eq:distB_BS}
\end{equation}

\begin{equation}
\omega_{n}=\omega(t_{n})=\frac{1}{\sigma}ln\frac{S_{t}}{S_{0}};\;\omega_{c}=b=\frac{1}{\sigma}ln\frac{B}{S_{0}}.\label{eq:defVariaveisB_BS}
\end{equation}

\medskip{}

\medskip{}

\section{Analytical expansion for non-Gaussian distributions with moving barrier
in the path-integral formalism\label{sec:Analytical-expansion-for}}

\medskip{}

\hspace*{0.25in}In this Section, the non-Gaussian distribution with
absorbing moving barrier is obtained from the path integral formulation.
As in the work of \cite{MagIV}, we present two alternative approaches
in the expansion: (i) first, the hypothesis of \cite{ST02}, which
states that instants $t_{i}<t_{n}$ are insignificant compared to
$t_{n}$ in derivatives higher than the first order and (ii) second,
barrier moves slowly. The latter we call ``adiabatic barriers''.

The accomplishment of this task involves expanding the non-Gaussian
distribution with moving barrier, $\Pi_{\epsilon\rightarrow0}\left(\omega_{0},\omega_{n};t_{n}\right)$,
in an expression of the form:

\medskip{}

\begin{equation}
\Pi_{\epsilon\rightarrow0}\left(\omega_{0},\omega_{n};t_{n}\right)=\Pi_{\epsilon\rightarrow0}^{mb}\left(\omega_{0},\omega_{n};t_{n}\right)+derivatives\;of\;\Pi_{\epsilon\rightarrow0}^{mb}\left(\omega_{0},\omega_{n};t_{n}\right),
\end{equation}

\medskip{}

\noindent where, in each approach (Sheth-Tormen and adiabatic barriers),
$\Pi_{\epsilon\rightarrow0}^{mb}\left(\omega_{0},\omega_{n};t_{n}\right)$
assumes different formats, both involving the Gaussian distribution
with fixed barrier (\ref{eq:distB_BS}), plus terms regarding moving
barriers.

\medskip{}

\subsection{The Sheth-Tormen approach\label{subsec:The-Sheth-Tormen-approach}}

\medskip{}

\hspace*{0.25in}Consider the expansion in cumulants (\ref{eq:20}),
in the case of a barrier that moves according to a deterministic rule
$B(t_{i}),\;i=1,...,n-1$:

\medskip{}



\begin{equation}
\Pi_{\epsilon\rightarrow0}\left(\omega_{0},\omega_{n};t_{n}\right)=\int_{-\infty}^{B(t_{1})}d\omega_{1}...\int_{-\infty}^{B(t_{n-1})}d\omega_{n-1}W\left(\omega_{0},\omega_{1},...,\omega_{n};t_{n}\right),
\end{equation}

\medskip{}

\noindent with $W\left(\omega_{0},\omega_{1},...,\omega_{n};t_{n}\right)$
given by (\ref{eq:19})






\medskip{}

Next, we assume that the barrier does not change significantly and
expand in a Taylor series around $B(t_{n})\equiv B_{n}$. Therefore,

\medskip{}

\begin{equation}
B(t_{i})=B(t_{n})+\stackrel[p=1]{\infty}{\sum}\frac{B_{n}^{(p)}}{p!}\left(t_{i}-t_{n}\right)^{p}\label{eq:39}
\end{equation}

\begin{equation}
B_{n}^{(p)}=\frac{d^{p}B\left(t_{n}\right)}{dt_{n}^{p}}.
\end{equation}


\medskip{}

Redefining the variables $\omega_{i}$, $i=1,...,n-1$:

\medskip{}

\[
\varpi_{i}\equiv\omega_{i}-\stackrel[p=1]{\infty}{\sum}\frac{B_{n}^{(p)}}{p!}\left(t_{i}-t_{n}\right)^{p}
\]

\[
\therefore\varpi_{i}=\omega_{i}-\left(B(t_{i})-B(t_{n})\right)
\]

\begin{equation}
d\varpi_{i}=d\omega_{i}.
\end{equation}

\medskip{}






Thus,

\medskip{}

\begin{equation}
\Pi_{\epsilon\rightarrow0}\left(\omega_{0}=0,\varpi_{n};t_{n}\right)=\int_{-\infty}^{B_{n}}d\varpi_{1}...\int_{-\infty}^{B_{n}}d\varpi_{n-1}\int_{-\infty}^{\infty}Du\cdot e^{Z}
\end{equation}

\[
Z=i\stackrel[i=1]{n}{\sum}u_{i}\varpi_{i}+i\stackrel[i=1]{n-1}{\sum}u_{i}\stackrel[p=1]{\infty}{\sum}\frac{B_{n}^{(p)}}{p!}\left(t_{i}-t_{n}\right)^{p}
\]

\[
-i\stackrel[i=1]{n}{\sum}u_{i}\kappa_{i}-\frac{1}{2}\stackrel[i,j=1]{n}{\sum}u_{i}u_{j}\kappa_{ij}
\]

\begin{equation}
+\frac{\left(-i\right)^{3}}{3!}\stackrel[i,j,k=1]{n}{\sum}u_{i}u_{j}u_{k}\kappa_{ijk}+\frac{\left(-i\right)^{4}}{4!}\stackrel[i,j,k,l=1]{n}{\sum}u_{i}u_{j}u_{k}u_{l}\kappa_{ijkl}+...
\end{equation}

\medskip{}

Since $\varpi_{i}$ is a dummy variable, we will use the notation
$\omega_{i}$ again. We work with the expansion until the 5th order,
generalizing it later.\medskip{}

\[
Z=i\stackrel[i=1]{n}{\sum}u_{i}\omega_{i}-i\stackrel[i=1]{n}{\sum}u_{i}\kappa_{i}-\frac{1}{2}\stackrel[i,j=1]{n}{\sum}u_{i}u_{j}\kappa_{ij}
\]

\[
+\frac{\left(-i\right)^{3}}{3!}\stackrel[i,j,k=1]{n}{\sum}u_{i}u_{j}u_{k}\kappa_{ijk}+\frac{\left(-i\right)^{4}}{4!}\stackrel[i,j,k,l=1]{n}{\sum}u_{i}u_{j}u_{k}u_{l}\kappa_{ijkl}
\]

\begin{equation}
+\frac{\left(-i\right)^{5}}{5!}\stackrel[i,j,k,l,m=1]{n}{\sum}u_{i}u_{j}u_{k}u_{l}u_{m}\kappa_{ijklm}+...+i\stackrel[i=1]{n-1}{\sum}u_{i}\stackrel[p=1]{\infty}{\sum}\frac{B_{n}^{(p)}}{p!}\left(t_{i}-t_{n}\right)^{p}.\label{eq:46}
\end{equation}

\medskip{}

The first line of this equation is the Gaussian term. Applying the
Taylor expansion to the exponential term of the non-Gaussian part
(2nd and 3rd lines of (\ref{eq:46})), one can write:



\medskip{}

\[
\Pi_{\epsilon\rightarrow0}\left(\omega_{0}=0,\omega_{n};t_{n}\right)=\int_{-\infty}^{B_{n}}d\omega_{1}...\int_{-\infty}^{B_{n}}d\omega_{n-1}\int_{-\infty}^{\infty}Du\cdot
\]

\[
exp\left(i\stackrel[i=1]{n}{\sum}u_{i}\omega_{i}-i\stackrel[i=1]{n}{\sum}u_{i}\kappa_{i}-\frac{1}{2}\stackrel[i,j=1]{n}{\sum}u_{i}u_{j}\kappa_{ij}+i\stackrel[i=1]{n-1}{\sum}u_{i}\stackrel[p=1]{\infty}{\sum}\frac{B_{n}^{(p)}}{p!}\left(t_{i}-t_{n}\right)^{p}\right)
\]

\[
+\int_{-\infty}^{B_{n}}d\omega_{1}...\int_{-\infty}^{B_{n}}d\omega_{n-1}\int_{-\infty}^{\infty}Du\cdot
\]

\[
exp\left(i\stackrel[i=1]{n}{\sum}u_{i}\omega_{i}-i\stackrel[i=1]{n}{\sum}u_{i}\kappa_{i}-\frac{1}{2}\stackrel[i,j=1]{n}{\sum}u_{i}u_{j}\kappa_{ij}+i\stackrel[i=1]{n-1}{\sum}u_{i}\stackrel[p=1]{\infty}{\sum}\frac{B_{n}^{(p)}}{p!}\left(t_{i}-t_{n}\right)^{p}\right)\cdot
\]

\[
\left(\frac{\left(-i\right)^{3}}{3!}\stackrel[i,j,k=1]{n}{\sum}u_{i}u_{j}u_{k}\kappa_{ijk}+\frac{\left(-i\right)^{4}}{4!}\stackrel[i,j,k,l=1]{n}{\sum}u_{i}u_{j}u_{k}u_{l}\kappa_{ijkl}\right.
\]

\begin{equation}
\left.+\frac{\left(-i\right)^{5}}{5!}\stackrel[i,j,k,l,m=1]{n}{\sum}u_{i}u_{j}u_{k}u_{l}u_{m}\kappa_{ijklm}+....\right).\label{eq:50}
\end{equation}

\medskip{}

The summation term involving the barrier can also be expanded in Taylor
series. We also consider up to second order:

\medskip{}

\[
exp\left(i\stackrel[i=1]{n-1}{\sum}u_{i}\stackrel[p=1]{\infty}{\sum}\frac{B_{n}^{(p)}}{p!}\left(t_{i}-t_{n}\right)^{p}\right)\simeq1+i\stackrel[i=1]{n-1}{\sum}u_{i}\stackrel[p=1]{\infty}{\sum}\frac{B_{n}^{(p)}}{p!}\left(t_{i}-t_{n}\right)^{p}
\]

\begin{equation}
-\frac{1}{2}\stackrel[i,j=1]{n-1}{\sum}u_{i}u_{j}\stackrel[p,q=1]{\infty}{\sum}\frac{B_{n}^{(p)}B_{n}^{(q)}}{p!q!}\left(t_{i}-t_{n}\right)^{p}\left(t_{j}-t_{n}\right)^{q}+...\label{eq:51}
\end{equation}

\medskip{}

(\ref{eq:50}) can be rewritten as:

\medskip{}

\[
\Pi_{\epsilon\rightarrow0}\left(\omega_{0}=0,\omega_{n};t_{n}\right)=\int_{-\infty}^{B_{n}}d\omega_{1}...\int_{-\infty}^{B_{n}}d\omega_{n-1}\int_{-\infty}^{\infty}Du\cdot
\]

\[
exp\left(i\stackrel[i=1]{n}{\sum}u_{i}\omega_{i}-i\stackrel[i=1]{n}{\sum}u_{i}\kappa_{i}-\frac{1}{2}\stackrel[i,j=1]{n}{\sum}u_{i}u_{j}\kappa_{ij}\right)
\]

\[
+\int_{-\infty}^{B_{n}}d\omega_{1}...\int_{-\infty}^{B_{n}}d\omega_{n-1}\int_{-\infty}^{\infty}Du\cdot
\]

\[
exp\left(i\stackrel[i=1]{n}{\sum}u_{i}\omega_{i}-i\stackrel[i=1]{n}{\sum}u_{i}\kappa_{i}-\frac{1}{2}\stackrel[i,j=1]{n}{\sum}u_{i}u_{j}\kappa_{ij}\right)\cdot
\]

\[
\left(i\stackrel[i=1]{n-1}{\sum}u_{i}\stackrel[p=1]{\infty}{\sum}\frac{B_{n}^{(p)}}{p!}\left(t_{i}-t_{n}\right)^{p}\right)
\]

\[
+\int_{-\infty}^{B_{n}}d\omega_{1}...\int_{-\infty}^{B_{n}}d\omega_{n-1}\int_{-\infty}^{\infty}Du\cdot
\]

\[
exp\left(i\stackrel[i=1]{n}{\sum}u_{i}\omega_{i}-i\stackrel[i=1]{n}{\sum}u_{i}\kappa_{i}-\frac{1}{2}\stackrel[i,j=1]{n}{\sum}u_{i}u_{j}\kappa_{ij}\right)\cdot
\]

\[
\left(-\frac{1}{2}\stackrel[i,j=1]{n-1}{\sum}u_{i}u_{j}\stackrel[p,q=1]{\infty}{\sum}\frac{B_{n}^{(p)}B_{n}^{(q)}}{p!q!}\left(t_{i}-t_{n}\right)^{p}\left(t_{j}-t_{n}\right)^{q}\right)
\]

\[
+\int_{-\infty}^{B_{n}}d\omega_{1}...\int_{-\infty}^{B_{n}}d\omega_{n-1}\int_{-\infty}^{\infty}Du\cdot
\]

\[
exp\left(i\stackrel[i=1]{n}{\sum}u_{i}\omega_{i}-i\stackrel[i=1]{n}{\sum}u_{i}\kappa_{i}-\frac{1}{2}\stackrel[i,j=1]{n}{\sum}u_{i}u_{j}\kappa_{ij}+i\stackrel[i=1]{n-1}{\sum}u_{i}\stackrel[p=1]{\infty}{\sum}\frac{B_{n}^{(p)}}{p!}\left(t_{i}-t_{n}\right)^{p}\right)\cdot
\]

\[
\left(\frac{\left(-i\right)^{3}}{3!}\stackrel[i,j,k=1]{n}{\sum}u_{i}u_{j}u_{k}\kappa_{ijk}+\frac{\left(-i\right)^{4}}{4!}\stackrel[i,j,k,l=1]{n}{\sum}u_{i}u_{j}u_{k}u_{l}\kappa_{ijkl}\right.
\]

\begin{equation}
\left.+\frac{\left(-i\right)^{5}}{5!}\stackrel[i,j,k,l,m=1]{n}{\sum}u_{i}u_{j}u_{k}u_{l}u_{m}\kappa_{ijklm}+....\right).\label{eq:52}
\end{equation}

\medskip{}

Using $\Pi_{\epsilon\rightarrow0}^{gm}\left(\omega_{0},\omega_{n};t_{n}\right)$
as given by (\ref{eq:PiGaussianaBarreira}), we can decompose $\Pi_{\epsilon\rightarrow0}\left(\omega_{0}=0,\omega_{n};t_{n}\right)$
as\medskip{}

\[
\Pi_{\epsilon\rightarrow0}\left(\omega_{0}=0,\omega_{n};t_{n}\right)=\Pi_{\epsilon\rightarrow0}^{gm}\left(\omega_{0},\omega_{n};t_{n}\right)+\Pi_{\epsilon\rightarrow0}^{(1)}\left(\omega_{n},t_{n}\right)+\Pi_{\epsilon\rightarrow0}^{(2)}\left(\omega_{n},t_{n}\right)
\]

\[
+\int_{-\infty}^{B_{n}}d\omega_{1}...\int_{-\infty}^{B_{n}}d\omega_{n-1}\int_{-\infty}^{\infty}Du\cdot
\]

\[
exp\left(i\stackrel[i=1]{n}{\sum}u_{i}\omega_{i}-i\stackrel[i=1]{n}{\sum}u_{i}\kappa_{i}-\frac{1}{2}\stackrel[i,j=1]{n}{\sum}u_{i}u_{j}\kappa_{ij}+i\stackrel[i=1]{n-1}{\sum}u_{i}\stackrel[p=1]{\infty}{\sum}\frac{B_{n}^{(p)}}{p!}\left(t_{i}-t_{n}\right)^{p}\right)\cdot
\]

\[
\left(\frac{\left(-i\right)^{3}}{3!}\stackrel[i,j,k=1]{n}{\sum}u_{i}u_{j}u_{k}\kappa_{ijk}+\frac{\left(-i\right)^{4}}{4!}\stackrel[i,j,k,l=1]{n}{\sum}u_{i}u_{j}u_{k}u_{l}\kappa_{ijkl}\right.
\]

\begin{equation}
\left.+\frac{\left(-i\right)^{5}}{5!}\stackrel[i,j,k,l,m=1]{n}{\sum}u_{i}u_{j}u_{k}u_{l}u_{m}\kappa_{ijklm}+....\right),\label{eq:195}
\end{equation}
where the gaussian markovian piece $\Pi_{\epsilon\rightarrow0}^{gm}\left(\omega_{0},\omega_{n};t_{n}\right)$
was already given in Eq.(\ref{eq:distB_BS}).

\medskip{}
 The remainder of this Section is devoted to the computation of the
different terms in Equation (\ref{eq:195}).

\medskip{}
 Consider (\ref{eq:Wgaussiana}):

\medskip{}

\[
W^{gm}\left(\omega_{0},\omega_{1},...,\omega_{n};t_{n}\right)=\int_{-\infty}^{\infty}Du\cdot exp\left(i\stackrel[j=1]{n}{\sum}u_{j}\omega_{j}-i\stackrel[i=1]{n}{\sum}u_{i}\kappa_{i}-\frac{1}{2}\stackrel[i,j=1]{n}{\sum}u_{i}u_{j}\kappa_{ij}\right)
\]

\begin{equation}
\equiv\int_{-\infty}^{\infty}Du\cdot exp\left(Z^{gm}\right)
\end{equation}
\medskip{}

\noindent with

\medskip{}

\begin{equation}
Z^{gm}=i\stackrel[j=1]{n}{\sum}u_{j}\omega_{j}-i\stackrel[i=1]{n}{\sum}u_{i}\kappa_{i}-\frac{1}{2}\stackrel[i,j=1]{n}{\sum}u_{i}u_{j}\kappa_{ij}.
\end{equation}

\medskip{}

Defining $\partial/\partial\omega_{i}=\partial_{i}$, we note that

\medskip{}

\begin{equation}
iu_{i}e^{iu_{i}\omega_{i}}=\partial_{i}e^{iu_{i}\omega_{i}}.\label{eq:55}
\end{equation}

\medskip{}

Then, the second term of (\ref{eq:51}), in the first term of (\ref{eq:50})
can be rewritten as\medskip{}

\[
\Pi_{\epsilon\rightarrow0}^{(1)}\left(\omega_{n},t_{n}\right)=\int_{-\infty}^{B_{n}}d\omega_{1}...\int_{-\infty}^{B_{n}}d\omega_{n-1}
\]

\[
\left[\int_{-\infty}^{\infty}Du\cdot\left(i\stackrel[i=1]{n-1}{\sum}u_{i}\stackrel[p=1]{\infty}{\sum}\frac{B_{n}^{(p)}}{p!}\left(t_{i}-t_{n}\right)^{p}\cdot\right.\right.
\]

\[
\left.\left.exp\left(i\stackrel[i=1]{n}{\sum}u_{i}\omega_{i}-i\stackrel[i=1]{n}{\sum}u_{i}\kappa_{i}-\frac{1}{2}\stackrel[i,j=1]{n}{\sum}u_{i}u_{j}\kappa_{ij}\right)\right)\right]
\]

\begin{equation}
=\int_{-\infty}^{B_{n}}d\omega_{1}...\int_{-\infty}^{B_{n}}d\omega_{n-1}\left[\stackrel[i=1]{n-1}{\sum}\stackrel[p=1]{\infty}{\sum}\frac{B_{n}^{(p)}}{p!}\left(t_{i}-t_{n}\right)^{p}\cdot\partial_{i}W^{gm}\left(\omega_{0},\omega_{1},...,\omega_{n};t_{n}\right)\right].\label{eq:56}
\end{equation}

\medskip{}

The third term of (\ref{eq:51}) also in the first term of (\ref{eq:50}),
observing the rule (\ref{eq:55}),

\medskip{}

\[
\Pi_{\epsilon\rightarrow0}^{(2)}\left(\omega_{n},t_{n}\right)=\int_{-\infty}^{B_{n}}d\omega_{1}...\int_{-\infty}^{B_{n}}d\omega_{n-1}
\]

\[
\left[\int_{-\infty}^{\infty}Du\cdot\left(-\frac{1}{2}\stackrel[i,j=1]{n-1}{\sum}u_{i}u_{j}\stackrel[p,q=1]{\infty}{\sum}\frac{B_{n}^{(p)}B_{n}^{(q)}}{p!q!}\left(t_{i}-t_{n}\right)^{p}\left(t_{j}-t_{n}\right)^{q}\cdot\right.\right.
\]

\[
\left.\left.exp\left(i\stackrel[i=1]{n}{\sum}u_{i}\omega_{i}-i\stackrel[i=1]{n}{\sum}u_{i}\kappa_{i}-\frac{1}{2}\stackrel[i,j=1]{n}{\sum}u_{i}u_{j}\kappa_{ij}\right)\right)\right]
\]

\[
=\int_{-\infty}^{B_{n}}d\omega_{1}...\int_{-\infty}^{B_{n}}d\omega_{n-1}\cdot
\]

\begin{equation}
\left[\left(\frac{1}{2}\stackrel[i,j=1]{n-1}{\sum}\stackrel[p,q=1]{\infty}{\sum}\frac{B_{n}^{(p)}B_{n}^{(q)}}{p!q!}\left(t_{i}-t_{n}\right)^{p}\left(t_{j}-t_{n}\right)^{q}\cdot\left.\partial_{j}\partial_{i}W^{gm}\left(\omega_{0},\omega_{1},...,\omega_{n};t_{n}\right)\right)\right.\right].\label{eq:57}
\end{equation}

\medskip{}

To evaluate (\ref{eq:56}), we use (\ref{eq:100}) and the transformation
(\ref{eq:133}):

\medskip{}

\begin{equation}
\Pi_{\epsilon\rightarrow0}^{(1)}\left(\omega_{n},t_{n}\right)=\frac{1}{\epsilon}\int_{0}^{t_{n}}dt_{i}\left[\stackrel[p=1]{\infty}{\sum}\frac{B_{n}^{(p)}}{p!}\left(t_{i}-t_{n}\right)^{p}\cdot\left(\Pi_{\epsilon}^{gm}\left(\omega_{0},\omega_{c};t_{i}\right)\Pi_{\epsilon}^{gm}\left(\omega_{c},\omega_{n};t_{n}-t_{i}\right)\right)\right].
\end{equation}

\medskip{}

Now we use (\ref{eq:143}) and (\ref{eq:144}), with $\omega_{0}=0$:

\medskip{}

\[
\Pi_{\epsilon\rightarrow0}^{(1)}\left(\omega_{n},t_{n}\right)=\frac{1}{\epsilon}\int_{0}^{t_{n}}dt_{i}\left[\stackrel[p=1]{\infty}{\sum}\frac{B_{n}^{(p)}}{p!}\left(t_{i}-t_{n}\right)^{p}\cdot\right.
\]

\selectlanguage{brazil}%
\[
\left(\sqrt{\epsilon}\frac{1}{\sqrt{\pi}}e^{\alpha\left(\omega_{n}-B_{n}\right)}\frac{B_{n}-\omega_{0}}{t_{i}^{3/2}}e^{-\frac{\left[B_{n}-\omega_{0}-\alpha t_{i}\right]^{2}}{2t_{i}}}\right)\cdot
\]

\[
\left.\left(\sqrt{\epsilon}\frac{1}{\sqrt{\pi}}e^{\alpha\left(\omega_{n}-B_{n}\right)}\frac{B_{n}-\omega_{n}}{\left(t_{n}-t_{i}\right)^{3/2}}e^{-\frac{\left[\left(B_{n}-\omega_{n}\right)-\alpha\left(t_{n}-t_{i}\right)\right]^{2}}{2\left(t_{n}-t_{i}\right)}}\right)\right]
\]

\[
=\frac{\left(B_{n}-\omega_{n}\right)\left(B_{n}-\omega_{0}\right)}{\pi}e^{2\alpha\left(\omega_{n}-B_{n}\right)}\stackrel[p=1]{\infty}{\sum}\frac{\left(-1\right)^{p}B_{n}^{(p)}}{p!}\cdot
\]

\begin{equation}
\cdot\int_{0}^{t_{n}}dt_{i}\frac{\left(t_{n}-t_{i}\right)^{p-\frac{3}{2}}}{t_{i}^{3/2}}e^{-\frac{\left[\left(B_{n}-\omega_{0}\right)-\alpha t_{i}\right]^{2}}{2t_{i}}}e^{-\frac{\left[\left(B_{n}-\omega_{n}\right)-\alpha\left(t_{n}-t_{i}\right)\right]^{2}}{2\left(t_{n}-t_{i}\right)}}\label{eq:171}
\end{equation}

\selectlanguage{english}%
To solve $\Pi_{\epsilon\rightarrow0}^{(1)}\left(\omega_{n},t_{n}\right)$
and $\Pi_{\epsilon\rightarrow0}^{(2)}\left(\omega_{n},t_{n}\right)$,
we adopt at this point an approximation due to \cite{ST02}, abbreviated
by ``ST'', which implies that $t_{n}\gg t_{i}$ in higher than first
order derivatives in (\ref{eq:39}):

\medskip{}

\begin{equation}
\left(t_{n}-t_{i}\right)^{p-1}\simeq\left(t_{n}\right)^{p-1}\label{eq:172}
\end{equation}
to obtain






\medskip{}

\begin{equation}
\Pi_{\epsilon\rightarrow0}^{(1)}\left(\omega_{n},t_{n}\right)=\frac{2\left(B_{n}-\omega_{n}\right)}{\sqrt{2\pi}t_{n}^{3/2}}e^{-\frac{1}{2t_{n}}\left[\alpha t_{n}-\left(2B_{n}-\omega_{0}-\omega_{n}\right)\right]^{2}}e^{2\alpha\left(\omega_{n}-B_{n}\right)}\stackrel[p=1]{\infty}{\sum}\frac{\left(-1\right)^{p}B_{n}^{(p)}}{p!}.\label{eq:174}
\end{equation}

\medskip{}

We develop now (\ref{eq:57}), $\Pi_{\epsilon\rightarrow0}^{(2)}\left(\omega_{n},t_{n}\right)$,
also using (ST), (\ref{eq:172}). To do so, we will use (\ref{eq:107}),
with (\ref{eq:109}):\medskip{}

\[
\stackrel[i,j=1]{n-1}{\sum}\partial_{i}\partial_{j}=2\underset{i<j}{\sum}\partial_{i}\partial_{j}+\stackrel[i=1]{n-1}{\sum}\partial_{i}^{2}
\]

\begin{equation}
=2\stackrel[j=2]{n-1}{\sum}\stackrel[i=1]{j-1}{\sum}\partial_{i}\partial_{j}+\stackrel[i=1]{n-1}{\sum}\partial_{i}^{2}
\end{equation}

\medskip{}

As demonstrated in \ref{sec:An=00003D00003D0000E1lise-de-termos},
there are divergent terms in (\ref{eq:109}), which cancel with the
second term of RHS (\ref{eq:107}). Specifically, there is a divergent
term, when $t_{i}=t_{j}$, at the very beginning of the summation
$\stackrel[j=1]{n-1}{\sum}$, making the denominator zero, but its
contribution is canceled by the second term of the RHS (\ref{eq:107}),
which is fully divergent, that is, it is not just one part of it that
diverges. As a consequence, we can write:

\medskip{}

\begin{equation}
\stackrel[i,j=1]{n-1}{\sum}\partial_{i}\partial_{j}\rightarrow2\stackrel[i=1]{n-2}{\sum}\stackrel[j=i+1]{n-1}{\sum}\rightarrow2\frac{1}{\epsilon^{2}}\intop_{0}^{t_{n}}dt_{i}\intop_{t_{i}}^{t_{n}}dt_{j}.\label{eq:182}
\end{equation}

\medskip{}

Using (\ref{eq:103}), (\ref{eq:148}), (\ref{eq:144}) and (\ref{eq:143})
to compute (\ref{eq:57}):

\medskip{}

\[
\int_{-\infty}^{B_{n}}d\omega_{1}...\int_{-\infty}^{B_{n}}d\omega_{n-1}\cdot
\]

\[
\left[\left(\frac{1}{2}\stackrel[i,j=1]{n-1}{\sum}\stackrel[p,q=1]{\infty}{\sum}\frac{B_{n}^{(p)}B_{n}^{(q)}}{p!q!}\left(t_{i}-t_{n}\right)^{p}\left(t_{j}-t_{n}\right)^{q}\cdot\left.\partial_{j}\partial_{i}W^{gm}\left(\omega_{0},\omega_{1},...,\omega_{n};t_{n}\right)\right)\right.\right]
\]

\[
=\int_{0}^{t_{n}}dt_{i}\int_{t_{i}}^{t_{n}}dt_{j}\stackrel[p,q=1]{\infty}{\sum}\frac{B_{n}^{(p)}B_{n}^{(q)}}{p!q!}\left(t_{i}-t_{n}\right)^{p}\left(t_{j}-t_{n}\right)^{q}\frac{\left(t_{n}-t_{i}\right)}{\left(t_{n}-t_{i}\right)}\frac{\left(t_{n}-t_{j}\right)}{\left(t_{n}-t_{j}\right)}\cdot
\]

\[
\frac{1}{\pi}\frac{1}{\sqrt{2\pi}}\frac{B_{n}-\omega_{0}}{t_{i}^{3/2}}\frac{e^{-\frac{\alpha^{2}\left(t_{j}-t_{i}\right)}{2}}}{\left(t_{j}-t_{i}\right)^{3/2}}\frac{B_{n}-\omega_{n}}{\left(t_{n}-t_{j}\right)^{3/2}}e^{-\frac{\left[\left(B_{n}-\omega_{0}\right)-\alpha t_{i}\right]^{2}}{2t_{i}}}e^{2\alpha\left(\omega_{n}-B_{n}\right)}e^{-\frac{\left[\left(B_{n}-\omega_{n}\right)-\alpha\left(t_{n}-t_{j}\right)\right]^{2}}{2\left(t_{n}-t_{j}\right)}}
\]

\[
=\frac{\left(B_{n}-\omega_{0}\right)\left(B_{n}-\omega_{n}\right)}{\pi\sqrt{2\pi}}\stackrel[p,q=1]{\infty}{\sum}\frac{B_{n}^{(p)}B_{n}^{(q)}}{p!q!}\left(-t_{n}\right)^{p-1}\left(-t_{n}\right)^{q-1}e^{2\alpha\left(\omega_{n}-B_{n}\right)}\cdot
\]

\[
\int_{0}^{t_{n}}dt_{i}\frac{e^{\frac{\alpha^{2}t_{i}}{2}}\left(t_{n}-t_{i}\right)e^{-\frac{\left[\left(B_{n}-\omega_{0}\right)-\alpha t_{i}\right]^{2}}{2t_{i}}}}{t_{i}^{3/2}}\int_{t_{i}}^{t_{n}}dt_{j}\frac{e^{-\frac{\alpha^{2}t_{j}}{2}}e^{-\frac{\left[\left(B_{n}-\omega_{n}\right)-\alpha\left(t_{n}-t_{j}\right)\right]^{2}}{2\left(t_{n}-t_{j}\right)}}}{\left(t_{j}-t_{i}\right)^{3/2}\left(t_{n}-t_{j}\right)^{1/2}}.
\]

\[
=\frac{\left(B_{n}-\omega_{0}\right)\left(B_{n}-\omega_{n}\right)}{\pi\sqrt{2\pi}}\stackrel[p,q=1]{\infty}{\sum}\frac{B_{n}^{(p)}B_{n}^{(q)}}{p!q!}\left(-t_{n}\right)^{p-1}\left(-t_{n}\right)^{q-1}e^{\alpha\left(\omega_{n}-B_{n}\right)}e^{\alpha\left(B_{n}-\omega_{0}\right)}e^{-\frac{\alpha^{2}t_{n}}{2}}\cdot
\]

\begin{equation}
\int_{0}^{t_{n}}dt_{i}\frac{\left(t_{n}-t_{i}\right)e^{-\frac{\left(B_{n}-\omega_{0}\right)^{2}}{2t_{i}}}}{t_{i}^{3/2}}\int_{t_{i}}^{t_{n}}dt_{j}\frac{e^{-\frac{\left(B_{n}-\omega_{n}\right)^{2}}{2\left(t_{n}-t_{j}\right)}}}{\left(t_{j}-t_{i}\right)^{3/2}\left(t_{n}-t_{j}\right)^{1/2}}.\label{eq:185}
\end{equation}
\medskip{}

Thus, 

\medskip{}

\[
\Pi_{\epsilon\rightarrow0}^{(2)}\left(\omega_{n},t_{n}\right)=
\]

\begin{equation}
\frac{-2\left(B_{n}-\omega_{n}\right)^{2}}{\sqrt{2\pi}t_{n}^{5/2}}e^{-\frac{\left[\left(2B_{n}-\omega_{0}-\omega_{n}\right)-\alpha t_{n}\right]^{2}}{2t_{n}}}e^{2\alpha\left(\omega_{n}-B_{n}\right)}\left[\stackrel[p=1]{\infty}{\sum}\frac{\left(-t_{n}\right)^{p}}{p!}B_{n}^{(p)}\right]^{2}.\label{eq:193}
\end{equation}

\medskip{}

Denoting by $W^{mb}$ the following expression, where$"mb"$ refers
to \emph{moving barrier}:

\medskip{}

\[
W^{mb}\left(\omega_{0},\omega_{1},...,\omega_{n};t_{n}\right)=\int_{-\infty}^{\infty}Du\cdot
\]

\begin{equation}
exp\left(i\stackrel[i=1]{n}{\sum}u_{i}\omega_{i}-i\stackrel[i=1]{n}{\sum}u_{i}\kappa_{i}-\frac{1}{2}\stackrel[i,j=1]{n}{\sum}u_{i}u_{j}\kappa_{ij}+i\stackrel[i=1]{n-1}{\sum}u_{i}\stackrel[p=1]{\infty}{\sum}\frac{B_{n}^{(p)}}{p!}\left(t_{i}-t_{n}\right)^{p}\right).\label{eq:196}
\end{equation}

\medskip{}

(\ref{eq:196}) is in the second term $\Pi_{\epsilon\rightarrow0}^{(1)}\left(\omega_{n},t_{n}\right)$
of (\ref{eq:52}), that we have just computed. We denote the first
line of (\ref{eq:195}) by:

\medskip{}

\[
\Pi_{\epsilon\rightarrow0}^{mb}=\int_{-\infty}^{B_{n}}d\omega_{1}...\int_{-\infty}^{B_{n}}d\omega_{n-1}W^{mb}\left(\omega_{0},\omega_{1},...,\omega_{n};t_{n}\right)=\Pi_{\epsilon\rightarrow0}^{gm}\left(\omega_{0},\omega_{n};t_{n}\right)+
\]

\begin{equation}
\Pi_{\epsilon\rightarrow0}^{(1)}\left(\omega_{n},t_{n}\right)+\Pi_{\epsilon\rightarrow0}^{(2)}\left(\omega_{n},t_{n}\right).\label{eq:197}
\end{equation}

\medskip{}

Now consider the remaining term of (\ref{eq:195}):

\medskip{}

\[
\int_{-\infty}^{B_{n}}d\omega_{1}...\int_{-\infty}^{B_{n}}d\omega_{n-1}\int_{-\infty}^{\infty}Du\cdot
\]

\[
exp\left(i\stackrel[i=1]{n}{\sum}u_{i}\omega_{i}-i\stackrel[i=1]{n}{\sum}u_{i}\kappa_{i}-\frac{1}{2}\stackrel[i,j=1]{n}{\sum}u_{i}u_{j}\kappa_{ij}+i\stackrel[i=1]{n-1}{\sum}u_{i}\stackrel[p=1]{\infty}{\sum}\frac{B_{n}^{(p)}}{p!}\left(t_{i}-t_{n}\right)^{p}\right)\cdot
\]

\[
\left(\frac{\left(-i\right)^{3}}{3!}\stackrel[i,j,k=1]{n}{\sum}u_{i}u_{j}u_{k}\kappa_{ijk}+\frac{\left(-i\right)^{4}}{4!}\stackrel[i,j,k,l=1]{n}{\sum}u_{i}u_{j}u_{k}u_{l}\kappa_{ijkl}\right.
\]

\begin{equation}
\left.+\frac{\left(-i\right)^{5}}{5!}\stackrel[i,j,k,l,m=1]{n}{\sum}u_{i}u_{j}u_{k}u_{l}u_{m}\kappa_{ijklm}+....\right)\label{eq:198}
\end{equation}

\medskip{}

\noindent with the relations\medskip{}

\[
\left(i\right)^{3}u_{i}u_{j}u_{k}exp\left(i\stackrel[i=1]{n}{\sum}u_{i}\omega_{i}\right)=\partial_{i}\partial_{j}\partial_{k}exp\left(i\stackrel[i=1]{n}{\sum}u_{i}\omega_{i}\right)
\]

\[
\left(i\right)^{4}u_{i}u_{j}u_{k}u_{l}exp\left(i\stackrel[i=1]{n}{\sum}u_{i}\omega_{i}\right)=\partial_{i}\partial_{j}\partial_{k}\partial_{l}exp\left(i\stackrel[i=1]{n}{\sum}u_{i}\omega_{i}\right)
\]

\begin{equation}
\left(i\right)^{5}u_{i}u_{j}u_{k}u_{l}u_{m}exp\left(i\stackrel[i=1]{n}{\sum}u_{i}\omega_{i}\right)=\partial_{i}\partial_{j}\partial_{k}\partial_{l}\partial_{m}exp\left(i\stackrel[i=1]{n}{\sum}u_{i}\omega_{i}\right),
\end{equation}
\medskip{}

\noindent remembering that $\partial_{i}=\partial/\partial\omega_{i}$.

\medskip{}

\[
\Pi_{\epsilon\rightarrow0}\left(\omega_{0}=0,\omega_{n};t_{n}\right)=\Pi_{\epsilon\rightarrow0}^{gm}\left(\omega_{0},\omega_{n};t_{n}\right)+\Pi_{\epsilon\rightarrow0}^{(1)}\left(\omega_{n},t_{n}\right)+\Pi_{\epsilon\rightarrow0}^{(2)}\left(\omega_{n},t_{n}\right)
\]

\[
-\frac{1}{3!}\int_{-\infty}^{B_{n}}d\omega_{1}...\int_{-\infty}^{B_{n}}d\omega_{n-1}\stackrel[i,j,k=1]{n}{\sum}\kappa_{ijk}\partial_{i}\partial_{j}\partial_{k}W^{mb}\left(\omega_{0},\omega_{1},...,\omega_{n};t_{n}\right)
\]

\[
+\frac{1}{4!}\int_{-\infty}^{B_{n}}d\omega_{1}...\int_{-\infty}^{B_{n}}d\omega_{n-1}\stackrel[i,j,k,l=1]{n}{\sum}\kappa_{ijkl}\partial_{i}\partial_{j}\partial_{k}\partial_{l}W^{mb}\left(\omega_{0},\omega_{1},...,\omega_{n};t_{n}\right)
\]

\begin{equation}
-\frac{1}{5!}\int_{-\infty}^{B_{n}}d\omega_{1}...\int_{-\infty}^{B_{n}}d\omega_{n-1}\stackrel[i,j,k,l,m=1]{n}{\sum}\kappa_{ijklm}\partial_{i}\partial_{j}\partial_{k}\partial_{l}\partial_{m}W^{mb}\left(\omega_{0},\omega_{1},...,\omega_{n};t_{n}\right)+....,\label{eq:200}
\end{equation}

\medskip{}

\noindent where $W^{mb}$ is given by (\ref{eq:196}).

At this point, we will take a step that is compatible with the stationarity
of the time series. The cumulants $\kappa_{ijk}$, $\kappa_{ijkl}$,
... refer to time-dependent variables and, without the stationarity
hypothesis, change over the interval $\left[0,t_{n}\right]$. If $t_{n}$
is small, we can expand the cumulant in a Taylor series around $t_{i}=t_{j}=...=t_{n}$.
For instance, $\kappa_{ijk}$ is expanded around $\kappa_{nnn}$,
which we denote by $\kappa_{3}$. Thus, we define the derivative in
the Taylor expansion of the cumulant (here we show the 5th one):

\medskip{}

\begin{equation}
G_{5}^{(p,q,r,s,t)}\left(t_{n}\right)=\frac{d^{p}}{dt_{i}^{p}}\frac{d^{q}}{dt_{j}^{q}}\frac{d^{r}}{dt_{k}^{r}}\frac{d^{s}}{dt_{l}^{s}}\frac{d^{t}}{dt_{m}^{t}}\left.\kappa_{ijklm}\right|_{i=j=k=l=m=n}.\label{eq:201}
\end{equation}
\medskip{}

\selectlanguage{brazil}%
and write:

\selectlanguage{english}%
\noindent \medskip{}

\[
\kappa_{ijklm}=\stackrel[p,q,r,s,t=0]{\infty}{\sum}\frac{\left(-1\right)^{p+q+r+s+t}}{p!q!r!s!t!}\left(t_{n}-t_{i}\right)^{p}\left(t_{n}-t_{j}\right)^{q}\left(t_{n}-t_{k}\right)^{r}\cdot
\]

\begin{equation}
\left(t_{n}-t_{l}\right)^{s}\left(t_{n}-t_{m}\right)^{t}G_{5}^{(p,q,r,s,t)}\left(t_{n}\right).\label{eq:202}
\end{equation}

\medskip{}

The relevant contribution is $p=q=r=s=t=0$. Then, the summation of
(\ref{eq:202}) reduces to $\kappa_{5}$ and the summation

\medskip{}

\begin{equation}
\stackrel[i,j,k,l,m=1]{n}{\sum}\kappa_{ijklm}\partial_{i}\partial_{j}\partial_{k}\partial_{l}\partial_{m}
\end{equation}

\medskip{}

\noindent becomes

\medskip{}

\begin{equation}
\kappa_{5}\stackrel[i,j,k,l,m=1]{n}{\sum}\partial_{i}\partial_{j}\partial_{k}\partial_{l}\partial_{m}.\label{eq:204}
\end{equation}

\medskip{}

The other summation terms in cumulants in (\ref{eq:200}) have analogous
expressions to (\ref{eq:204}). To sum up, what we have done is to
approximate time-varying cumulants to their final values. This converges
to the stationary case, where moments (cumulants) are constant, just
depending on the size of the analyzed period.

As (\ref{eq:204}) sugests, we need to value objects such as:

\medskip{}

\begin{equation}
\begin{array}{cc}
cumulant & object\\
3 & \stackrel[i,j,k=1]{n}{\sum}\partial_{i}\partial_{j}\partial_{k}\\
4 & \stackrel[i,j,k,l=1]{n}{\sum}\partial_{i}\partial_{j}\partial_{k}\partial_{l}\\
5 & \stackrel[i,j,k,l,m=1]{n}{\sum}\partial_{i}\partial_{j}\partial_{k}\partial_{l}\partial_{m}\\
\vdots & \vdots
\end{array}\label{eq:205}
\end{equation}

\medskip{}

These summations can be expressed by components whose coefficients
are given by the Pascal triangle. For the cumulants we have made explicit,
we have:

\medskip{}

\begin{equation}
\stackrel[i,j,k=1]{n}{\sum}\partial_{i}\partial_{j}\partial_{k}=\partial_{n}^{3}+3\stackrel[i,j=1]{n-1}{\sum}\partial_{i}\partial_{j}\partial_{n}+3\stackrel[i=1]{n-1}{\sum}\partial_{i}\partial_{n}^{2}+\stackrel[i,j,k=1]{n-1}{\sum}\partial_{i}\partial_{j}\partial_{k}\label{eq:206}
\end{equation}

\[
\stackrel[i,j,k,l=1]{n}{\sum}\partial_{i}\partial_{j}\partial_{k}\partial_{l}=\partial_{n}^{4}+4\stackrel[i,j,k=1]{n-1}{\sum}\partial_{i}\partial_{j}\partial_{k}\partial_{n}
\]

\begin{equation}
+6\stackrel[i,j=1]{n-1}{\sum}\partial_{i}\partial_{j}\partial_{n}^{2}+4\stackrel[i=1]{n-1}{\sum}\partial_{i}\partial_{n}^{3}+\stackrel[i,j,k,l=1]{n-1}{\sum}\partial_{i}\partial_{j}\partial_{k}\partial_{l}\label{eq:207}
\end{equation}

\[
\stackrel[i,j,k,l,m=1]{n}{\sum}\partial_{i}\partial_{j}\partial_{k}\partial_{l}\partial_{m}=\partial_{n}^{5}+5\stackrel[i,j,k,l=1]{n-1}{\sum}\partial_{i}\partial_{j}\partial_{k}\partial_{l}\partial_{n}+10\stackrel[i,j,k=1]{n-1}{\sum}\partial_{i}\partial_{j}\partial_{k}\partial_{n}^{2}
\]

\begin{equation}
+10\stackrel[i,j=1]{n-1}{\sum}\partial_{i}\partial_{j}\partial_{n}^{3}+5\stackrel[i=1]{n-1}{\sum}\partial_{i}\partial_{n}^{4}+\stackrel[i,j,k,l,m=1]{n-1}{\sum}\partial_{i}\partial_{j}\partial_{k}\partial_{l}\partial_{m}.\label{eq:208}
\end{equation}

\medskip{}

As in (\ref{eq:106-0}) e (\ref{eq:111}), we get:

\medskip{}

\begin{equation}
\stackrel[i=1]{n-1}{\sum}\int_{-\infty}^{Bn}d\omega_{1}...d\omega_{n-1}\partial_{i}W^{mb}\left(\omega_{0},\omega_{1},...,\omega_{n};t_{n}\right)=\frac{\partial\Pi_{\epsilon\rightarrow0}^{mb}}{\partial B_{n}}\left(\omega_{0}=0,\omega_{n};t_{n}\right),\label{eq:214}
\end{equation}

\[
\vdots
\]

\[
\stackrel[i,j,k,l,m=1]{n-1}{\sum}\int_{-\infty}^{Bn}d\omega_{1}...d\omega_{n-1}\partial_{i}\partial_{j}\partial_{k}\partial_{l}\partial_{m}W^{mb}\left(\omega_{0},\omega_{1},...,\omega_{n};t_{n}\right)=
\]

\begin{equation}
\frac{\partial^{5}\Pi_{\epsilon\rightarrow0}^{mb}}{\partial B_{n}^{5}}\left(\omega_{0}=0,\omega_{n};t_{n}\right).\label{eq:215}
\end{equation}
\medskip{}

\noindent where $\Pi_{\epsilon\rightarrow0}^{mb}$ is given by (\ref{eq:197}).
Returning to (\ref{eq:200}),

\medskip{}

\[
\Pi_{\epsilon\rightarrow0}\left(\omega_{0}=0,\omega_{n};t_{n}\right)=\Pi_{\epsilon\rightarrow0}^{gm}\left(\omega_{0},\omega_{n};t_{n}\right)+\Pi_{\epsilon\rightarrow0}^{(1)}\left(\omega_{n},t_{n}\right)+\Pi_{\epsilon\rightarrow0}^{(2)}\left(\omega_{n},t_{n}\right)
\]

\[
-\frac{1}{3!}\kappa_{3}\int_{-\infty}^{B_{n}}d\omega_{1}...\int_{-\infty}^{B_{n}}d\omega_{n-1}\left\{ \left[\partial_{n}^{3}+3\stackrel[i,j=1]{n-1}{\sum}\partial_{i}\partial_{j}\partial_{n}\right.\right.
\]

\[
\left.\left.+3\stackrel[i=1]{n-1}{\sum}\partial_{i}\partial_{n}^{2}+\stackrel[i,j,k=1]{n-1}{\sum}\partial_{i}\partial_{j}\partial_{k}\right]W^{mb}\left(\omega_{0},\omega_{1},...,\omega_{n};t_{n}\right)\right\} 
\]

\[
+\frac{1}{4!}\kappa_{4}\int_{-\infty}^{B_{n}}d\omega_{1}...\int_{-\infty}^{B_{n}}d\omega_{n-1}\left\{ \left[\partial_{n}^{4}+4\stackrel[i,j,k=1]{n-1}{\sum}\partial_{i}\partial_{j}\partial_{k}\partial_{n}\right.\right.
\]

\[
\left.\left.+6\stackrel[i,j=1]{n-1}{\sum}\partial_{i}\partial_{j}\partial_{n}^{2}+4\stackrel[i=1]{n-1}{\sum}\partial_{i}\partial_{n}^{3}+\stackrel[i,j,k,l=1]{n-1}{\sum}\partial_{i}\partial_{j}\partial_{k}\partial_{l}\right]W^{mb}\left(\omega_{0},\omega_{1},...,\omega_{n};t_{n}\right)\right\} 
\]

\[
-\frac{1}{5!}\kappa_{5}\int_{-\infty}^{B_{n}}d\omega_{1}...\int_{-\infty}^{B_{n}}d\omega_{n-1}\left\{ \left[\partial_{n}^{5}+5\stackrel[i,j,k,l=1]{n-1}{\sum}\partial_{i}\partial_{j}\partial_{k}\partial_{l}\partial_{n}+10\stackrel[i,j,k=1]{n-1}{\sum}\partial_{i}\partial_{j}\partial_{k}\partial_{n}^{2}\right.\right.
\]

\begin{equation}
\left.\left.+10\stackrel[i,j=1]{n-1}{\sum}\partial_{i}\partial_{j}\partial_{n}^{3}+5\stackrel[i=1]{n-1}{\sum}\partial_{i}\partial_{n}^{4}+\stackrel[i,j,k,l,m=1]{n-1}{\sum}\partial_{i}\partial_{j}\partial_{k}\partial_{l}\partial_{m}\right]W^{mb}\left(\omega_{0},\omega_{1},...,\omega_{n};t_{n}\right)\right\} +...\label{eq:216}
\end{equation}

\medskip{}

The derivative operators acting in $W^{mb}\left(\omega_{0},\omega_{1},...,\omega_{n};t_{n}\right)$
follow the relations (\ref{eq:214}) and their analogues, noticing
that the operator $\partial_{n}^{...}$ exits the integral

\medskip{}

\[
\int_{-\infty}^{B_{n}}d\omega_{1}...\int_{-\infty}^{B_{n}}d\omega_{n-1}W^{mb}\left(\omega_{0},\omega_{1},...,\omega_{n};t_{n}\right)
\]

\medskip{}

\noindent because $\omega_{n}$ is not part of the integration. Thus,
making explicit some second order crossed terms, recovering (\ref{eq:50}),

\medskip{}

\[
\Pi_{\epsilon\rightarrow0}\left(\omega_{0}=0,\omega_{n};t_{n}\right)=\Pi_{\epsilon\rightarrow0}^{mb}\left(\omega_{n},t_{n}\right)
\]

\[
-\frac{1}{3!}\kappa_{3}\left\{ \frac{\partial^{3}}{\partial\omega_{n}^{3}}\Pi_{\epsilon\rightarrow0}^{mb}\left(\omega_{n},t_{n}\right)+3\frac{\partial^{2}}{\partial\omega_{n}^{2}}\frac{\partial}{\partial B_{n}}\Pi_{\epsilon\rightarrow0}^{mb}\left(\omega_{n},t_{n}\right)\right.
\]

\[
\left.+3\frac{\partial}{\partial\omega_{n}}\frac{\partial^{2}}{\partial B_{n}^{2}}\Pi_{\epsilon\rightarrow0}^{mb}\left(\omega_{n},t_{n}\right)+\frac{\partial^{3}}{\partial B_{n}^{3}}\Pi_{\epsilon\rightarrow0}^{mb}\left(\omega_{n},t_{n}\right)\right\} 
\]

\[
+\frac{1}{4!}\kappa_{4}\left\{ \frac{\partial^{4}}{\partial\omega_{n}^{4}}\Pi_{\epsilon\rightarrow0}^{mb}\left(\omega_{n},t_{n}\right)+4\frac{\partial^{3}}{\partial\omega_{n}^{3}}\frac{\partial}{\partial B_{n}}\Pi_{\epsilon\rightarrow0}^{mb}\left(\omega_{n},t_{n}\right)+6\frac{\partial^{2}}{\partial\omega_{n}^{2}}\frac{\partial^{2}}{\partial B_{n}^{2}}\Pi_{\epsilon\rightarrow0}^{mb}\left(\omega_{n},t_{n}\right)\right.
\]

\[
\left.+4\frac{\partial}{\partial\omega_{n}}\frac{\partial^{3}}{\partial B_{n}^{3}}\Pi_{\epsilon\rightarrow0}^{mb}\left(\omega_{n},t_{n}\right)+\frac{\partial^{4}}{\partial B_{n}^{4}}\Pi_{\epsilon\rightarrow0}^{mb}\left(\omega_{n},t_{n}\right)\right\} 
\]

\[
-\frac{1}{5!}\kappa_{5}\left\{ \frac{\partial^{5}}{\partial\omega_{n}^{5}}\Pi_{\epsilon\rightarrow0}^{mb}\left(\omega_{n},t_{n}\right)+5\frac{\partial^{4}}{\partial\omega_{n}^{4}}\frac{\partial}{\partial B_{n}}\Pi_{\epsilon\rightarrow0}^{mb}\left(\omega_{n},t_{n}\right)+10\frac{\partial^{3}}{\partial\omega_{n}^{3}}\frac{\partial^{2}}{\partial B_{n}^{2}}\Pi_{\epsilon\rightarrow0}^{mb}\left(\omega_{n},t_{n}\right)\right.
\]

\[
\left.+10\frac{\partial^{2}}{\partial\omega_{n}^{2}}\frac{\partial^{3}}{\partial B_{n}^{3}}\Pi_{\epsilon\rightarrow0}^{mb}\left(\omega_{n},t_{n}\right)+5\frac{\partial}{\partial\omega_{n}}\frac{\partial^{4}}{\partial B_{n}^{4}}\Pi_{\epsilon\rightarrow0}^{mb}\left(\omega_{n},t_{n}\right)+\frac{\partial^{5}}{\partial B_{n}^{5}}\Pi_{\epsilon\rightarrow0}^{mb}\left(\omega_{n},t_{n}\right)\right\} +...
\]

\[
+\left[\frac{1}{2}\cdot\left(\frac{1}{3!}\kappa_{3}\right)^{2}+\frac{1}{6!}\kappa_{6}\right]\left\{ \frac{\partial^{6}}{\partial\omega_{n}^{6}}\Pi_{\epsilon\rightarrow0}^{mb}\left(\omega_{n},t_{n}\right)\right.
\]

\[
\left.+6\frac{\partial^{5}}{\partial\omega_{n}^{5}}\frac{\partial}{\partial B_{n}}\Pi_{\epsilon\rightarrow0}^{mb}\left(\omega_{n},t_{n}\right)+...+\frac{\partial^{6}}{\partial B_{n}^{6}}\Pi_{\epsilon\rightarrow0}^{mb}\left(\omega_{n},t_{n}\right)\right\} 
\]

\[
-\left[\left(\frac{1}{3!}\kappa_{3}\right)\left(\frac{1}{4!}\kappa_{4}\right)+\frac{1}{7!}\kappa_{7}\right]\left\{ \frac{\partial^{7}}{\partial\omega_{n}^{7}}\Pi_{\epsilon\rightarrow0}^{mb}\left(\omega_{n},t_{n}\right)\right.
\]

\[
\left.+7\frac{\partial^{6}}{\partial\omega_{n}^{6}}\frac{\partial}{\partial B_{n}}\Pi_{\epsilon\rightarrow0}^{mb}\left(\omega_{n},t_{n}\right)+...+\frac{\partial^{7}}{\partial B_{n}^{7}}\Pi_{\epsilon\rightarrow0}^{mb}\left(\omega_{n},t_{n}\right)\right\} 
\]

\[
+\left[\frac{1}{2}\cdot\left(\frac{1}{4!}\kappa_{4}\right)^{2}-\left(\frac{1}{3!}\kappa_{3}\right)\left(\frac{1}{5!}\kappa_{5}\right)+\frac{1}{8!}\kappa_{8}\right]\left\{ \frac{\partial^{8}}{\partial\omega_{n}^{8}}\Pi_{\epsilon\rightarrow0}^{mb}\left(\omega_{n},t_{n}\right)\right.
\]

\[
\left.+8\frac{\partial^{7}}{\partial\omega_{n}^{7}}\frac{\partial}{\partial B_{n}}\Pi_{\epsilon\rightarrow0}^{mb}\left(\omega_{n},t_{n}\right)+...+\frac{\partial^{8}}{\partial B_{n}^{8}}\Pi_{\epsilon\rightarrow0}^{mb}\left(\omega_{n},t_{n}\right)\right\} 
\]

\[
-\left[\left(\frac{1}{4!}\kappa_{4}\right)\left(\frac{1}{5!}\kappa_{5}\right)+\left(\frac{1}{3!}\kappa_{3}\right)\left(\frac{1}{6!}\kappa_{6}\right)+\frac{1}{9!}\kappa_{9}\right]\left\{ \frac{\partial^{9}}{\partial\omega_{n}^{9}}\Pi_{\epsilon\rightarrow0}^{mb}\left(\omega_{n},t_{n}\right)\right.
\]

\begin{equation}
\left.+9\frac{\partial^{8}}{\partial\omega_{n}^{8}}\frac{\partial}{\partial B_{n}}\Pi_{\epsilon\rightarrow0}^{mb}\left(\omega_{n},t_{n}\right)+...+\frac{\partial^{9}}{\partial B_{n}^{9}}\Pi_{\epsilon\rightarrow0}^{mb}\left(\omega_{n},t_{n}\right)\right\} +...\label{eq:217-1}
\end{equation}

\medskip{}

We remember that $\Pi_{\epsilon\rightarrow0}^{mb}\left(\omega_{n},t_{n}\right)$
and is given by (\ref{eq:197}) and depends on $\Pi_{\epsilon\rightarrow0}^{gm}\left(\omega_{n},t_{n}\right)$,
$\Pi_{\epsilon\rightarrow0}^{(1)}\left(\omega_{n},t_{n}\right)$ e
$\Pi_{\epsilon\rightarrow0}^{(2)}\left(\omega_{n},t_{n}\right)$,
computed according to (\ref{eq:PiGaussianaBarreira}), (\ref{eq:174})
and (\ref{eq:193}), respectively.

If we keep up to the $\kappa_{3}$ term in (\ref{eq:217-1}), and
remove the drift, that is, impose $\alpha=0$ in the components $\Pi_{\epsilon\rightarrow0}^{mb}\left(\omega_{n},t_{n}\right)$,
we recover the results of \cite{MagIV}.

Instead of the (ST) hypothesis, next we develop alternative forms
for $\Pi_{\epsilon\rightarrow0}^{(1)}\left(\omega_{n},t_{n}\right)$
and $\Pi_{\epsilon\rightarrow0}^{(2)}\left(\omega_{n},t_{n}\right)$,
under the alternative hypothesis that the moving barrier evolves slowly
through time.

\medskip{}

\subsection{Slowly moving barrier\label{subsec:Slow-moving-barrier}}

\medskip{}

\hspace*{0.25in}In this section, the hypothesis is that the barrier
moves slowly with $t$. For future references we may call it \emph{abiabatic
barrier}. Under this assumption, we may discuss which terms in (\ref{eq:39})
are relevant. We will work with up to second order terms in the barrier
time derivatives: $\partial^{2}B_{n}/\partial t_{n}^{2}$ e $\left(\partial B_{n}/\partial t_{n}\right)^{2}$.
In a certain way, this hypothesis and the ST one work in the same
direction: in (\ref{eq:39}), the (ST) acts in the terms $\left(t_{i}-t_{n}\right)^{p}$,
while the slow varying barrier is related to the derivatives.

Under these conditions, (\ref{eq:56}) breakes into two terms, denoted
by the following expressions:\medskip{}

\begin{equation}
\Pi_{\epsilon\rightarrow0}^{(a)}\left(\omega_{n},t_{n}\right)=\stackrel[i=1]{n-1}{\sum}B_{n}^{'}\left(t_{i}-t_{n}\right)\int_{-\infty}^{B_{n}}d\omega_{1}...\int_{-\infty}^{B_{n}}d\omega_{n-1}\partial_{i}W^{gm}\left(\omega_{0},\omega_{1},...,\omega_{n};t_{n}\right)\label{eq:218-1}
\end{equation}

\begin{equation}
\Pi_{\epsilon\rightarrow0}^{(b)}\left(\omega_{n},t_{n}\right)=\frac{1}{2}\stackrel[i=1]{n-1}{\sum}B_{n}^{''}\left(t_{i}-t_{n}\right)^{2}\int_{-\infty}^{B_{n}}d\omega_{1}...\int_{-\infty}^{B_{n}}d\omega_{n-1}\partial_{i}W^{gm}\left(\omega_{0},\omega_{1},...,\omega_{n};t_{n}\right).\label{eq:219-1}
\end{equation}

\medskip{}

Equation (\ref{eq:57}) enables us to write:

\medskip{}

\[
\Pi_{\epsilon\rightarrow0}^{(c)}\left(\omega_{n},t_{n}\right)=\frac{1}{2}\stackrel[i=1]{n-1}{\sum}\left(B_{n}^{'}\right)^{2}\left(t_{i}-t_{n}\right)\left(t_{j}-t_{n}\right)\cdot
\]

\begin{equation}
\int_{-\infty}^{B_{n}}d\omega_{1}...\int_{-\infty}^{B_{n}}d\omega_{n-1}\partial_{i}\partial_{j}W^{gm}\left(\omega_{0},\omega_{1},...,\omega_{n};t_{n}\right),\label{eq:220-1}
\end{equation}

\medskip{}
 \medskip{}

\noindent where $B_{n}^{'}=\partial B\left(t_{n}\right)/\partial t_{n}$
and $B_{n}^{''}=\partial^{2}B\left(t_{n}\right)/\partial t_{n}^{2}$.

Starting by $\Pi_{\epsilon\rightarrow0}^{(a)}\left(\omega_{n},t_{n}\right)$,
from $\eqref{eq:171}$, with $p=1$,

\medskip{}

\[
\Pi_{\epsilon\rightarrow0}^{(a)}\left(\omega_{n},t_{n}\right)=
\]

\selectlanguage{brazil}%
\[
=-\frac{\left(B_{n}-\omega_{n}\right)\left(B_{n}-\omega_{0}\right)}{\pi}e^{2\alpha\left(\omega_{n}-B_{n}\right)}\cdot\frac{dB_{n}}{dt_{n}}
\]

\begin{equation}
\cdot\int_{0}^{t_{n}}dt_{i}\frac{\left(t_{n}-t_{i}\right)^{-\frac{1}{2}}}{t_{i}^{3/2}}e^{-\frac{\left[\left(B_{n}-\omega_{0}\right)-\alpha t_{i}\right]^{2}}{2t_{i}}}e^{-\frac{\left[\left(B_{n}-\omega_{n}\right)-\alpha\left(t_{n}-t_{i}\right)\right]^{2}}{2\left(t_{n}-t_{i}\right)}}.\label{eq:221-1}
\end{equation}

\selectlanguage{english}%
\medskip{}

\begin{equation}
\therefore\Pi_{\epsilon\rightarrow0}^{(a)}\left(\omega_{n},t_{n}\right)=-\sqrt{\frac{2}{\pi}}\frac{dB_{n}}{dt_{n}}\frac{\left(B_{n}-\omega_{n}\right)}{t_{n}^{1/2}}e^{2\alpha\left(\omega_{n}-B_{n}\right)}e^{-\frac{\left[\left(2B_{n}-\omega_{0}-\omega_{n}\right)-\alpha t_{n}\right]^{2}}{2t_{n}}}.\label{eq:228-1}
\end{equation}

\medskip{}

We now value $\Pi_{\epsilon\rightarrow0}^{(b)}\left(\omega_{n},t_{n}\right)$.
In (\ref{eq:171}), with $p=2$,\medskip{}

\[
\Pi_{\epsilon\rightarrow0}^{(b)}\left(\omega_{n},t_{n}\right)=
\]

\selectlanguage{brazil}%
\[
=\frac{\left(B_{n}-\omega_{n}\right)\left(B_{n}-\omega_{0}\right)}{\pi}e^{2\alpha\left(\omega_{n}-B_{n}\right)}\frac{1}{2}\cdot\frac{d^{2}B_{n}}{dt_{n}^{2}}
\]

\begin{equation}
\cdot\int_{0}^{t_{n}}dt_{i}\frac{\left(t_{n}-t_{i}\right)^{\frac{1}{2}}}{t_{i}^{3/2}}e^{-\frac{\left[\left(B_{n}-\omega_{0}\right)-\alpha t_{i}\right]^{2}}{2t_{i}}}e^{-\frac{\left[\left(B_{n}-\omega_{n}\right)-\alpha\left(t_{n}-t_{i}\right)\right]^{2}}{2\left(t_{n}-t_{i}\right)}}.\label{eq:229-1}
\end{equation}

\selectlanguage{english}%
\medskip{}

\[
\therefore\Pi_{\epsilon\rightarrow0}^{(b)}\left(\omega_{n},t_{n}\right)=
\]

\[
\frac{1}{2\pi}\frac{d^{2}B_{n}}{dt_{n}^{2}}\left(B_{n}-\omega_{n}\right)e^{2\alpha\left(\omega_{n}-B_{n}\right)}e^{\alpha\left(2B_{n}-\omega_{0}-\omega_{n}\right)-\frac{\alpha^{2}t_{n}}{2}}\cdot
\]

\begin{equation}
\left[\sqrt{2\pi}t_{n}^{1/2}e^{-\frac{\left(2B_{n}-\omega_{0}-\omega_{n}\right)^{2}}{2t_{n}}}-\pi\left(B_{n}-\omega_{0}\right)Erfc\left(\frac{2B_{n}-\omega_{0}-\omega_{n}}{\sqrt{2t_{n}}}\right)\right].\label{eq:233-1}
\end{equation}

\medskip{}

Finally, now computing $\Pi_{\epsilon\rightarrow0}^{(c)}\left(\omega_{n},t_{n}\right)$,
in (\ref{eq:220-1}), which refers to $\Pi_{\epsilon\rightarrow0}^{(2)}\left(\omega_{n},t_{n}\right)$,
from (\ref{eq:185}), with $p=q=1$:

\medskip{}

\[
\Pi_{\epsilon\rightarrow0}^{(c)}\left(\omega_{n},t_{n}\right)=\frac{\left(B_{n}-\omega_{0}\right)\left(B_{n}-\omega_{n}\right)}{\pi\sqrt{2\pi}}\left(\frac{dB_{n}}{dt_{n}}\right)^{2}e^{\alpha\left(\omega_{n}-B_{n}\right)}e^{\alpha\left(B_{n}-\omega_{0}\right)}e^{-\frac{\alpha^{2}t_{n}}{2}}\cdot
\]

\begin{equation}
\int_{0}^{t_{n}}dt_{i}\frac{\left(t_{n}-t_{i}\right)e^{-\frac{\left(B_{n}-\omega_{0}\right)^{2}}{2t_{i}}}}{t_{i}^{3/2}}\int_{t_{i}}^{t_{n}}dt_{j}\frac{e^{-\frac{\left(B_{n}-\omega_{n}\right)^{2}}{2\left(t_{n}-t_{j}\right)}}}{\left(t_{j}-t_{i}\right)^{3/2}\left(t_{n}-t_{j}\right)^{1/2}}.\label{eq:234-1}
\end{equation}

\medskip{}

\begin{equation}
\therefore\Pi_{\epsilon\rightarrow0}^{(c)}\left(\omega_{n},t_{n}\right)=-\sqrt{\frac{2}{\pi}}\frac{\left(B_{n}-\omega_{n}\right)^{2}}{t_{n}^{1/2}}\left(\frac{dB_{n}}{dt_{n}}\right)^{2}e^{2\alpha\left(\omega_{n}-B_{n}\right)}e^{-\frac{\left[\left(2B_{n}-\omega_{0}-\omega_{n}\right)-\alpha t_{n}\right]^{2}}{2t_{n}}}.\label{eq:235-1}
\end{equation}

\medskip{}

Therefore, in the hypothesis of this section, the equivalent to (\ref{eq:197})
is

\medskip{}

\[
\Pi_{\epsilon\rightarrow0}^{mb}\left(\omega_{n},t_{n}\right)=\int_{-\infty}^{B_{n}}d\omega_{1}...\int_{-\infty}^{B_{n}}d\omega_{n-1}W^{mb}\left(\omega_{0},\omega_{1},...,\omega_{n};t_{n}\right)=\Pi_{\epsilon\rightarrow0}^{gm}\left(\omega_{n};t_{n}\right)+
\]

\begin{equation}
\Pi_{\epsilon\rightarrow0}^{(a)}\left(\omega_{n},t_{n}\right)+\Pi_{\epsilon\rightarrow0}^{(b)}\left(\omega_{n},t_{n}\right)+\Pi_{\epsilon\rightarrow0}^{(c)}\left(\omega_{n},t_{n}\right),\label{eq:236-1}
\end{equation}

\medskip{}

\noindent with $\Pi_{\epsilon\rightarrow0}^{gm}\left(\omega_{n};t_{n}\right)$,
$\Pi_{\epsilon\rightarrow0}^{(a)}\left(\omega_{n},t_{n}\right)$,
$\Pi_{\epsilon\rightarrow0}^{(b)}\left(\omega_{n},t_{n}\right)$ and
$\Pi_{\epsilon\rightarrow0}^{(b)}\left(\omega_{n},t_{n}\right)$ given
by (\ref{eq:PiGaussianaBarreira}), (\ref{eq:228-1}), (\ref{eq:233-1})
and (\ref{eq:235-1}), respectively, taking $\omega_{0}=0$. (\ref{eq:217-1})
is still valid, with this specification for $\Pi_{\epsilon\rightarrow0}^{mb}\left(\omega_{n},t_{n}\right)$.

Therefore, in the presence of fixed or moving barriers, we have expressed
a distribution in terms of a cumulant expansion based on the Gaussian
distribution with fixed barrier.

\medskip{}

\medskip{}

\subsection{Non-Gaussian distribution with constant barrier}

\medskip{}

\hspace*{0.25in}The constant barrier $B_{n}=b=\frac{1}{\sigma}ln\frac{B}{S_{0}}$,
in a non-Gaussian distribution implies time derivatives of the barrier
equal to zero, in both (ST) and adiabatic barrier hypothesis:

\medskip{}

\begin{equation}
\underset{\frac{dB_{n}}{dt_{n}}\rightarrow0}{lim}\underset{\frac{d^{2}B_{n}}{dt_{n}^{2}}\rightarrow0}{lim}\Pi_{\epsilon\rightarrow0}^{mb}\left(\omega_{n},t_{n}\right)=\Pi_{\epsilon\rightarrow0}^{gm}\left(\omega_{n};t_{n}\right),\label{eq:238-1}
\end{equation}

\medskip{}

\noindent where $\Pi_{\epsilon\rightarrow0}^{gm}\left(\omega_{n};t_{n}\right)$
is given by (\ref{eq:PiGaussianaBarreira}). This distribution is
used in (\ref{eq:217-1}) to get $\Pi_{\epsilon\rightarrow0}\left(\omega_{0}=0,\omega_{n};t_{n}\right)$
related to this case. Non-Gaussian corrections emerge from the derivatives
related to $\omega_{n}$ and $B_{n}$.

\medskip{}

\medskip{}

\subsection{Linearly moving barrier}

\medskip{}

\hspace*{0.25in}The application to the moving barrier case depends
on derivatives, which appear in $\Pi_{\epsilon\rightarrow0}^{mb}\left(\omega_{n},t_{n}\right)$
components, with respect to the value of the barrier at the final
instant $t_{n}$. We will consider the case of a barrier tha evolves
linearly, to exemplify the use of the notation:

\medskip{}

\begin{equation}
B_{i}=B(t_{i})=B_{0}+\xi t_{i}.
\end{equation}

\medskip{}

\selectlanguage{brazil}%
We first find the value at $t_{n}$:

\selectlanguage{english}%
\medskip{}

\begin{equation}
B_{n}=B_{0}+\xi t_{n}
\end{equation}

\medskip{}

\noindent and write the derivatives. In our case, only the first order
one takes non-null value:

\medskip{}

\begin{equation}
B_{n}^{(p)}=\frac{d^{p}B_{n}}{d\left(t_{n}\right)^{p}}=\left\{ \begin{array}{ccc}
\xi &  & p=1\\
\\
0 &  & p>1
\end{array}\right..
\end{equation}

\medskip{}

\subsection{Non-Gaussian distribution in the absence of barriers}

\medskip{}

\hspace*{0.25in}The absence of barriers is a particular case of the
probability densities of sections (\ref{subsec:The-Sheth-Tormen-approach})
or (\ref{subsec:Slow-moving-barrier}). Specifically, add an extra
limit in (\ref{eq:238-1}), setting the barrier at infinity:

\medskip{}

\begin{equation}
\underset{B_{n}\rightarrow\infty}{lim}\underset{\frac{dB_{n}}{dt_{n}}\rightarrow0}{lim}\underset{\frac{d^{2}B_{n}}{dt_{n}^{2}}\rightarrow0}{lim}\Pi_{\epsilon\rightarrow0}^{mb}\left(\omega_{n},t_{n}\right)=\Pi_{\epsilon\rightarrow0}^{gm}\left(\omega_{n},t_{n}\right).\label{eq:239-1}
\end{equation}

\medskip{}

\begin{equation}
\Pi_{\epsilon\rightarrow0}^{gm}\left(\omega_{n},t_{n}\right)=\frac{1}{\sqrt{2\pi t_{n}}}e^{\alpha\omega_{n}-\frac{1}{2}\alpha^{2}t_{n}}e^{-\frac{\omega_{n}^{2}}{2t_{n}}}.\label{eq:239-2}
\end{equation}

\medskip{}

Second, in (\ref{eq:217-1}), since the derivative related to the
barrier $\left(\partial^{i}/\partial B_{n}^{i}\right)$ are limit
operations, they can be applied after those of (\ref{eq:239-1}).
At the infinity, the distribution converges to the Gaussian density,
without barrier, according to (\ref{eq:239-1}). As a consequence,
the derivative operator $\partial^{i}/\partial B_{n}^{i}$ nullifies
the term. In the case of first order derivative,

\medskip{}

\[
\underset{B_{n}\rightarrow\infty}{lim}\underset{\frac{dB_{n}}{dt_{n}}\rightarrow0}{lim}\underset{\frac{d^{2}B_{n}}{dt_{n}^{2}}\rightarrow0}{lim}\Pi_{\epsilon\rightarrow0}^{mb}\left(\omega_{n},t_{n}\right)=\underset{B_{n}\rightarrow\infty}{lim}\underset{\frac{dB_{n}}{dt_{n}}\rightarrow0}{lim}\underset{\frac{d^{2}B_{n}}{dt_{n}^{2}}\rightarrow0}{lim}\underset{\Delta B_{n}\rightarrow\infty}{lim}\frac{\Delta\Pi_{\epsilon\rightarrow0}^{mb}\left(\omega_{n},t_{n}\right)}{\Delta B_{n}}
\]

\[
=\underset{\Delta B_{n}\rightarrow\infty}{lim}\frac{\Delta\left[\underset{B_{n}\rightarrow\infty}{lim}\underset{\frac{dB_{n}}{dt_{n}}\rightarrow0}{lim}\underset{\frac{d^{2}B_{n}}{dt_{n}^{2}}\rightarrow0}{lim}\Pi_{\epsilon\rightarrow0}^{mb}\left(\omega_{n},t_{n}\right)\right]}{\Delta B_{n}}
\]

\begin{equation}
=\underset{\Delta B_{n}\rightarrow\infty}{lim}\frac{\Delta\Pi_{\epsilon\rightarrow0}^{gm}\left(\omega_{n},t_{n}\right)}{\Delta B_{n}}=0;\label{eq:240-1}
\end{equation}

\medskip{}

\noindent and so on, for higher order derivatives in the barrier.
Nevertheless, in (\ref{eq:217-1}) remain the derivative terms $\partial^{i}/\partial\omega_{n}^{i}$,
which assure the survival of non-Gaussian terms of the expansion.

\selectlanguage{brazil}%
In a nutshell, the non-barrier case of non-Gaussian distribution is
given by substituting $\Pi_{\epsilon\rightarrow0}^{mb}\left(\omega_{n},t_{n}\right)\rightarrow\Pi_{\epsilon\rightarrow0}^{gm}\left(\omega_{n},t_{n}\right)$
in (\ref{eq:217-1}), and excluding barrier derivatives of the distribution.
We rewrite (\ref{eq:217-1}) in this case:

\selectlanguage{english}%
\medskip{}

\[
\Pi_{\epsilon\rightarrow0}^{inf}\left(\omega_{0}=0,\omega_{n};t_{n}\right)=\Pi_{\epsilon\rightarrow0}^{gm}\left(\omega_{n},t_{n}\right)
\]

\[
-\frac{1}{3!}\kappa_{3}\frac{\partial^{3}}{\partial\omega_{n}^{3}}\Pi_{\epsilon\rightarrow0}^{gm}\left(\omega_{n},t_{n}\right)+\frac{1}{4!}\kappa_{4}\frac{\partial^{4}}{\partial\omega_{n}^{4}}\Pi_{\epsilon\rightarrow0}^{gm}\left(\omega_{n},t_{n}\right)
\]

\[
-\frac{1}{5!}\kappa_{5}\frac{\partial^{5}}{\partial\omega_{n}^{5}}\Pi_{\epsilon\rightarrow0}^{gm}\left(\omega_{n},t_{n}\right)+\left[\frac{1}{2}\cdot\left(\frac{1}{3!}\kappa_{3}\right)^{2}+\frac{1}{6!}\kappa_{6}\right]\frac{\partial^{6}}{\partial\omega_{n}^{6}}\Pi_{\epsilon\rightarrow0}^{gm}\left(\omega_{n},t_{n}\right)
\]

\[
-\left[\left(\frac{1}{3!}\kappa_{3}\right)\left(\frac{1}{4!}\kappa_{4}\right)+\frac{1}{7!}\kappa_{7}\right]\frac{\partial^{7}}{\partial\omega_{n}^{7}}\Pi_{\epsilon\rightarrow0}^{gm}\left(\omega_{n},t_{n}\right)
\]

\[
+\left[\frac{1}{2}\cdot\left(\frac{1}{4!}\kappa_{4}\right)^{2}-\left(\frac{1}{3!}\kappa_{3}\right)\left(\frac{1}{5!}\kappa_{5}\right)+\frac{1}{8!}\kappa_{8}\right]\frac{\partial^{8}}{\partial\omega_{n}^{8}}\Pi_{\epsilon\rightarrow0}^{gm}\left(\omega_{n},t_{n}\right)
\]

\begin{equation}
-\left[\left(\frac{1}{4!}\kappa_{4}\right)\left(\frac{1}{5!}\kappa_{5}\right)+\left(\frac{1}{3!}\kappa_{3}\right)\left(\frac{1}{6!}\kappa_{6}\right)+\frac{1}{9!}\kappa_{9}\right]\frac{\partial^{9}}{\partial\omega_{n}^{9}}\Pi_{\epsilon\rightarrow0}^{gm}\left(\omega_{n},t_{n}\right)+....\label{eq:240}
\end{equation}

\medskip{}

\noindent where

\medskip{}

\begin{equation}
\Pi_{\epsilon\rightarrow0}^{gm}\left(\omega_{0},\omega_{n};t_{n}\right)=\frac{1}{\sqrt{2\pi t_{n}}}e^{\alpha\left(\omega-\omega_{0}\right)-\frac{1}{2}\alpha^{2}t_{n}}e^{-\frac{\left(\omega-\omega_{0}\right)^{2}}{2t_{n}}}.\label{eq:241}
\end{equation}

\medskip{}

From now on, when we refer to the vanilla case under non-Gaussian
distribution, we will deal with (\ref{eq:240}) and (\ref{eq:241}),
the infinite barrier case.

\medskip{}

\subsection{Martingale condition for drift}

\medskip{}

\hspace*{0.25in}Under the risk-neutral measure $\mathbb{Q}$, the
martingale condition establishes the non-arbitrage drift condition

\medskip{}

\begin{equation}
e^{-r\left(t_{n}-t_{n-1}\right)}E^{\mathbb{Q}}\left[S_{n}|\mathcal{F}_{t_{n-1}}\right]=S_{n-1}.\label{eq:237-1}
\end{equation}

\medskip{}

In terms of the distribution (\ref{eq:240}),\medskip{}

\[
S_{0}=e^{-rt_{n}}\int_{-\infty}^{\infty}S_{0}e^{\sigma\omega_{n}}\Pi_{\epsilon\rightarrow0}^{inf}\left(\omega_{n},t_{n}\right)d\omega_{n},
\]

\begin{equation}
\therefore1=e^{-rt_{n}}\int_{-\infty}^{\infty}e^{\sigma\omega_{n}}\Pi_{\epsilon\rightarrow0}^{inf}\left(\omega_{n},t_{n}\right)d\omega_{n}.\label{eq:242}
\end{equation}
\medskip{}

When we keep up to the 15th order in derivatives in (\ref{eq:240}),
we get the equation (\ref{eq:315}) of \ref{sec:DriftDistrNG}.\medskip{}

\subsection{Analytical option pricing}

\medskip{}

\hspace*{0.25in}To price call vanilla european options, under
the probability density $\Pi_{\epsilon\rightarrow0}^{inf}\left(\omega_{n},t_{n}\right)$,

\medskip{}

\begin{equation}
P(S_0,K,t_n,\alpha,\kappa_3,\kappa_4,...)=\int_{-\infty}^{\infty}max\left[\left(S_{0}e^{\sigma\omega_{n}}-K\right),0\right]\Pi_{\epsilon\rightarrow0}^{inf}\left(\omega_{n},t_{n}\right)d\omega_{n},\label{eq:246-1}
\end{equation}

\medskip{}

Barrier options are priced with $\Pi_{\epsilon\rightarrow0}\left(\omega_{n},t_{n}\right)$
in (\ref{eq:217-1}), specifying $\Pi_{\epsilon\rightarrow0}^{mb}\left(\omega_{n},t_{n}\right)$
for fixed (\ref{eq:238-1}) or moving barriers, either in (ST), or
in adiabatic barrier approximations. The price of a \emph{knock-up-and-out}
call is given by:

\medskip{}

\begin{equation}
P=\int_{k}^{b}\left(S_{0}e^{\sigma\omega_{n}}-K\right)\Pi_{\epsilon\rightarrow0}\left(\omega_{n},t_{n}\right)d\omega_{n}.\label{eq:249-1}
\end{equation}
\medskip{}

\noindent and that of a \emph{knock-up-and-out }put by\footnote{The valuation of these integrals result in closed-form expressions.
However, if a high number of cumulants is included in the specification,
numerical valuation may become less costly.}:\medskip{}

\begin{equation}
P=\int_{-\infty}^{k}\left(K-S_{0}e^{\sigma\omega_{n}}\right)\Pi_{\epsilon\rightarrow0}\left(\omega_{n},t_{n}\right)d\omega_{n}.\label{eq:250-1}
\end{equation}
\medskip{}

\medskip{}

\section{Calibration}

\medskip{}

\hspace*{0.25in}In order to price barrier options, the parameters
$\sigma$ and $\kappa_{i}$ are calibrated with european vanilla call
options, for each maturity, meaning a piecewise constant set. In our
example, the daily data consist of foreign exchange (FX) european
call options of Brazilian Real against US dollar (BRL/USD), ranging
from 05/2009 to 05/2014. Each day includes implied volatility rates
corresponding to five standard delta values $\left\{ 10\%,25\%,50\%,75\%,90\%\right\} $,
for twenty-four months $\left\{ 1,2,3,...,24\right\} $. Therefore,
a volatility $\sigma_{ijk}$ is indexed by the date $t_{i}$ , delta
$\Delta_{j}$ and maturity $T_{k}$: $\sigma_{ijk}=\sigma\left(t_{i},\Delta_{j},T_{k}\right)$;
$i=1,...,1295$; $j=1,...,5$; $k=1,...,24.$ Deltas are converted
to strikes in the usual manner.

\begin{figure}[H]
\includegraphics[width=0.5\columnwidth]{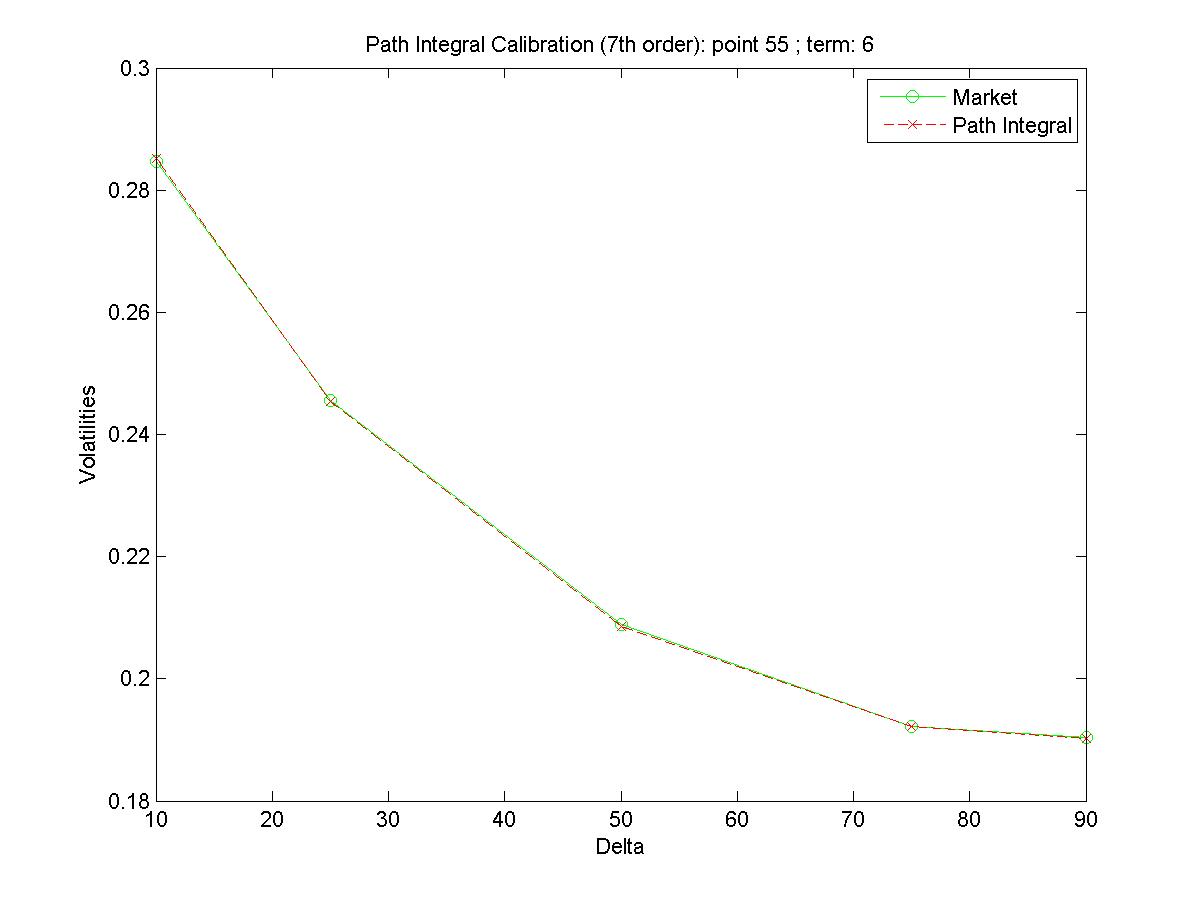}\includegraphics[width=0.5\columnwidth]{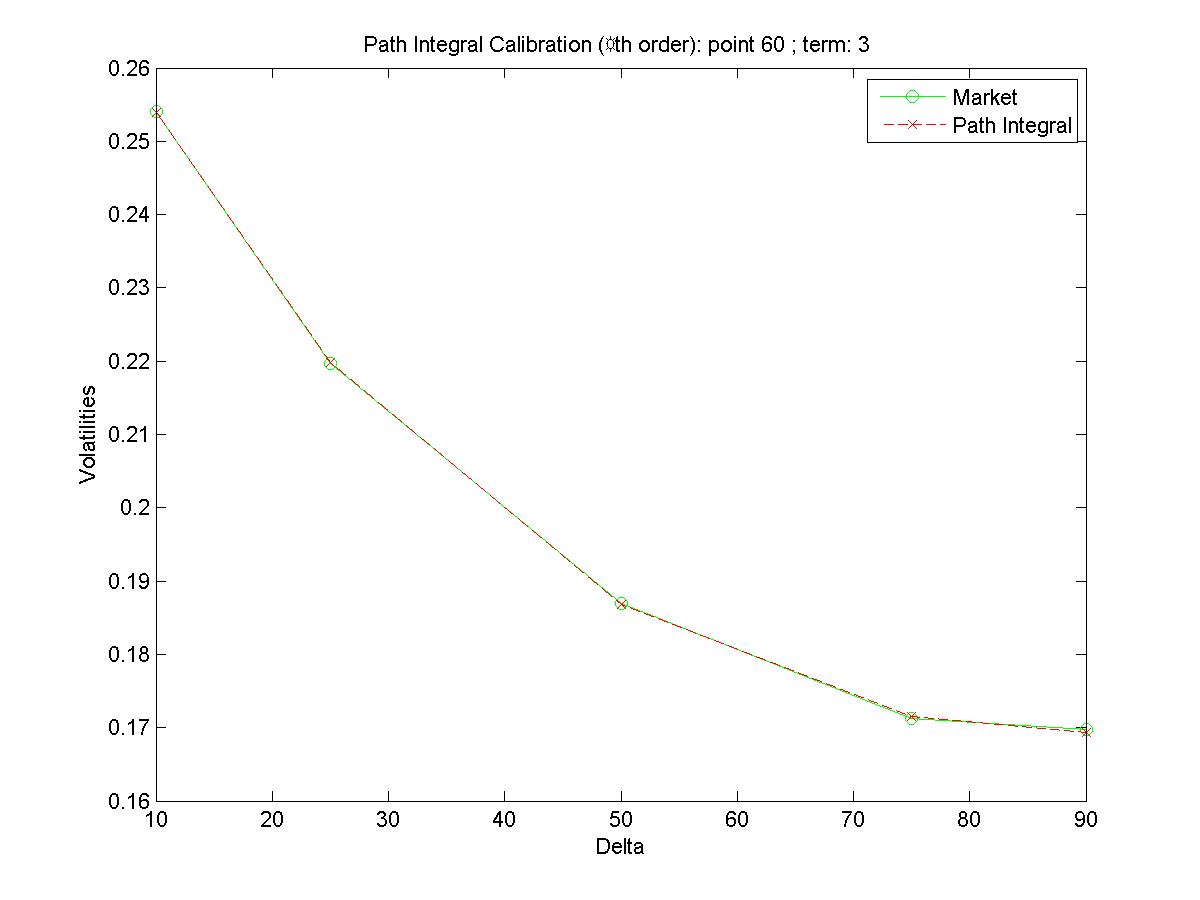}\caption{\label{fig:Calibracao Smile}Smile calibration for vanillas. The first
graph uses up to the 7th cumulant, while the second uses up to the
15th one.}
\end{figure}

\begin{figure}[H]
\includegraphics[width=1.0\columnwidth]{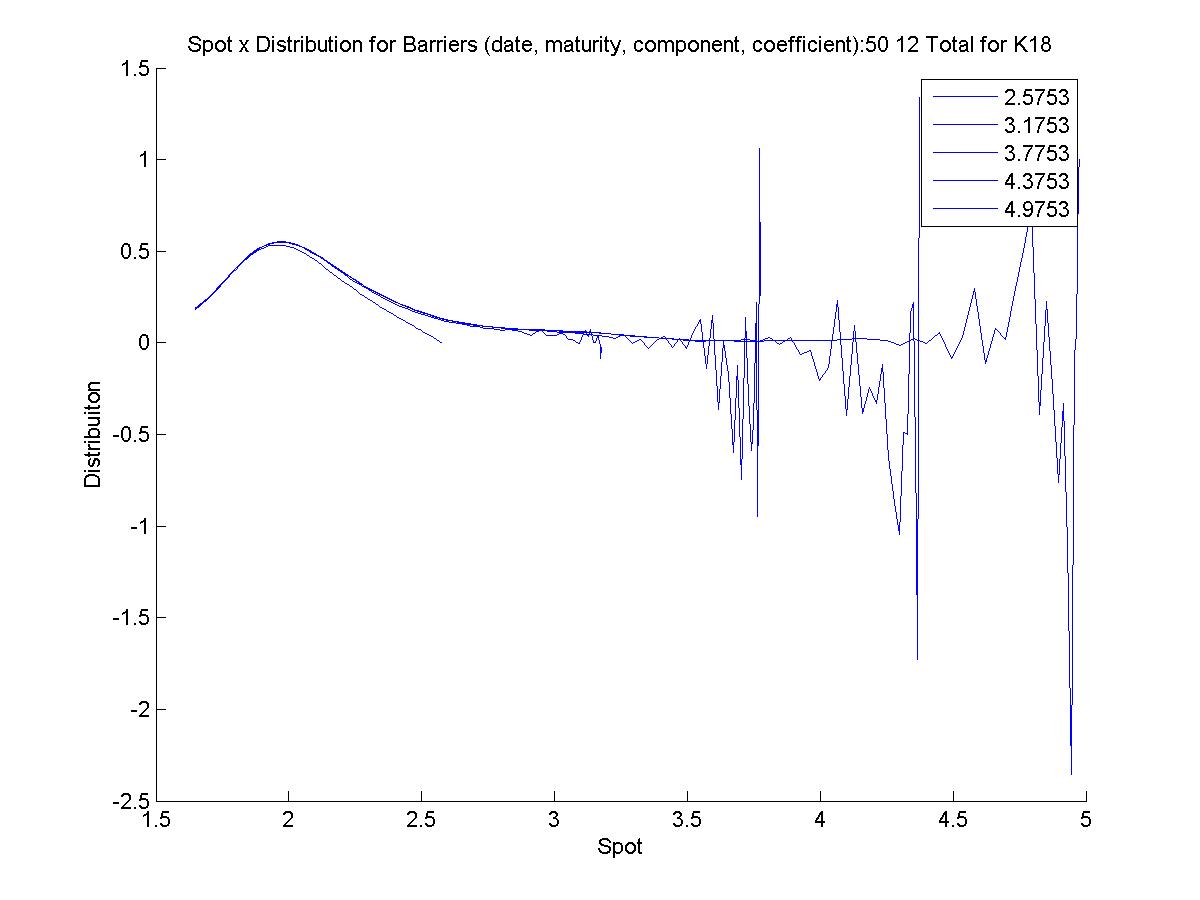} \\\includegraphics[width=1.0\columnwidth]{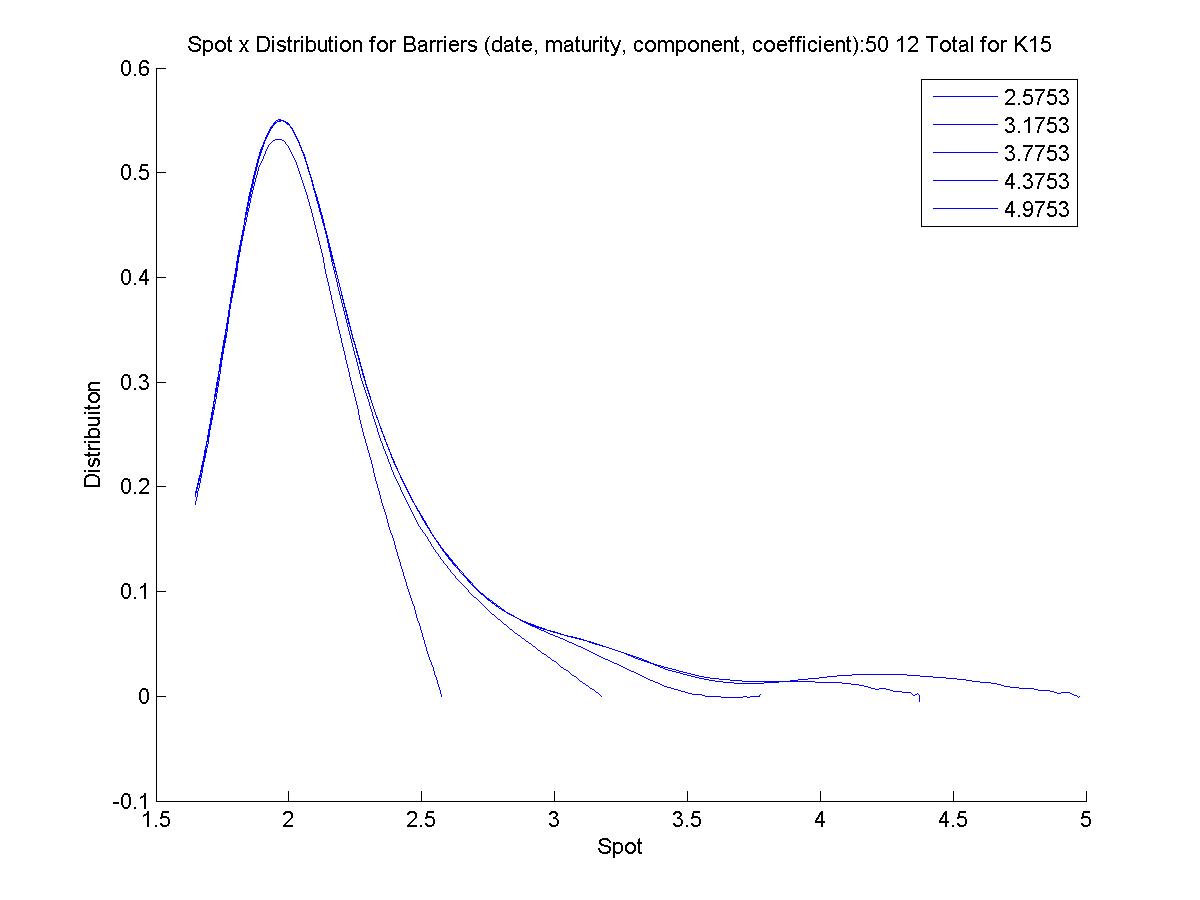}
\caption{
\label{fig:Polinomial components}Behaviour of non-Gaussian distribution
in the barrier region under inclusion of second order combinations
of parameters. Different values of the barrier are shown in the boxes.}
\end{figure}

Concerning the model specification, the number of parameters in the
case of vanilla options depends on the ability to fit the smile delta
range. On the left side of figure \ref{fig:Calibracao Smile}, the
maximum order was in $\kappa_{7}$, while, on the right side of the
figure, the maximum was $\kappa_{15}$. However, higher order corrections
gained with higher derivative orders in the expansion are done at
the expense of adding more parameters to the model. Such improvements
are necessary near the barrier region in pricing barrier options. Further, just including
higher derivative orders is not enough if combinations of parameters
are not considered (we took up to second order combinations, $\kappa_{i}$
$\kappa_{j}$, with $i+j=n$, $n=$ order of derivative correction
in the term). For instance, in the upper plot in figure \ref{fig:Polinomial components},
we used 18th order in derivatives for different values of the barrier, but didn't include the second order
combinations. When we included the second order combinations of parameters,
the 15th order was enough to improve precision in the barrier region,
as shown in the lower plot.

\medskip{}
 \medskip{}

Thus, in order to price knock-up-and-out calls, according to the integration
limits in the pricing equation (\ref{eq:249-1}), one should guarantee
a good distribution fitting in the interval $\left[k,b]\right]$.
Therefore, although price calibration is possible, because of the
large amount of terms in the pricing formula, we calibrated the parameters
of the model density (\ref{eq:240}) by fitting the probability density
retrieved from the \cite{BL78} theorem, taking into account the smile, as in \cite{Sh93}:

\begin{equation}
\Pi_{\epsilon\rightarrow0}^{inf}\left(\omega_{n}=k,t_{n}\equiv T\right)=e^{rT}K\sigma\frac{d^{2}C}{dK^{2}}.\label{eq:273}
\end{equation}

\medskip{}

\begin{figure}[H]
\includegraphics[width=1\columnwidth]{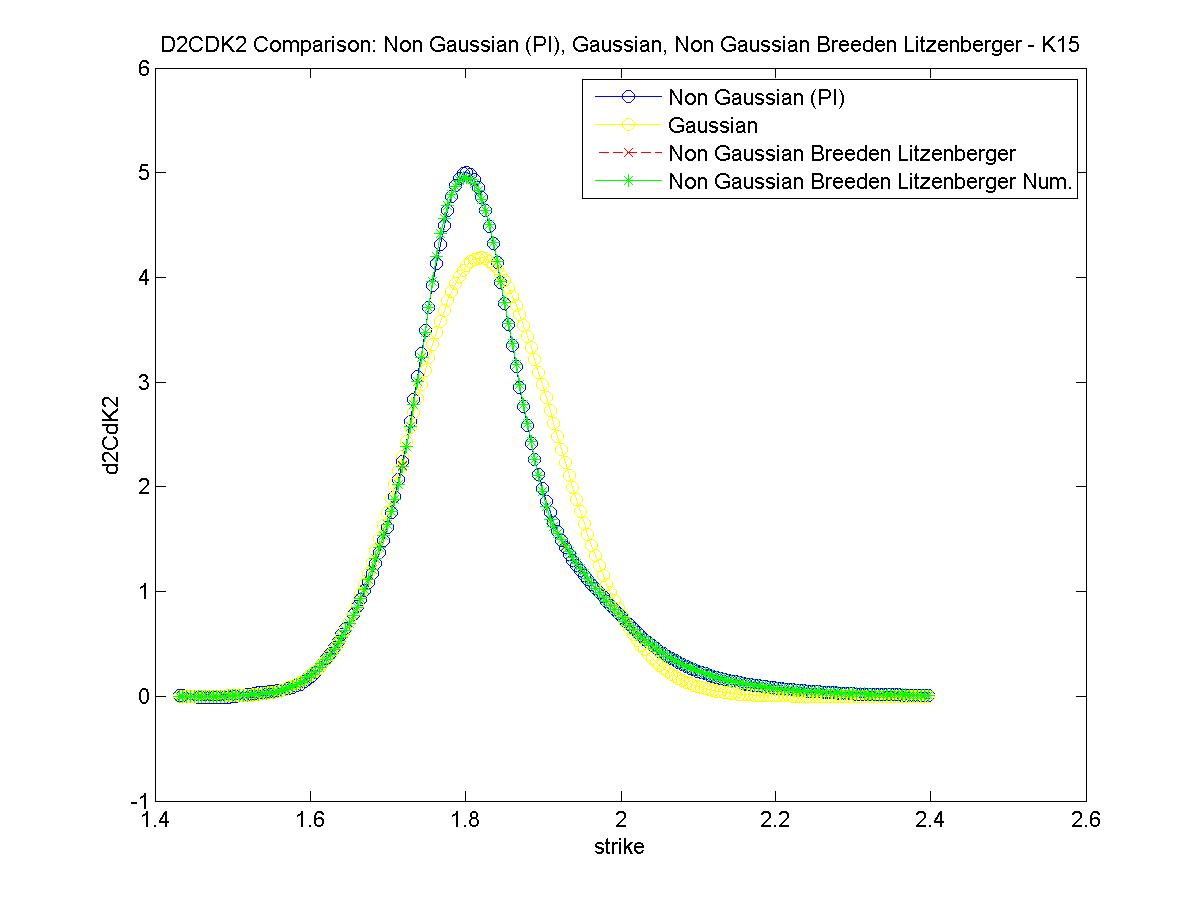}\caption{\label{fig:Densidade-Calibracao} Plot representing (i) the non-Gaussian
density in the path-integral (``non-Gaussian (PI)''); (ii) the distribution
according to the \cite{BL78} theorem, where derivatives are computed
with cubic spline (red) and numerically (green); and (iii) Gaussian
density using the at-the-money volatility. Data for 1-month maturity
of the 85th sample point.}
\end{figure}

The total derivative $d^{2}C/dK^{2}$ was computed numerically and
analitically, using the parameters of cubic spline interpolation,
both producing the same results. An example of calibration is given
in figure \ref{fig:Densidade-Calibracao}.\medskip{}

\medskip{}

\medskip{}

\medskip{}

The fitting of model prices $P_{model}$ to 300 vanilla call option
market prices $P_{market}$ can be summarized by fitting the linear
regression ($e$ is the residual):

\medskip{}

\begin{equation}
P_{Model}=a_{p}\cdot P_{market}+b_{p}+e,
\end{equation}

\begin{figure}[H]
\includegraphics[width=1\columnwidth]{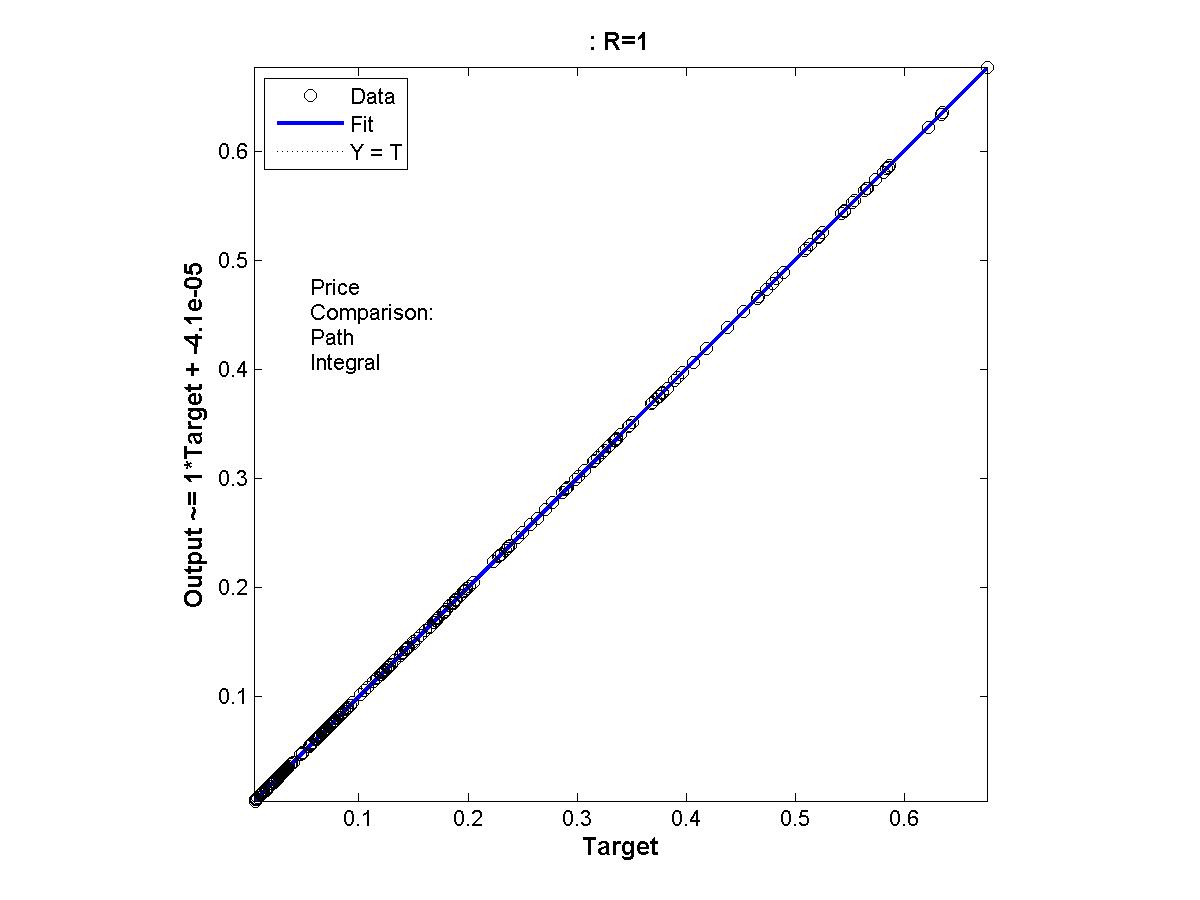}\caption{\label{fig:Regressao Vanilla}Price regression: dependent variable
is the model price and the independent one is the market prices of
vanilla call options.}
\end{figure}

\medskip{}

\noindent where we hope to get $a_{p}=1$ and $b_{p}=0$. The result
is in figure \ref{fig:Regressao Vanilla}, which displays $R^{2}\sim1$,
and $a_{p}=1$, $b_{p}=-4 \times 10^{-5}$.

\medskip{}

\medskip{}

\medskip{}

\medskip{}

\section{Barrier option pricing}

\medskip{}

\hspace*{0.25in}To analyze in a standardized way the effect of an
absorbing up barrier on the call option price, according to its proximity
to the initial underlying price or to the strike, we set fixed barrier
levels defined by a group of multiplicative factors $\Theta=\left\{ 1.1;\;1.2;\;1.3;\;1.5\right\} $,
to be applied to data base strikes, which cover the deltas from 10
to 90 in each maturity. In order to assure that the price does not
start deactivated by the barrier, when multiplication results in a
barrier below initial underlying future price, we change the rule,
setting the fixed barrier to a multiplicative factor regarding the
future price, given maturity. Thus, to each strike $i$ and maturity
$j$, the barrier $B_{k}$ is defined, resulting in price

\medskip{}

\medskip{}

\begin{equation}
P_{ijk}=P_{ijk}\left(B_{k},K_{i},T_{j}\right);B_{k}=\left\{ \begin{array}{ccc}
\Theta_{k}\cdot K_{i} & , & \Theta_{k}\cdot K_{i}>F_{j}\\
\Theta_{k}\cdot F_{j} & , & \Theta_{k}\cdot K_{i}\leq F_{j}
\end{array}\right.
\end{equation}

\medskip{}

\medskip{}

\noindent where $k=1,...,4;$ and $F_{j}$ is the future value related
to the maturity $T_{j}$.

\begin{figure}[H]
\includegraphics[width=1\columnwidth]{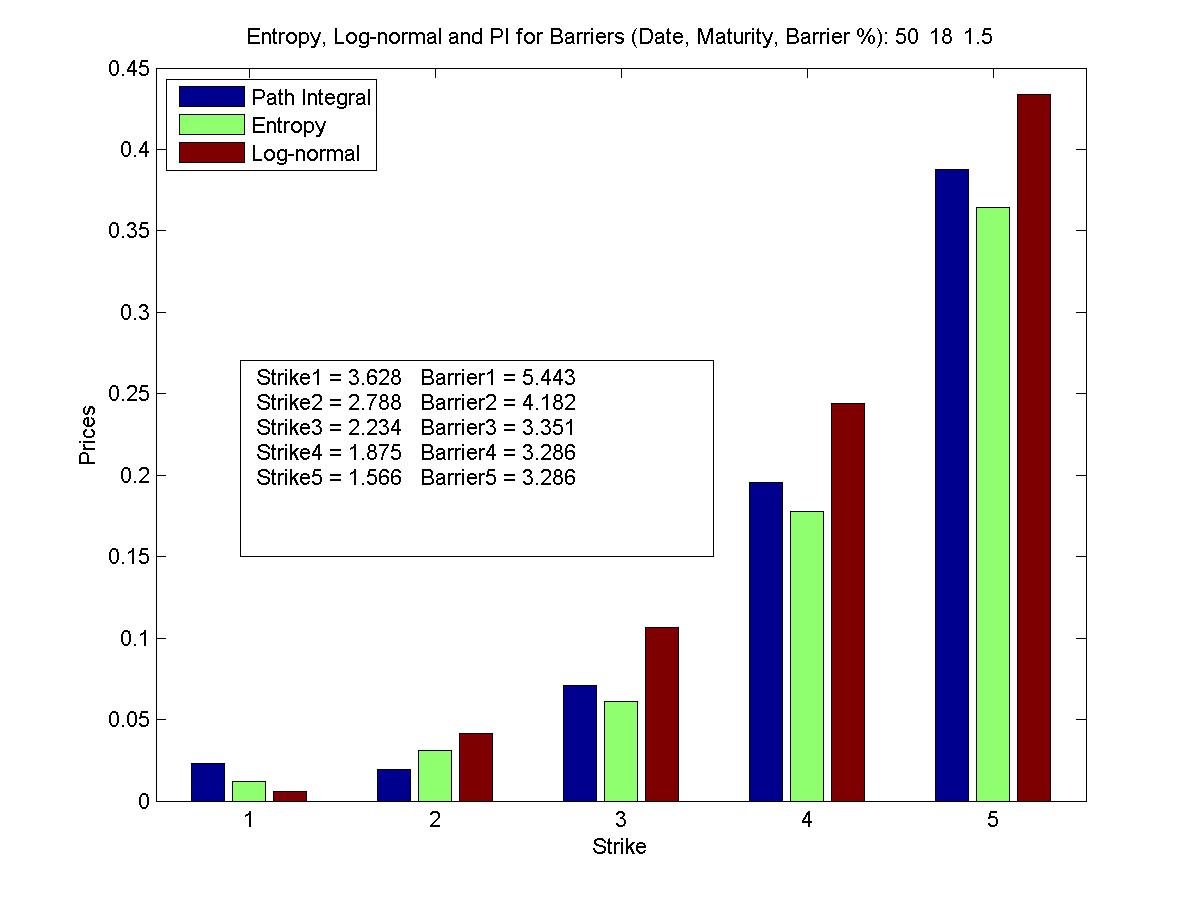}

\caption{\label{fig:barGraphBarreira18meses}18-month model comparison: path
integral, relative entropy and Black-Scholes.}
\end{figure}

\begin{figure}[H]
\includegraphics[width=0.5\columnwidth]{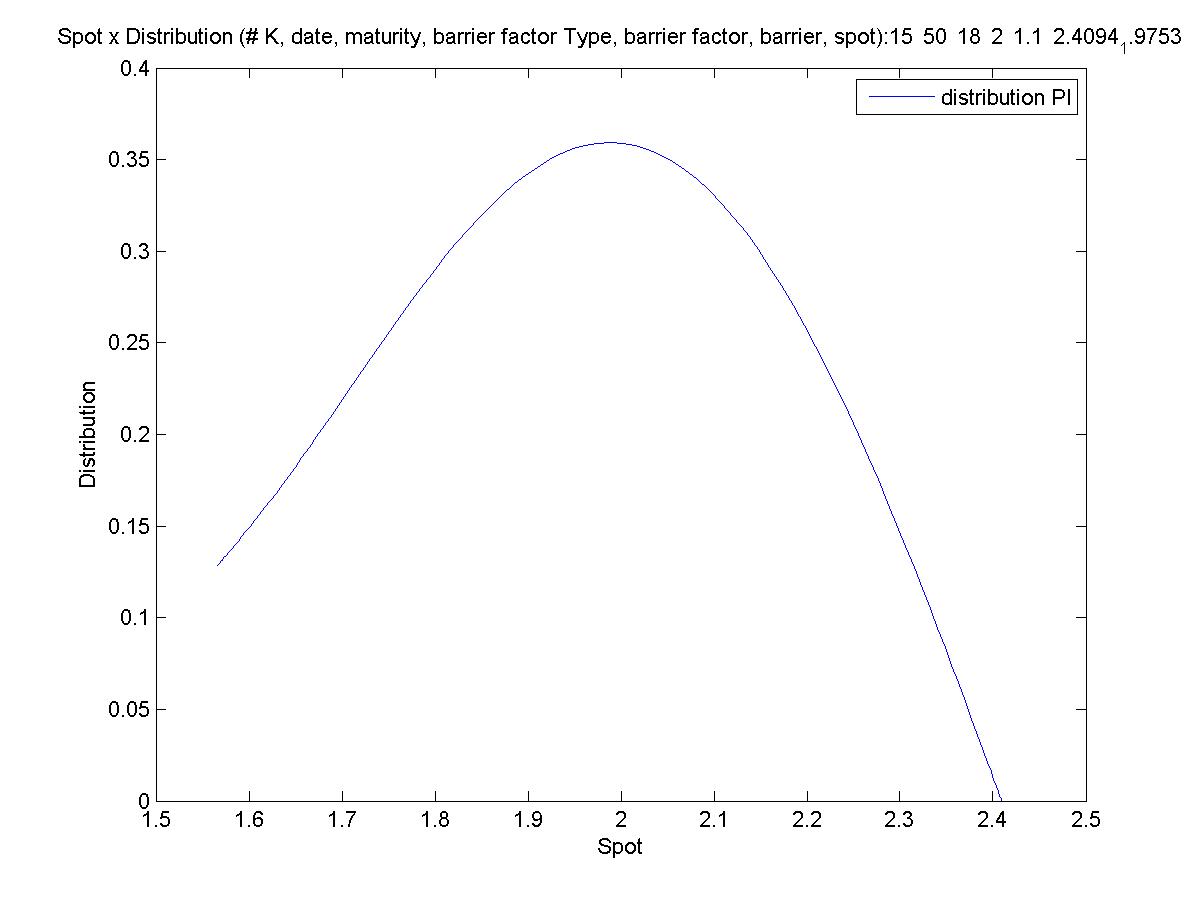}\includegraphics[width=0.5\columnwidth]{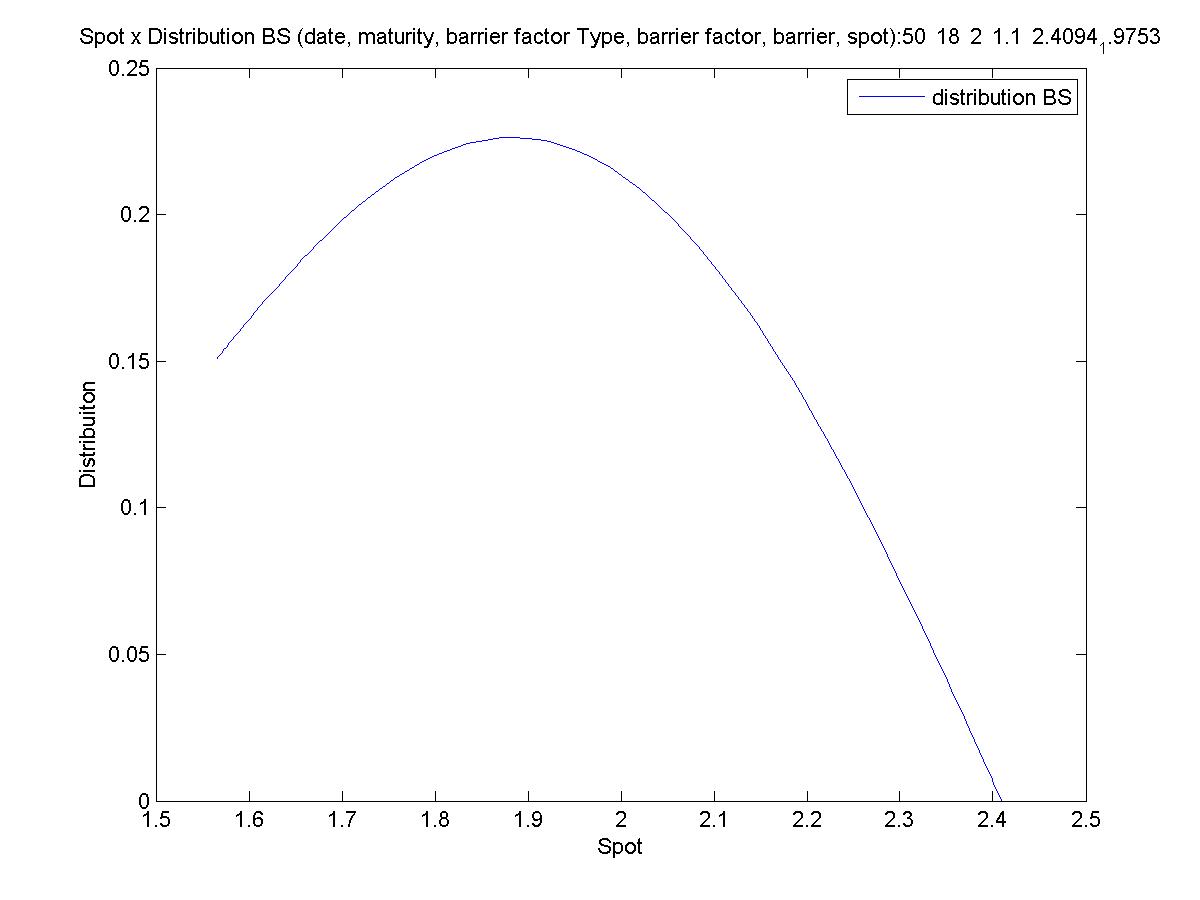}\caption{\label{fig: densidade barreira}Comparison between non-Gaussian (path
integral model) and Gaussian (Black-Scholes model) densities with
absorbing barrier. }
\end{figure}

Usually, data providers of barrier option prices rely on market-to-model
values. Therefore, we have chosen to compare ours results to the relative
entropy model of \cite{Ave01}. When the absorbing up barrier is near
the strike of the call or the inital underlying value, that is, when
$\Theta_{k}=1.1$, prices are closer to zero. In this situation, in
longer maturities, we have found greater divergence comparing the
path-integral approach, the lognormal model (Black-Scholes, with strike-related
volatility) and the relative entropy model, as in figure \ref{fig:barGraphBarreira18meses}.
In figure \ref{fig: densidade barreira}, in the 18-month case, we
notice that, although the average underlying price is higher in the
path-integral model, the Black-Scholes model presents a greater dispersion,
meaning that the barrier region has greater chance of being reached,
thus displaying lower prices in the case of barrier close to initial
underlying value. In shorter maturities, such as 1-month, the underlying
process has less time to reach, the closer the barrier; consequently,
prices are closer in the model comparison. Therefore, difference between
path-integral and entropy models is expected, since the first approach
includes the density behaviour in the vicinity of the barrier.

\medskip{}
 \medskip{}

\begin{figure}

\includegraphics[width=1\columnwidth]{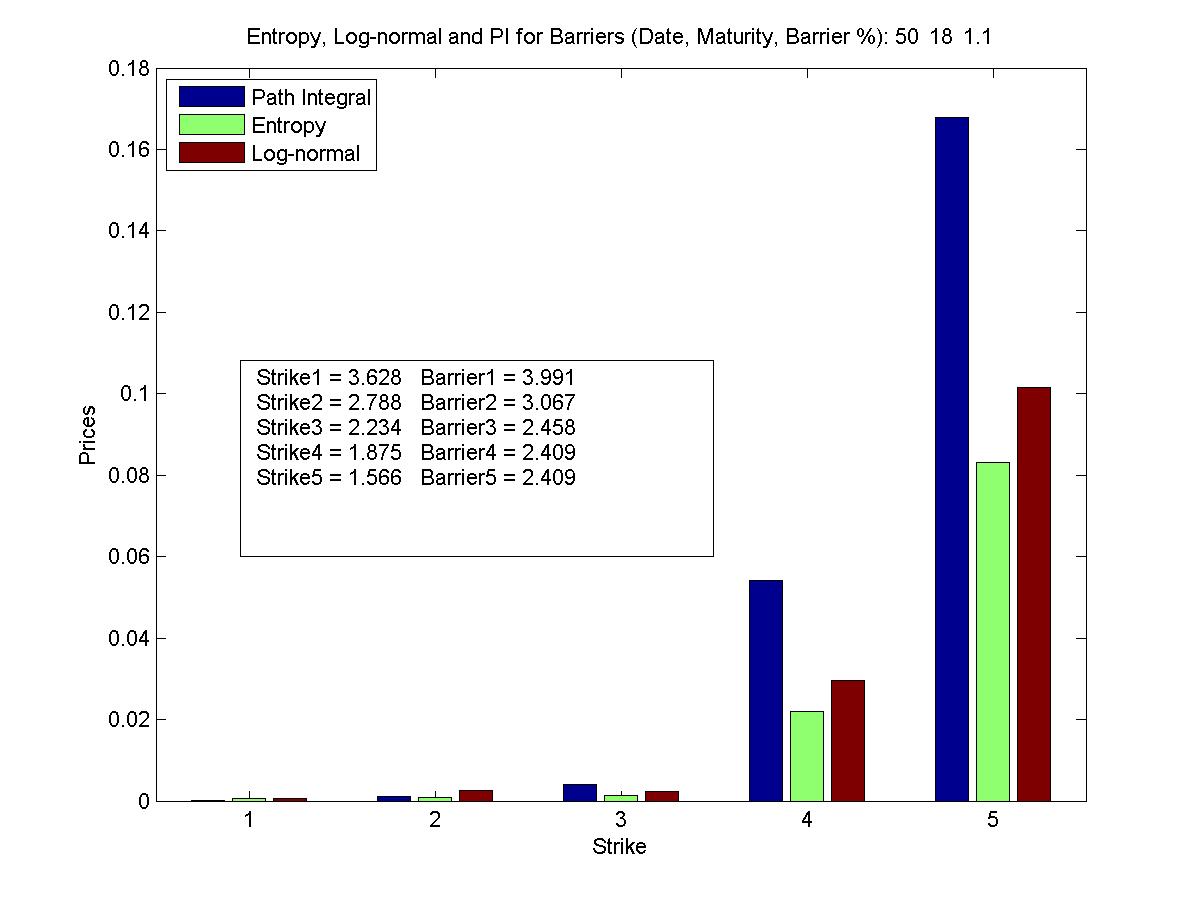}

\end{figure}

\begin{figure}

\includegraphics[width=1\columnwidth]{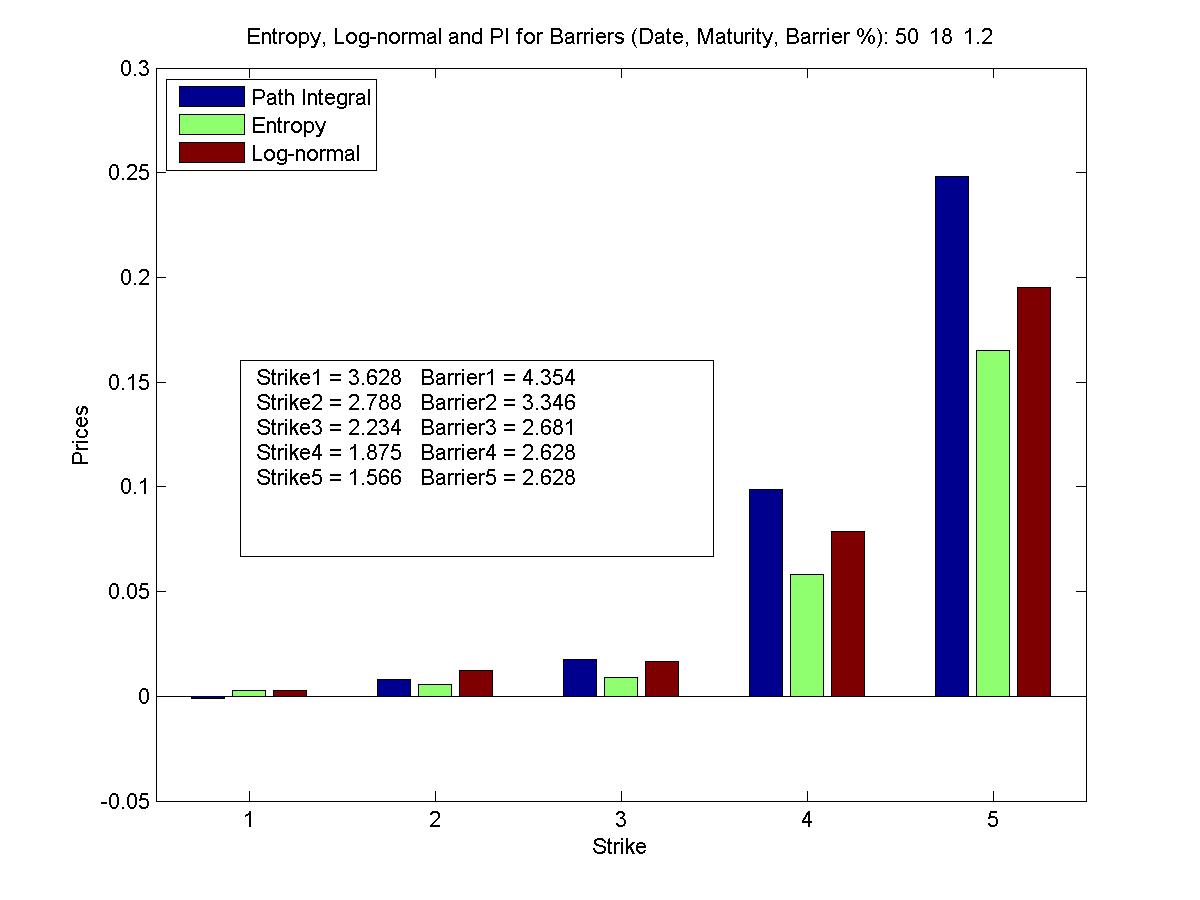}

\end{figure}
\begin{figure}
\includegraphics[width=1\columnwidth]{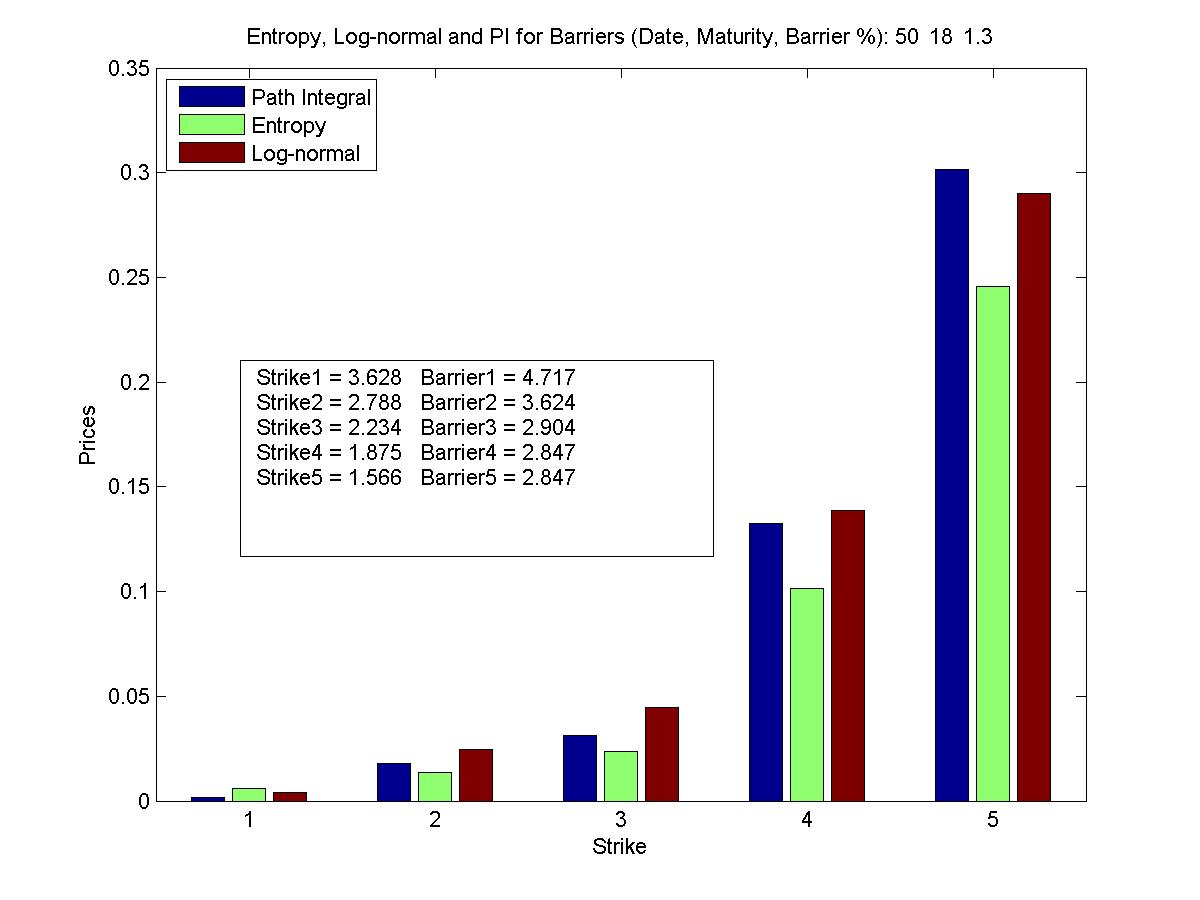}

\end{figure}

\medskip{}
 \medskip{}

\medskip{}
 \medskip{}

\begin{figure}
\includegraphics[width=1\columnwidth]{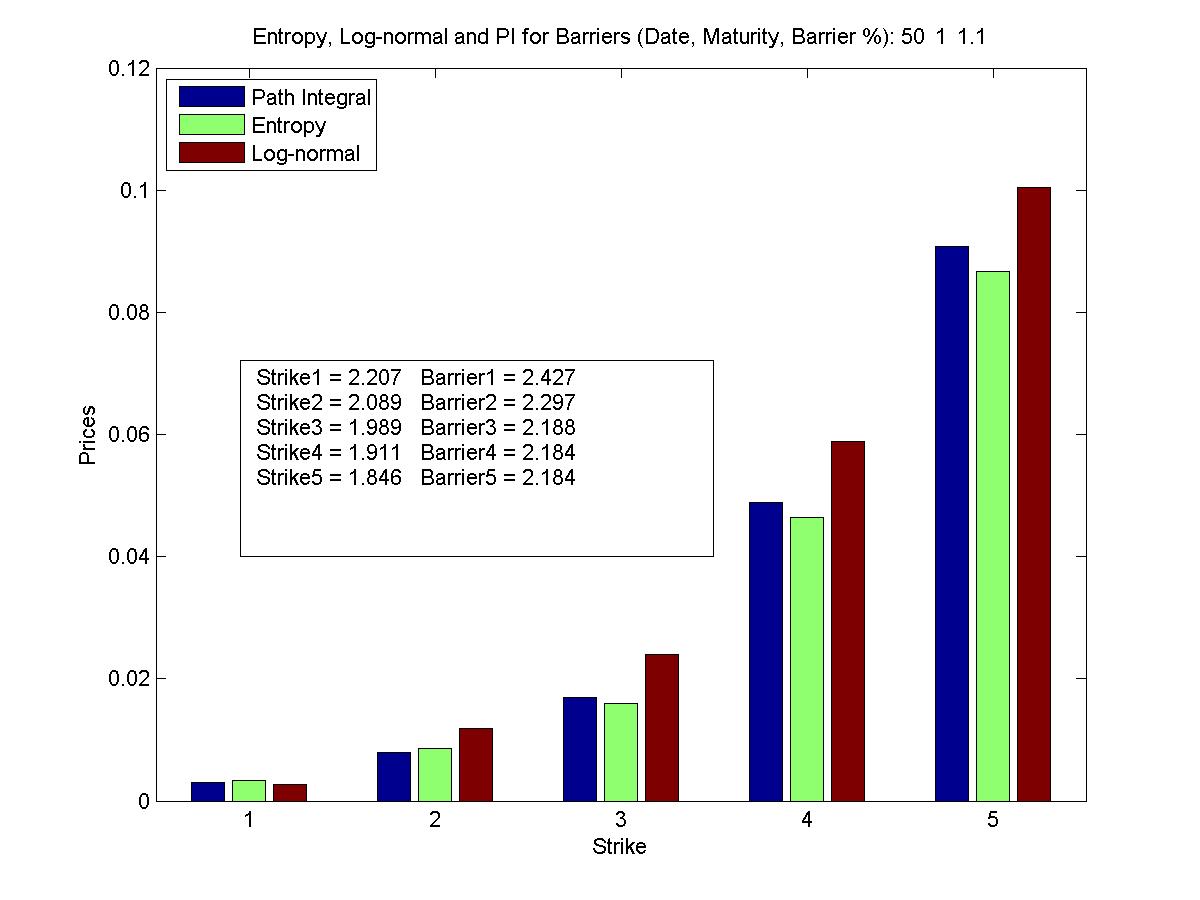}

\end{figure}

\begin{figure}
\includegraphics[width=1\columnwidth]{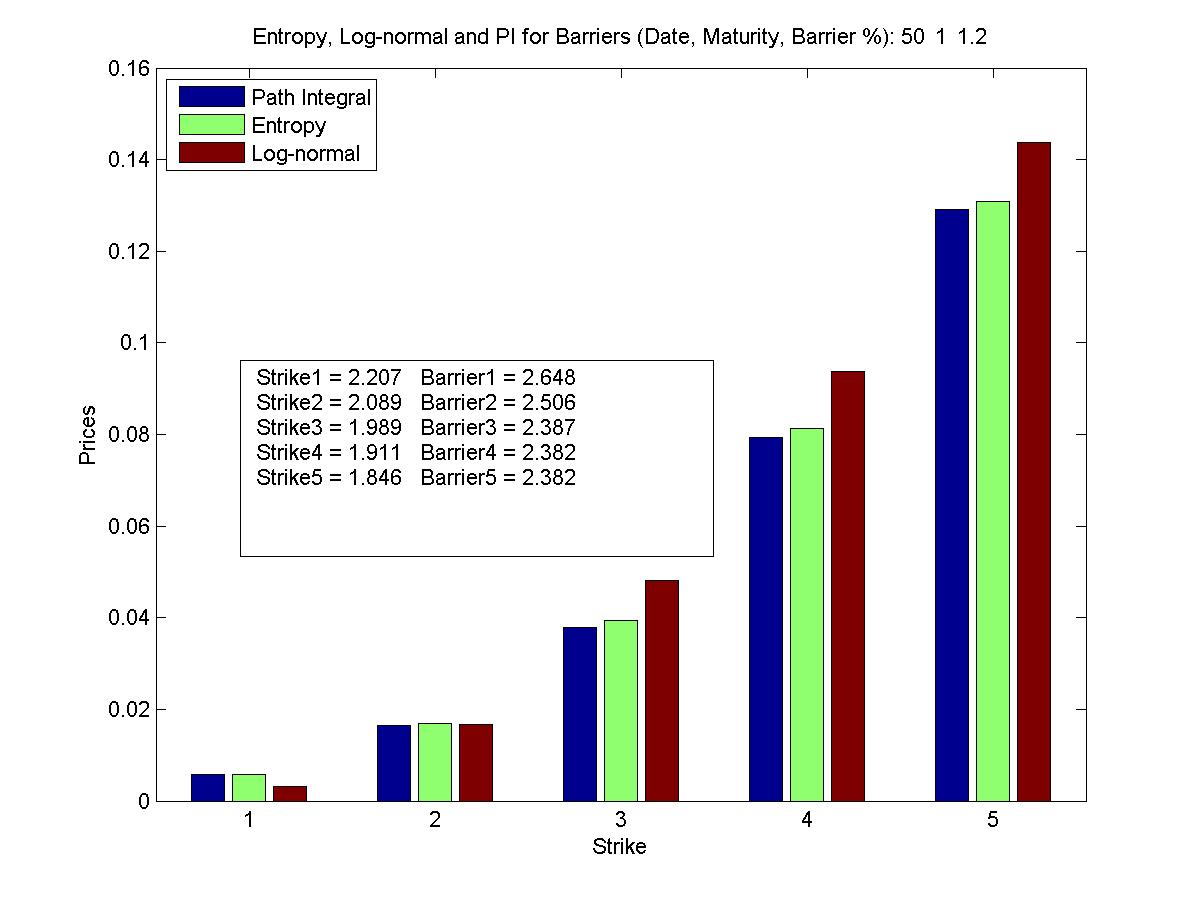}
\end{figure}

\begin{figure}

\includegraphics[width=1\columnwidth]{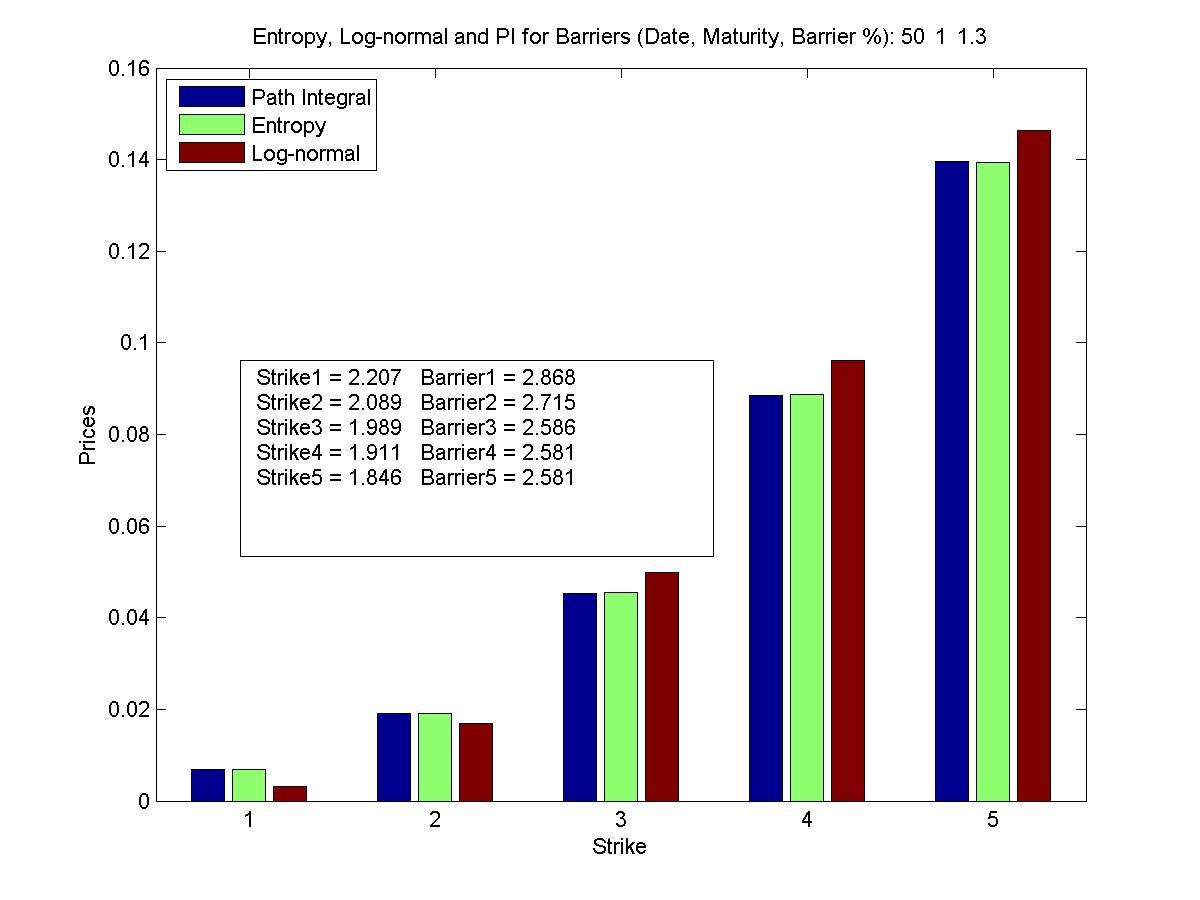}

\end{figure}

\begin{figure}[H]
\includegraphics[width=1\columnwidth]{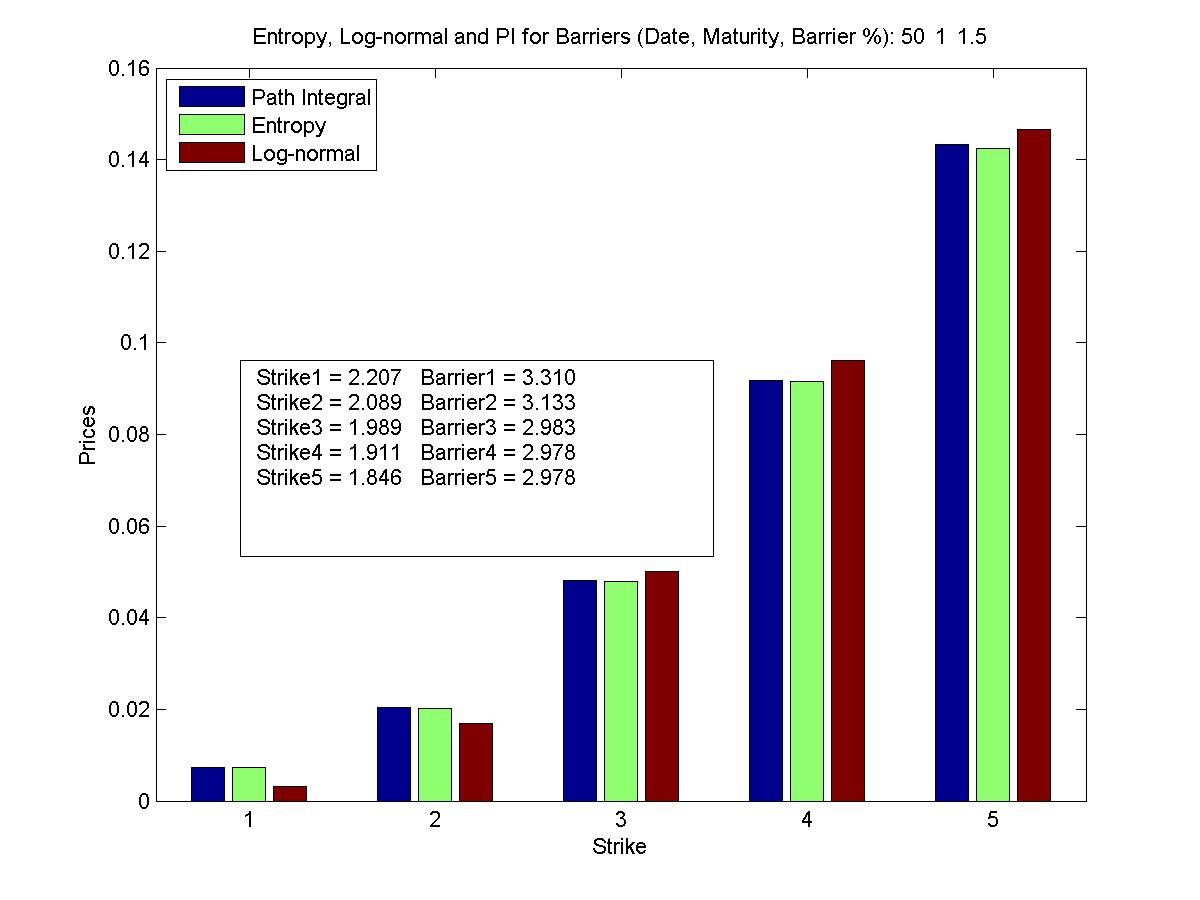}

\caption{\label{fig: bar GraphBarreira1mes}1-month model comparison: path
integral, relative entropy and Black-Scholes.}
\end{figure}

\medskip{}
 \medskip{}

In the same way as vanilla pricing, we summarize pricing differences
between the path integral and the relative entropy models by fitting
a linear regression:

\medskip{}

\begin{equation}
P_{Path\;integral}=a_{p}\cdot P_{Entropy}+b_{p}+e,
\end{equation}

\medskip{}

\begin{figure}[H]
\includegraphics[width=1\columnwidth]{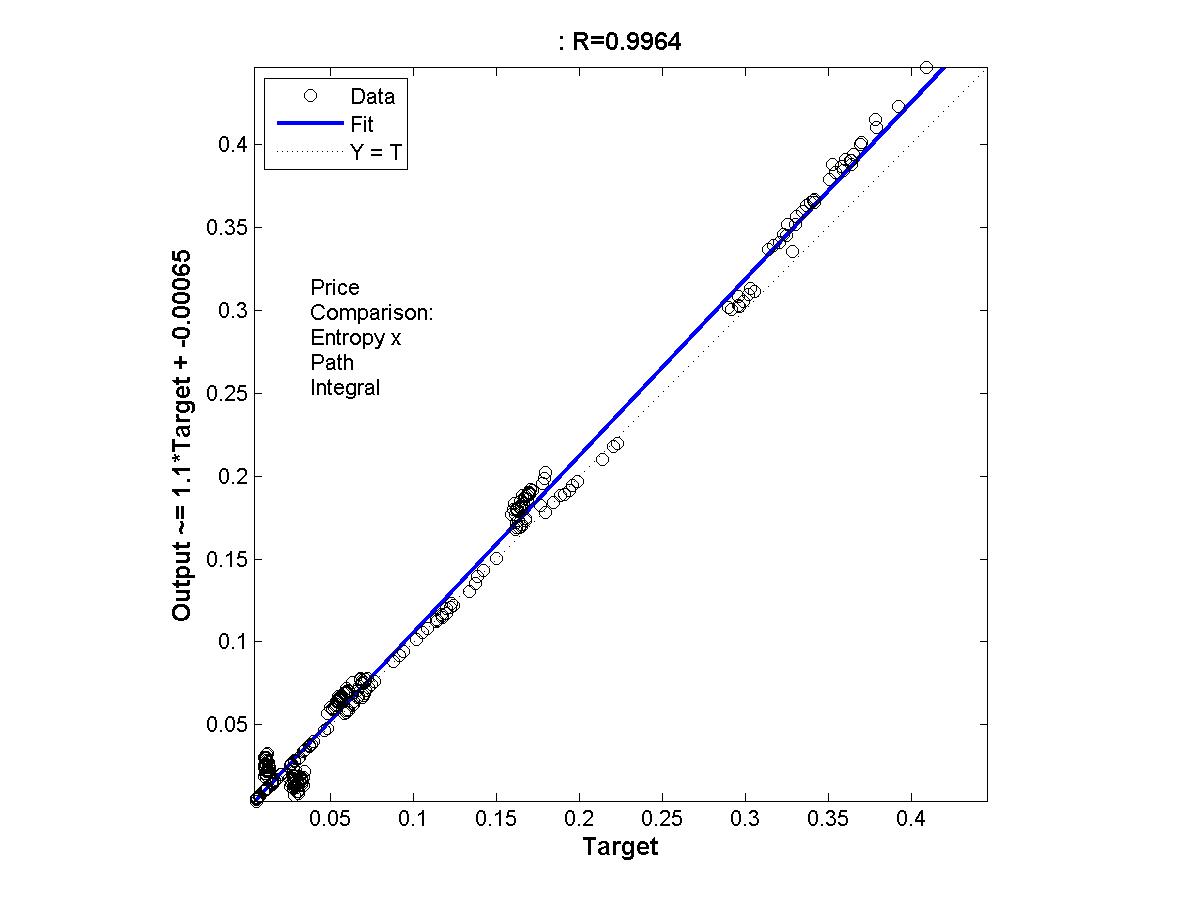}\caption{\label{fig:Regressao linear barreira tipo 2}Linear regression between
the relative entropy model (independent variable) and the path integral
model (dependent variable) of barrier option pricing.}
\end{figure}

In figure \ref{fig:Regressao linear barreira tipo 2}, we see that
the higher divergence in lower barrier levels at longer maturities
corresponds to the dispersion in the region of low prices in the graph
(target $<0.05$ and output $<0.05$).

\medskip{}

\medskip{}

\medskip{}

\medskip{}

\section{Conclusion}

\medskip{}

\hspace*{0.25in}In this article we presented a non-gaussian probability
distribution model, based on cumulant expansion in the the well-known
path integral formalism of Statistical Mechanics, including an absorbing,
deterministically moving, barrier. The idea relies on the work of
\cite{MagIV} for galaxy formation, in Cosmology, which we extend
to include drift and more cumulants than the original authors do.
In applying the model to option pricing in Finance, we find the condition
for a risk-neutral drift, and present an analytical method to price
deterministically moving absorbing barrier options. The development
encompasses analisys of the behaviour of the distribution in the vicinity
of the barrier. Usually, general distribution models used in such
products' pricing demand numerical and simulation methods; the work
of \cite{Kun92} being a case of closed-form solution, but under the
lognormal distribution hypothesis of \cite{BS73}.

In the case of constant barrier, we obtain an analytic non-gaussian
pricing model to price standard knock-up-and-out (KUO) barrier call
options. And, in the limit of infinite barrier, it becomes a non-gaussian
probability distribution model to price vanilla options. Since the
model parameters, volatility and cumulants, belong to both barrier
and vanilla versions of the model, we calibrate them with vanilla
option data and then price constant barrier KUO calls. Given that
barrier option pricing data contributors often provide market-to-model
values, we compare the constant barrier option prices of the path
integral model with the ones obtained from the relative entropy model. We adapted the approach of \cite{Ave01} 
to analytical constant barrier option pricing.

The results demonstrate that our model reproduces those obtained by the Entropy model.
The KUO barrier call option pricing presents larger differences when,
in long maturity contracts, the barrier is set close the initial underlying
price, and the delta is near 90\% (low strikes). However, in such
situation, the underlying process has enough time to reach the barrier,
deactivating the contract and, therefore, prices are expected to be
rather small in such combination, even more when there is the additional
condition that the call strike is high. So, these larger differences
between models refer to small prices; and it does not happen, for
instance, when the barrier, in a short maturity contract, is close
to the initial underlying price, since there is not enough time to
reach the barrier. In addition, we notice that such discrepancies
between the relative entropy and the path integral model we presented
also happened in the region of higher deltas, where the relative entropy
model did not fit properly to vanillas and, as we have emphasized, it is important to have a good calibration in the delta region associated to the barrier position.

Another point is that barrier option pricing requires the choice of
a larger number of cumulants than vanilla option pricing do, because
polynomial fine-tune corrections are important near the barrier.

Finally, as a future development, the model might include double barrier
contracts, stochastic barriers, and might also be extended to other
classes of products, such as interest rates.

\medskip{}
 \medskip{}

\pagebreak{}

\appendix


\section{Useful relations involving the path integral\label{sec:Rela=00003D00003D0000E7=00003D00003D0000F5es-=00003D00003D0000FAteis-envolvendo}}

\selectlanguage{english}%
\medskip{}

\hspace*{0.25in}Let $F\left(t_{i}\right)$ be a function. Consider
objects such as

\medskip{}

\begin{equation}
\stackrel[i=1]{n-1}{\sum}F\left(t_{i}\right)\int_{-\infty}^{\omega_{c}}d\omega_{1}...d\omega_{n-1}\partial_{i}W^{gm}\left(\omega_{0},\omega_{1},...,\omega_{n};t_{n}\right),\label{eq:96}
\end{equation}

\medskip{}

\noindent where $F$ is a generic function and $\partial_{i}\equiv\partial/\partial\omega_{i}$.
Because $W^{gm}$ is Gaussian, it is limited and tends to zero at
$-\infty$. Integrating,

\medskip{}

\[
\int_{-\infty}^{\omega_{c}}d\omega_{1}...d\omega_{n-1}\partial_{i}W^{gm}\left(\omega_{0},\omega_{1},...,\omega_{n};t_{n}\right)=
\]

\begin{equation}
\int_{-\infty}^{\omega_{c}}d\omega_{1}...d\hat{\omega}_{i}...d\omega_{n-1}W^{gm}\left(\omega_{0},\omega_{1},...,\omega_{i}=\omega_{c},...,\omega_{n-1},\omega_{n};t_{n}\right),\label{eq:97}
\end{equation}

\medskip{}

\noindent in which $\hat{\omega}_{i}$ denotes a variable that is
no longer in the integral, because it was integrated. The density
$W^{gm}$, by (\ref{eq:separacao}), satisfies

\medskip{}

\[
W^{gm}\left(\omega_{0},\omega_{1},...,\omega_{i}=\omega_{c},...,\omega_{n-1},\omega_{n};t_{n}\right)=
\]

\begin{equation}
W^{gm}\left(\omega_{0},\omega_{1},...,\omega_{i-1},\omega_{i}=\omega_{c};t_{i}\right)W^{gm}\left(\omega_{i}=\omega_{c},\omega_{i+1},...,\omega_{n-1},\omega_{n};t_{n}-t_{i}\right).
\end{equation}

\medskip{}

In (\ref{eq:97}),

\medskip{}

\[
\int_{-\infty}^{\omega_{c}}d\omega_{1}...d\omega_{i-1}\int_{-\infty}^{\omega_{c}}d\omega_{i+1}d\hat{\omega}_{i}...d\omega_{n-1}W^{gm}\left(\omega_{0},\omega_{1},...,\omega_{i-1},\omega_{i}=\omega_{c};t_{i}\right)\cdot
\]

\begin{equation}
W^{gm}\left(\omega_{i}=\omega_{c},\omega_{i+1},...,\omega_{n-1},\omega_{n};t_{n}-t_{i}\right)=\Pi_{\epsilon}^{gm}\left(\omega_{0},\omega_{c};t_{i}\right)\Pi_{\epsilon}^{gm}\left(\omega_{c},\omega_{n};t_{n}-t_{i}\right).\label{eq:99}
\end{equation}

\medskip{}

In (\ref{eq:96}),

\medskip{}

\[
\stackrel[i=1]{n-1}{\sum}F\left(t_{i}\right)\int_{-\infty}^{\omega_{c}}d\omega_{1}...d\omega_{n-1}\partial_{i}W^{gm}\left(\omega_{0},\omega_{1},...,\omega_{n};t_{n}\right)
\]

\begin{equation}
=\stackrel[i=1]{n-1}{\sum}F\left(t_{i}\right)\Pi_{\epsilon}^{gm}\left(\omega_{0},\omega_{c};t_{i}\right)\Pi_{\epsilon}^{gm}\left(\omega_{c},\omega_{n};t_{n}-t_{i}\right).\label{eq:100}
\end{equation}
\medskip{}

Another relationship, analogous to (\ref{eq:97}), is

\medskip{}

\[
\int_{-\infty}^{\omega_{c}}d\omega_{1}...d\omega_{n-1}\partial_{ij}W^{gm}\left(\omega_{0},\omega_{1},...,\omega_{n};t_{n}\right)
\]

\[
=\int_{-\infty}^{\omega_{c}}d\omega_{1}...d\omega_{i-1}\int_{-\infty}^{\omega_{c}}d\omega_{i+1}d\hat{\omega}_{i}...d\hat{\omega}_{j}...d\omega_{n-1}\cdot
\]

\begin{equation}
W^{gm}\left(\omega_{0},\omega_{1},...,\omega_{i-1},\omega_{i}=\omega_{c},...,\omega_{j-1},\omega_{j}=\omega_{c},...,\omega_{n-1},\omega_{n};t_{n}\right).\label{eq:101}
\end{equation}
\medskip{}

Because

\medskip{}

\[
W^{gm}\left(\omega_{0},\omega_{1},...,\omega_{i-1},\omega_{i}=\omega_{c},...,\omega_{j-1},\omega_{j}=\omega_{c},...,\omega_{n-1},\omega_{n};t_{,n}\right)
\]

\[
=W^{gm}\left(\omega_{0},\omega_{1},...,\omega_{i-1},\omega_{i}=\omega_{c};t_{i}\right)\cdot W^{gm}\left(\omega_{c},\omega_{i+1},...,\omega_{j-1},\omega_{j}=\omega_{c};t_{j}-t_{i}\right)\cdot
\]

\begin{equation}
W^{gm}\left(\omega_{c},\omega_{j+1},...,\omega_{n-1},\omega_{n};t_{n}-t_{j}\right),\label{eq:102}
\end{equation}

\medskip{}

\selectlanguage{brazil}%
we have:

\selectlanguage{english}%
\medskip{}

\[
\int_{-\infty}^{\omega_{c}}d\omega_{1}...d\omega_{n-1}\partial_{ij}W^{gm}\left(\omega_{0},\omega_{1},...,\omega_{n};t_{n}\right)
\]

\[
=\int_{-\infty}^{\omega_{c}}d\omega_{1}...d\omega_{i-1}\int_{-\infty}^{\omega_{c}}d\omega_{i+1}d\hat{\omega}_{i}...d\hat{\omega}_{j}...d\omega_{n-1}W^{gm}\left(\omega_{0},\omega_{1},...,\omega_{i-1},\omega_{i}=\omega_{c};t_{i}\right)\cdot
\]

\[
W^{gm}\left(\omega_{c},\omega_{i+1},...,\omega_{i-1},\omega_{i}=\omega_{c};t_{j}-t_{i}\right)\cdot W^{gm}\left(\omega_{c},\omega_{j+1},...,\omega_{n-1},\omega_{n};t_{n}-t_{j}\right)
\]

\begin{equation}
=\Pi_{\epsilon}^{gm}\left(\omega_{0},\omega_{c};t_{i}\right)\Pi_{\epsilon}^{gm}\left(\omega_{c},\omega_{c};t_{j}-t_{i}\right)\Pi_{\epsilon}^{gm}\left(\omega_{c},\omega_{n};t_{n}-t_{j}\right).\label{eq:103}
\end{equation}

\medskip{}

Besides, as in (\ref{eq:97}), we calculate the following integral:

\medskip{}

\[
\int_{-\infty}^{\omega_{c}}d\omega_{1}...d\omega_{n-1}\partial_{i}^{2}W^{gm}\left(\omega_{0},\omega_{1},...,\omega_{n};t_{n}\right)
\]

\[
=\int_{-\infty}^{\omega_{c}}d\omega_{1}...d\hat{\omega}_{i}...d\omega_{n-1}\partial_{i}W^{gm}\left(\omega_{0},\omega_{1},...,\omega_{i}=\omega_{c},...,\omega_{n-1},\omega_{n};t_{n}\right)
\]

\[
=\partial_{i}\int_{-\infty}^{\omega_{c}}d\omega_{1}...d\hat{\omega}_{i}...d\omega_{n-1}W^{gm}\left(\omega_{0},\omega_{1},...,\omega_{i}=\omega_{c},...,\omega_{n-1},\omega_{n};t_{n}\right)
\]

\begin{equation}
=\partial_{i}\left[\Pi_{\epsilon}^{gm}\left(\omega_{0},\omega_{i};t_{i}\right)\Pi_{\epsilon}^{gm}\left(\omega_{i},\omega_{n};t_{n}-t_{i}\right)\right]_{\omega_{i}=\omega_{c}}.\label{eq:104}
\end{equation}

\medskip{}

\selectlanguage{brazil}%
We also want to analyze the derivatives with respect to the barrier
$\omega_{c}$. \foreignlanguage{english}{Consider the derivative $\frac{\partial\Pi_{\epsilon}^{gm}}{\partial\omega_{c}}\left(\omega_{0}=0,\omega_{n};t_{n}\right)$:}

\selectlanguage{english}%
\medskip{}

\begin{equation}
\frac{\partial\Pi_{\epsilon}^{gm}}{\partial\omega_{c}}\left(\omega_{0}=0,\omega_{n};t_{n}\right)=\frac{\partial}{\partial\omega_{c}}\left(\int_{-\infty}^{\omega_{c}}d\omega_{1}...\int_{-\infty}^{\omega_{c}}d\omega_{n-1}W^{gm}\left(\omega_{0},\omega_{1},...,\omega_{n};t_{n}\right)\right),
\end{equation}

\medskip{}

\noindent where we used the definition (\ref{eq:densProbConfinamento}).
The limits of the integral depend on the variable with respect to
which we derive, $\omega_{c}$. By Leibniz rule:

\medskip{}

\begin{equation}
\frac{d}{dx}\int_{a(x)}^{b(x)}dtf(x,t)=f(b,t)\frac{db(x)}{dx}-f(a,t)\frac{da(x)}{dx}+\int_{a(x)}^{b(x)}dt\frac{d}{dx}f(x,t),\label{eq:105}
\end{equation}

\medskip{}

\[
\frac{\partial\Pi_{\epsilon}^{gm}}{\partial\omega_{c}}\left(\omega_{0}=0,\omega_{n};t_{n}\right)=\Pi_{\epsilon}^{gm}(\omega_{n}=\omega_{c},t_{n})\frac{\partial\omega_{c}}{\partial\omega_{c}}
\]

\selectlanguage{brazil}%
\begin{equation}
+\stackrel[i=1]{n-1}{\sum}\int_{-\infty}^{\omega_{c}}d\omega_{1}...d\omega_{n-1}\partial_{i}W^{gm}\left(\omega_{0},\omega_{1},...,\omega_{n};t_{n}\right).\label{eq:106-0}
\end{equation}

\selectlanguage{english}%
\medskip{}

\noindent where $\Pi_{\epsilon}^{gm}(\omega_{n}=\omega_{c},t_{n})$
is the Gaussian density with barrier, (\ref{eq:distB_BS}), which
becomes zero at the barrier. The first term on the RHS of (\ref{eq:106-0})
is zero, therefore. Then,

\selectlanguage{brazil}%
\[
\frac{\partial\Pi_{\epsilon}^{gm}}{\partial\omega_{c}}\left(\omega_{0}=0,\omega_{n};t_{n}\right)=\stackrel[i=1]{n-1}{\sum}\int_{-\infty}^{\omega_{c}}d\omega_{1}...d\omega_{n-1}\partial_{i}W^{gm}\left(\omega_{0},\omega_{1},...,\omega_{n};t_{n}\right)
\]

\[
=\stackrel[i=1]{n-1}{\sum}\int_{-\infty}^{\omega_{c}}d\omega_{1}...d\hat{\omega}_{i}...d\omega_{n-1}W^{gm}\left(\omega_{0},\omega_{1},...,\omega_{i}=\omega_{c},...,\omega_{n-1},\omega_{n};t_{n}\right)
\]

\begin{equation}
=\stackrel[i=1]{n-1}{\sum}\Pi_{\epsilon}^{gm}\left(\omega_{0},\omega_{c};t_{i}\right)\Pi_{\epsilon}^{gm}\left(\omega_{c},\omega_{n};t_{n}-t_{i}\right),\label{eq:106}
\end{equation}

\selectlanguage{english}%
\medskip{}

\noindent where we have used (\ref{eq:97}), (\ref{eq:99}).

The second derivative with respect to the barrier, $\frac{\partial^{2}\Pi_{\epsilon}^{gm}}{\partial\omega_{c}^{2}}\left(\omega_{0}=0,\omega_{n};t_{n}\right)$,
can also be computed. Initially, we note that

\medskip{}

\begin{equation}
\stackrel[i,j=1]{n-1}{\sum}\partial_{i}\partial_{j}=2\underset{i<j}{\sum}\partial_{i}\partial_{j}+\stackrel[i=1]{n-1}{\sum}\partial_{i}^{2}.\label{eq:107}
\end{equation}

\medskip{}

The first term in RHS (\ref{eq:107}) is equivalent to

\medskip{}

\begin{equation}
\stackrel[i=1]{n-2}{\sum}\stackrel[j=i+1]{n-1}{\sum}\partial_{i}\partial_{j}\label{eq:108}
\end{equation}

\medskip{}

\noindent or

\medskip{}

\begin{equation}
\stackrel[j=2]{n-1}{\sum}\stackrel[i=1]{j-1}{\sum}\partial_{i}\partial_{j}.\label{eq:109}
\end{equation}
\medskip{}

Thus, comparing to (\ref{eq:103}) and (\ref{eq:104}),

\medskip{}

\[
\frac{\partial^{2}\Pi_{\epsilon}^{gm}}{\partial\omega_{c}^{2}}\left(\omega_{0}=0,\omega_{n};t_{n}\right)=\stackrel[i,j=1]{n-1}{\sum}\int_{-\infty}^{\omega_{c}}d\omega_{1}...d\omega_{i-1}\int_{-\infty}^{\omega_{c}}d\omega_{i+1}d\hat{\omega}_{i}...d\hat{\omega}_{j}...d\omega_{n-1}\cdot
\]

\[
W^{gm}\left(\omega_{0},\omega_{1},...,\omega_{i-1},\omega_{i}=\omega_{c},...,\omega_{j-1},\omega_{j}=\omega_{c},...,\omega_{n-1},\omega_{n};t_{n}\right)
\]

\[
=2\stackrel[i=1]{n-2}{\sum}\stackrel[j=i+1]{n-1}{\sum}\int_{-\infty}^{\omega_{c}}d\omega_{1}...d\omega_{i-1}\int_{-\infty}^{\omega_{c}}d\omega_{i+1}d\hat{\omega}_{i}...d\hat{\omega}_{j}...d\omega_{n-1}\cdot
\]

\[
W^{gm}\left(\omega_{0},\omega_{1},...,\omega_{i-1},\omega_{i}=\omega_{c},...,\omega_{j-1},\omega_{j}=\omega_{c},...,\omega_{n-1},\omega_{n};t_{n}\right)
\]

\[
+\stackrel[i=1]{n-1}{\sum}\int_{-\infty}^{\omega_{c}}d\omega_{1}...d\hat{\omega}_{i}...d\omega_{n-1}\frac{\partial}{\partial\omega_{c}}W^{gm}\left(\omega_{0},\omega_{1},...,\omega_{i}=\omega_{c},...,\omega_{n-1},\omega_{n};t_{n}\right)
\]

\[
=2\stackrel[i=1]{n-2}{\sum}\stackrel[j=i+1]{n-1}{\sum}\int_{-\infty}^{\omega_{c}}d\omega_{1}...d\omega_{n-1}\partial_{i}\partial_{j}W^{gm}+\stackrel[i=1]{n-1}{\sum}\int_{-\infty}^{\omega_{c}}d\omega_{1}...d\omega_{n-1}\partial_{i}^{2}W^{gm}
\]

\begin{equation}
=\stackrel[i,j=1]{n-1}{\sum}\int_{-\infty}^{\omega_{c}}d\omega_{1}...d\omega_{n-1}\partial_{i}\partial_{j}W^{gm}
\end{equation}

\begin{equation}
\therefore\frac{\partial^{2}\Pi_{\epsilon}^{gm}}{\partial\omega_{c}^{2}}\left(\omega_{0}=0,\omega_{n};t_{n}\right)=\stackrel[i,j=1]{n-1}{\sum}\int_{-\infty}^{\omega_{c}}d\omega_{1}...d\omega_{n-1}\partial_{i}\partial_{j}W^{gm}.\label{eq:111}
\end{equation}

\medskip{}

Higher order derivatives follow analogous expressions.

\medskip{}


\section{Analysis of the Gaussian distribution near the barrier\label{sec:An=00003D00003D0000E1lise-quando-a}}

\selectlanguage{english}%
\medskip{}

\hspace*{0.25in}In this appendix we analyze, in the presence of barrier
$\omega_{c}$, the behaviour of the Gaussian density of probability,
given by (\ref{eq:distGaussiana}), when the variable $\omega_{n}$
approximates $\omega_{c}$. The first discussed situation is when
the variable starts at $\omega_{0}$, reaching the barrier $\omega_{c}$,
in $t_{n}$, $\Pi_{\epsilon\rightarrow0}^{gm}\left(\omega_{0},\omega_{n}=\omega_{c};t_{n}\right)$.
The second situation is when the variables starts in the barrier vicinity,
in $t_{0}$, and arrives at $\omega_{n}\neq\omega_{c}$ in $t_{n}$,
$\Pi_{\epsilon\rightarrow0}^{gm}\left(\omega_{0}=\omega_{c},\omega_{n};t_{n}\right)$.
The third situation corresponds to the case in which the variable
starts in the barrier vicinity and stays near it, $\Pi_{\epsilon\rightarrow0}^{gm}\left(\omega_{0}=\omega_{c},\omega_{n}=\omega_{c};t_{n}\right)$.
To do so, the behaviour of the Gaussian distribution near the barrier,
as the variable approaches it, is obtained by the Taylor expansion
in the element $\Delta\omega=\omega_{n}-\omega_{n-1}$. We will see
that there is a regime change in the relation of the probability with
time discretization $\epsilon$ as the variable tends to the barrier:
$\omega_{n}\rightarrow\omega_{c}$. To arrive at the expressions $\Pi_{\epsilon\rightarrow0}^{gm}\left(\omega_{0},\omega_{n}=\omega_{c};t_{n}\right)$,
$\Pi_{\epsilon\rightarrow0}^{gm}\left(\omega_{0}=\omega_{c},\omega_{n};t_{n}\right)$
and $\Pi_{\epsilon\rightarrow0}^{gm}\left(\omega_{0}=\omega_{c},\omega_{n}=\omega_{c};t_{n}\right)$,
we expand the Gaussian density in powers of $\sqrt{\epsilon}$.

\medskip{}

\medskip{}

\selectlanguage{brazil}%

\subsection{Behaviour of the Gaussian distribution near the barrier}

\selectlanguage{english}%
\medskip{}

\hspace*{0.25in}We start by getting some relations from $\Pi_{\epsilon}^{gm}\left(\omega_{0},\omega_{n};t_{n}\right)$
(\ref{eq:pi gm}). From (\ref{eq:deltaOmega}), $\omega_{n-1}=\omega_{n}-\Delta\omega$.
In (\ref{eq:pi gm}), changing the variable to $\Delta\omega$, $d\left(\Delta\omega\right)=-d\omega_{n-1}$,
for a given $\omega_{n}$, fixed. Thus, the limits of integration
are $\Delta\omega|_{1}=\omega_{n}-\omega_{n-1}|_{\omega_{C}}=\omega_{n}-\omega_{c}$
and $\Delta\omega|_{2}=\omega_{n}-\omega_{n-1}|_{-\infty}=\infty$;
inverting them, due to the negative sign coming from $d\left(\Delta\omega\right)$,
we have:

\medskip{}

\[
\Pi_{\epsilon}^{gm}\left(\omega_{0},\omega_{n};t_{n}=t_{n-1}+\epsilon\right)=-\int_{\infty}^{\omega_{n}-\omega_{c}}d\left(\Delta\omega\right)\Psi_{\epsilon}\left(\Delta\omega\right)\Pi_{\epsilon}^{gm}\left(\omega_{0},\omega_{n-1};t_{n-1}\right)
\]

\[
=\int_{\omega_{n}-\omega_{c}}^{\infty}d\left(\Delta\omega\right)\Psi_{\epsilon}\left(\Delta\omega\right)\Pi_{\epsilon}^{gm}\left(\omega_{0},\omega_{n-1};t_{n-1}\right)
\]

\begin{equation}
=\int_{\omega_{n}-\omega_{c}}^{\infty}d\left(\Delta\omega\right)\Psi_{\epsilon}\left(\Delta\omega\right)\Pi_{\epsilon}^{gm}\left(\omega_{0},\omega_{n}-\Delta\omega;t_{n-1}\right).\label{eq:PiGaussianaBarreira1}
\end{equation}

\medskip{}

Notice that

\medskip{}

\begin{equation}
\underset{\epsilon\rightarrow0}{lim}\Psi_{\epsilon}\left(\Delta\omega\right)=\delta\left(\Delta\omega\right).
\end{equation}

\medskip{}

If $\omega_{n}-\omega_{c}<0$ (that is, if the variable crosses the
barrier after $t_{n}$), it includes the support of the Dirac $\delta$
function. If $\omega_{n}-\omega_{c}>0$, the integral is zero because
it is outside the support. And, if $\omega_{n}=\omega_{c}$, with
the condition of the continuum limit $\epsilon\rightarrow0$, it goes
to zero because of the initial condition. Therefore,

\medskip{}

\begin{equation}
\Pi_{\epsilon\rightarrow0}^{gm}\left(\omega_{0},\omega_{n};t_{n}\right)=0,\;if\;\omega_{n}\geq\omega_{c}.\label{eq:AnaliseAcimaBarreira}
\end{equation}

\medskip{}

In the case $\omega_{n}<\omega_{c}$, we analyze (\ref{eq:PiGaussianaBarreira1}),
starting by expanding its LHS, in terms of $t_{n-1}$:

\medskip{}

\[
\Pi_{\epsilon}^{gm}\left(\omega_{0},\omega_{n};t_{n-1}+\epsilon\right)=\Pi_{\epsilon}^{gm}\left(\omega_{0},\omega_{n};t_{n-1}\right)
\]

\begin{equation}
+\epsilon\frac{\partial\Pi_{\epsilon}^{gm}\left(\omega_{0},\omega_{n};t_{n-1}\right)}{\partial t_{n-1}}+\frac{\epsilon^{2}}{2}\frac{\partial^{2}\Pi_{\epsilon}^{gm}\left(\omega_{0},\omega_{n};t_{n-1}\right)}{\partial t_{n-1}^{2}}+...
\end{equation}

\medskip{}

Expanding in Taylor series the RHS (\ref{eq:PiGaussianaBarreira1}),
in terms of $\Delta\omega$, in the region of $\Delta\omega=0$, that
is, when $\omega_{n}\rightarrow\omega_{c}$:

\medskip{}

\[
\int_{\omega_{n}-\omega_{c}}^{\infty}d\left(\Delta\omega\right)\Psi_{\epsilon}\left(\Delta\omega\right)\Pi_{\epsilon}^{gm}\left(\omega_{0},\omega_{n}-\Delta\omega;t_{n-1}\right)=
\]

\begin{equation}
=\stackrel[i=0]{\infty}{\sum}\frac{\left(-1\right)^{i}}{i!}\frac{\partial^{i}\Pi_{\epsilon}^{gm}\left(\omega_{0},\omega_{n};t_{n-1}\right)}{\partial\omega_{n}^{i}}\int_{\omega_{n}-\omega_{c}}^{\infty}d\left(\Delta\omega\right)\left(\Delta\omega\right)^{i}\Psi_{\epsilon}\left(\Delta\omega\right)\label{eq:expansao na integral}
\end{equation}

\medskip{}

\noindent where, when expanding in $\Delta\omega$, which involves
$\frac{\partial\Pi_{\epsilon}^{gm}\left(\omega_{0},\omega_{n}-\Delta\omega;t_{n-1}\right)}{\partial\Delta\omega}$,
we applied the chain rule (also considering the expansion in the region
$\Delta\omega=0$): \medskip{}

\begin{equation}
\frac{\partial}{\partial\Delta\omega}=\frac{\partial}{\partial\left(\omega_{n}-\Delta\omega\right)}\frac{\partial\left(\omega_{n}-\Delta\omega\right)}{\partial\Delta\omega}=\left.\frac{\partial}{\partial\left(\omega_{n}-\Delta\omega\right)}\cdot\left(-1\right)\right|_{\Delta\omega=0}=\frac{\partial}{\partial\omega_{n}}\cdot\left(-1\right).
\end{equation}

\medskip{}

For second order,

\medskip{}

\[
\frac{\partial^{2}}{\partial\Delta\omega^{2}}=\frac{\partial}{\partial\Delta\omega}\left(\frac{\partial}{\partial\Delta\omega}\right)=\frac{\partial}{\partial\left(\omega_{n}-\Delta\omega\right)}\cdot\left(\frac{\partial}{\partial\left(\omega_{n}-\Delta\omega\right)}\cdot\left(-1\right)\right)\frac{\partial\left(\omega_{n}-\Delta\omega\right)}{\partial\Delta\omega}
\]

\begin{equation}
=\left.\frac{\partial^{2}}{\partial\left(\omega_{n}-\Delta\omega\right)^{2}}\cdot\left(-1\right)\right|_{\Delta\omega=0}=\frac{\partial^{2}}{\partial\omega_{n}^{2}}\cdot\left(-1\right)^{2}.
\end{equation}

\medskip{}

Then, for $i-th$ order,\medskip{}

\begin{equation}
\frac{\partial^{i}}{\partial\Delta\omega^{i}}=\frac{\partial^{i}}{\partial\omega_{n}^{i}}\cdot\left(-1\right)^{i}.
\end{equation}

\medskip{}

Back to (\ref{eq:expansao na integral}), we see that, changing to
the variable $y=\frac{\Delta\omega}{\sqrt{2\epsilon}}$,

\medskip{}

\begin{equation}
\int_{\omega_{n}-\omega_{c}}^{\infty}d\left(\Delta\omega\right)\left(\Delta\omega\right)^{i}\Psi_{\epsilon}\left(\Delta\omega\right)=\frac{\left(2\epsilon\right)^{i/2}}{\sqrt{\pi}}\int_{-\frac{\left(\omega_{c}-\omega_{n}\right)}{\sqrt{2\epsilon}}}^{\infty}dy\cdot y^{i}e^{-y^{2}}.\label{eq:IntegralExpdw}
\end{equation}

\medskip{}

\selectlanguage{brazil}%
Consider the cases $i=0$ and $i=1$, with \foreignlanguage{english}{$\omega_{n}\rightarrow\omega_{c}$:}

\selectlanguage{english}%
\medskip{}

\begin{equation}
\left.\int_{\omega_{n}-\omega_{c}}^{\infty}d\left(\Delta\omega\right)\left(\Delta\omega\right)^{i}\Psi_{\epsilon}\left(\Delta\omega\right)\right|_{i=0,\omega_{n}\rightarrow\omega_{c}}\implies\int_{0}^{\infty}d\left(\Delta\omega\right)\Psi_{\epsilon}\left(\Delta\omega\right)=\frac{1}{2}\label{eq:integral Gaussiana 0 inf}
\end{equation}

\medskip{}

\selectlanguage{brazil}%
\[
\left.\int_{\omega_{n}-\omega_{c}}^{\infty}d\left(\Delta\omega\right)\left(\Delta\omega\right)^{i}\Psi_{\epsilon}\left(\Delta\omega\right)\right|_{i=1,\omega_{n}\rightarrow\omega_{c}}\implies\int_{0}^{\infty}d\left(\Delta\omega\right)\left(\Delta\omega\right)\Psi_{\epsilon}\left(\Delta\omega\right)
\]

\begin{equation}
=\frac{\left(2\epsilon\right)^{1/2}}{\sqrt{\pi}}\int_{0}^{\infty}dy\cdot y\cdot e^{-y^{2}}=\left(\frac{\epsilon}{2\pi}\right)^{1/2}\label{eq:integralExpdW1}
\end{equation}

\selectlanguage{english}%
\medskip{}

\noindent where we integrated by parts. Back to (\ref{eq:expansao na integral}),
taking, as mentioned, the terms $i=0$ and $i=1$,

\medskip{}

\[
\int_{0}^{\infty}d\left(\Delta\omega\right)\Psi_{\epsilon}\left(\Delta\omega\right)\Pi_{\epsilon}^{gm}\left(\omega_{0},\omega_{n}-\Delta\omega;t_{n-1}\right)
\]

\begin{equation}
=\frac{1}{2}\Pi_{\epsilon}^{gm}\left(\omega_{0},\omega_{c};t_{n-1}\right)-\left(\frac{\epsilon}{2\pi}\right)^{1/2}\left.\frac{\partial}{\partial\omega_{n}}\Pi_{\epsilon}^{gm}\left(\omega_{0},\omega_{n};t_{n-1}\right)\right|_{\omega_{n}=\omega_{c}}+...\label{eq:expansao Integral}
\end{equation}

\medskip{}

Therefore, when $\omega_{n}-\omega_{c}<0$, the behaviour when $\omega_{n}\rightarrow\omega_{c}$,
with $\epsilon\rightarrow0$, depends on $\sqrt{\epsilon}$. Consider
the case $\omega_{n}-\omega_{c}<0$, but when we are \textbf{not}
in the situation $\omega_{n}\rightarrow\omega_{c}$. In this case,
in the continuum $\epsilon\rightarrow0$, the inferior limit $-\frac{\left(\omega_{c}-\omega_{n}\right)}{\sqrt{2\epsilon}}$
of the integral (\ref{eq:IntegralExpdw}) becomes $-\infty$. Then,
the new integral (\ref{eq:IntegralExpdw}), with new limits, is

\medskip{}

\begin{equation}
\int_{\omega_{n}-\omega_{c}}^{\infty}d\left(\Delta\omega\right)\left(\Delta\omega\right)^{i}\Psi_{\epsilon}\left(\Delta\omega\right)=\frac{\left(2\epsilon\right)^{i/2}}{\sqrt{\pi}}\int_{-\frac{\left(\omega_{c}-\omega_{n}\right)}{\sqrt{2\epsilon}}}^{\infty}dy\cdot y^{i}e^{-y^{2}}\rightarrow\frac{\left(2\epsilon\right)^{i/2}}{\sqrt{\pi}}\int_{-\infty}^{\infty}dy\cdot y^{i}e^{-y^{2}}.
\end{equation}
\medskip{}

Because

\medskip{}

\selectlanguage{brazil}%
\begin{equation}
\int_{-\infty}^{\infty}dy\cdot y^{n}e^{-y^{2}}\cong\frac{1+\left(-1\right)^{n}}{2}\frac{\sqrt{\pi}}{2^{n/2}}\left(n-1\right)!!,
\end{equation}

\selectlanguage{english}%
\medskip{}

\noindent then

\medskip{}

\begin{equation}
\int_{\omega_{n}-\omega_{c}}^{\infty}d\left(\Delta\omega\right)\left(\Delta\omega\right)^{i}\Psi_{\epsilon}\left(\Delta\omega\right)\rightarrow\left\{ \begin{array}{ccc}
\epsilon^{i/2}\left(n-1\right)!! &  & i\;even\\
\\
0 &  & i\;odd
\end{array}\right..
\end{equation}
\medskip{}

In this case, the expansion \ref{eq:expansao na integral}, with the
analogous terms to (\ref{eq:integral Gaussiana 0 inf}) and (\ref{eq:integralExpdW1}),
respectively, for $i=0$ and $i=2$ are:

\medskip{}

\begin{equation}
\left.\int_{\omega_{n}-\omega_{c}}^{\infty}d\left(\Delta\omega\right)\left(\Delta\omega\right)^{i}\Psi_{\epsilon}\left(\Delta\omega\right)\right|_{i=0,\omega_{n}<\omega_{c}}\implies\int_{-\infty}^{\infty}d\left(\Delta\omega\right)\Psi_{\epsilon}\left(\Delta\omega\right)=1
\end{equation}

\selectlanguage{brazil}%
\begin{equation}
\left.\int_{\omega_{n}-\omega_{c}}^{\infty}d\left(\Delta\omega\right)\left(\Delta\omega\right)^{i}\Psi_{\epsilon}\left(\Delta\omega\right)\right|_{i=2,\omega_{n}<\omega_{c}}=\epsilon.
\end{equation}

\selectlanguage{english}%
\medskip{}

Then, in the case $\omega_{n}<\omega_{c}$, $\epsilon\rightarrow0$,
but $\omega_{n}$ not near $\omega_{c}$,

\medskip{}

\[
\int_{-\infty}^{\infty}d\left(\Delta\omega\right)\Psi_{\epsilon}\left(\Delta\omega\right)\Pi_{\epsilon}^{gm}\left(\omega_{0},\omega_{n}-\Delta\omega;t_{n-1}\right)
\]

\begin{equation}
=\Pi_{\epsilon}^{gm}\left(\omega_{0},\omega_{c};t_{n-1}\right)+\frac{\epsilon}{2}\left.\frac{\partial^{2}}{\partial\omega_{n}^{2}}\Pi_{\epsilon}^{gm}\left(\omega_{0},\omega_{n};t_{n-1}\right)\right|_{\omega_{n}=\omega_{c}}+...\label{eq:expansao Integral2}
\end{equation}
\medskip{}

Thus, there is a regime change with respect to $\epsilon$ in the
passage from $\omega_{n}<\omega_{c}$ to the case $\omega_{n}<\omega_{c}$,
with the extra condition $\omega_{n}\rightarrow\omega_{c}$: in the
first case, the next leading term of the expansion (\ref{eq:expansao Integral2})
behaves as $\epsilon$, while in the second, the next leading term
dominating the expansion (\ref{eq:expansao Integral}) follows $\sqrt{\epsilon}$.
Such transition is ruled by the inferior limit $\frac{\left(\omega_{c}-\omega_{n}\right)}{\sqrt{2\epsilon}}$
in integral (\ref{eq:IntegralExpdw}). To tackle this, we define

\medskip{}

\begin{equation}
\eta=\frac{\left(\omega_{c}-\omega_{n}\right)}{\sqrt{2\epsilon}}.\label{eq:def eta}
\end{equation}
\medskip{}

\selectlanguage{brazil}%
We write $\Pi_{\epsilon}^{gm}$ in the form

\selectlanguage{english}%
\medskip{}

\begin{equation}
\Pi_{\epsilon}^{gm}\left(\omega_{0},\omega_{n};t_{n}\right)=C_{\epsilon}\left(\omega_{0},\omega_{n};t_{n}\right)v\left(\eta\right).\label{eq:C*v}
\end{equation}
\medskip{}

\selectlanguage{brazil}%
Here, $C$ is the smooth part of the function, while $v\left(\eta\right)$
is responsible for the regime transition of the function $\Pi_{\epsilon}^{gm}$.
We must impose

\selectlanguage{english}%
\medskip{}

\begin{equation}
\underset{\eta\rightarrow\infty}{lim}v\left(\eta\right)=1\label{eq:171-1}
\end{equation}
\medskip{}

\noindent so that $C$ is the solution of $\Pi_{\epsilon}^{gm}$ when
$\omega_{c}-\omega_{n}$ is finite and positive. Consider the Gaussian
probability density, which, in the continuum limit ($\epsilon\rightarrow0)$,
in the presence of a barrier, is the solution (\ref{eq:distB_BS}):

\medskip{}

\begin{equation}
\Pi_{\epsilon\rightarrow0}^{gm}\left(\omega_{0},\omega_{n};t_{n}\right)=\frac{1}{\sqrt{2\pi t_{n}}}e^{\alpha\left(\omega_{n}-\omega_{0}\right)-\frac{1}{2}\alpha^{2}t_{n}}\left[e^{-\frac{\left(\omega_{n}-\omega_{0}\right)^{2}}{2t_{n}}}-e^{-\frac{\left(2\omega_{c}-\omega_{n}-\omega_{0}\right)^{2}}{2t_{n}}}\right].\label{eq:PiGaussianaBarreira}
\end{equation}
\medskip{}

Isolating $\omega_{n}$ in (\ref{eq:def eta}),

\medskip{}

\begin{equation}
\Pi_{\epsilon\rightarrow0}^{gm}\left(\omega_{0},\omega_{n};t_{n}\right)=\frac{1}{\sqrt{2\pi t_{n}}}e^{\alpha\left(\omega_{c}-\eta\sqrt{2\epsilon}-\omega_{0}\right)-\frac{1}{2}\alpha^{2}t_{n}}\left[e^{-\frac{\left(\omega_{c}-\eta\sqrt{2\epsilon}-\omega_{0}\right)^{2}}{2t_{n}}}-e^{-\frac{\left(\omega_{c}+\eta\sqrt{2\epsilon}-\omega_{0}\right)^{2}}{2t_{n}}}\right].
\end{equation}
\medskip{}

Expanding in powers of $\sqrt{\epsilon}$, which means that $\eta\rightarrow\infty$
and, consequently, \eqref{eq:171-1} :\medskip{}

\begin{equation}
exp\left(-\frac{1}{2t_{n}}\left(\omega_{c}\mp\eta\sqrt{2\epsilon}-\omega_{0}\right)^{2}\right)=e^{-\frac{1}{2t_{n}}\left(\omega_{c}-\omega_{0}\right)^{2}}\left[1\pm\frac{\eta\sqrt{2}}{2t_{n}}2\left(\omega_{c}-\omega_{0}\right)\sqrt{\epsilon}+\ldots\right]
\end{equation}

\begin{equation}
exp\left(\alpha\left(\omega_{c}-\eta\sqrt{2\epsilon}-\omega_{0}\right)\right)=e^{\alpha\left(\omega_{c}-\omega_{0}\right)}\left[1-\eta\sqrt{2}\alpha\sqrt{\epsilon}+\ldots\right]
\end{equation}

\[
\therefore C_{\epsilon}\left(\omega_{0}=0,\omega_{n},T\right)=\frac{1}{\sqrt{2\pi t_{n}}}e^{-\frac{1}{2}\alpha^{2}t_{n}}e^{\alpha\left(\omega_{c}-\omega_{0}\right)}e^{-\frac{1}{2t_{n}}\left(\omega_{c}-\omega_{0}\right)^{2}}\left[1-\eta\sqrt{2}\alpha\sqrt{\epsilon}+\ldots\right]\cdot
\]

\[
\cdot\left[1+\frac{\eta\sqrt{2}}{t_{n}}\left(\omega_{c}-\omega_{0}\right)\sqrt{\epsilon}-1+\frac{\eta\sqrt{2}}{t_{n}}\left(\omega_{c}-\omega_{0}\right)\sqrt{\epsilon}+\ldots\right]
\]

\[
=\frac{1}{\sqrt{2\pi t_{n}}}e^{-\frac{1}{2}\alpha^{2}t_{n}}e^{\alpha\left(\omega_{c}-\omega_{0}\right)}e^{-\frac{1}{2t_{n}}\left(\omega_{c}-\omega_{0}\right)^{2}}\left[1-\eta\sqrt{2}\alpha\sqrt{\epsilon}+\ldots\right]\left[2\frac{\eta\sqrt{2}}{t_{n}}\left(\omega_{c}-\omega_{0}\right)\sqrt{\epsilon}+\ldots\right]
\]

\begin{equation}
=\sqrt{\epsilon}\frac{2\eta}{\sqrt{\pi}}\frac{\left(\omega_{c}-\omega_{0}\right)}{t_{n}^{3/2}}e^{-\frac{1}{2}\alpha^{2}t_{n}}e^{\alpha\left(\omega_{c}-\omega_{0}\right)}e^{-\frac{1}{2t_{n}}\left(\omega_{c}-\omega_{0}\right)^{2}}+\mathcal{O}\left(\epsilon\right).\label{eq:127}
\end{equation}

\medskip{}

\selectlanguage{brazil}%
In (\ref{eq:C*v}), when $\omega_{n}\rightarrow\omega_{c}$,

\selectlanguage{english}%
\medskip{}

\selectlanguage{brazil}%
\begin{equation}
\Pi_{\epsilon}^{gm}\left(\omega_{0}=0,\omega_{n};t_{n}=T\right)=\sqrt{\epsilon}\gamma\frac{\left(\omega_{c}-\omega_{0}\right)}{t_{n}^{3/2}}e^{-\frac{1}{2}\alpha^{2}t_{n}}e^{\alpha\left(\omega_{c}-\omega_{0}\right)}e^{-\frac{1}{2t_{n}}\left(\omega_{c}-\omega_{0}\right)^{2}}+\mathcal{O}\left(\epsilon\right)\label{eq:128}
\end{equation}

\begin{equation}
\gamma=\frac{2}{\sqrt{\pi}}\underset{\eta\rightarrow0}{lim}\eta v\left(\eta\right).\label{eq:133-1}
\end{equation}

\selectlanguage{english}%
\medskip{}

In this equation, the limit is taken as $\eta\rightarrow0$, because
it corresponds to $\omega_{n}\rightarrow\omega_{c}$ in definition
(\ref{eq:def eta}). Next, we will show that

\medskip{}

\begin{equation}
\gamma=\frac{1}{\sqrt{\pi}}e^{\alpha\left(\omega_{n}-\omega_{c}\right)}.\label{eq:gama}
\end{equation}
\medskip{}

Now consider the derivative with respect to the barrier, given by
(\ref{eq:106}). In the limit $\epsilon\rightarrow0$,

\medskip{}

\begin{equation}
\stackrel[i=1]{n-1}{\sum}\rightarrow\frac{1}{\epsilon}\int_{0}^{t_{n}}dt_{i}.\label{eq:133}
\end{equation}

\medskip{}

In (\ref{eq:106}),

\medskip{}

\[
\frac{\partial\Pi_{\epsilon\rightarrow0}^{gm}}{\partial\omega_{c}}\left(\omega_{0}=0,\omega_{n};t_{n}\right)=\stackrel[i=1]{n-1}{\sum}\Pi_{\epsilon}^{gm}\left(\omega_{0},\omega_{c};t_{i}\right)\Pi_{\epsilon}^{gm}\left(\omega_{c},\omega_{n};t_{n}-t_{i}\right)
\]

\begin{equation}
=\int_{0}^{t_{n}}dt_{i}\underset{\epsilon\rightarrow0}{lim}\frac{1}{\epsilon}\Pi_{\epsilon}^{gm}\left(\omega_{0},\omega_{c};t_{i}\right)\Pi_{\epsilon}^{gm}\left(\omega_{c},\omega_{n};t_{n}-t_{i}\right).\label{eq:134}
\end{equation}
\medskip{}

\selectlanguage{brazil}%
The LHS of (\ref{eq:134}) is computed by (\ref{eq:PiGaussianaBarreira}).
In the case $\omega_{0}\neq0$,

\selectlanguage{english}%
\medskip{}

\[
\frac{\partial\Pi_{\epsilon\rightarrow0}^{gm}}{\partial\omega_{c}}\left(\omega_{0},\omega_{n};t_{n}\right)=\frac{2}{\sqrt{2\pi t_{n}}}e^{\alpha\left(\omega_{n}-\omega_{0}\right)-\frac{1}{2}\alpha^{2}t_{n}}\left[\frac{\left(2\omega_{c}-\omega_{n}-\omega_{0}\right)}{t_{n}}e^{-\frac{\left(2\omega_{c}-\omega_{n}-\omega_{0}\right)^{2}}{2t_{n}}}\right]
\]

\[
=\left(\frac{2}{\pi}\right)^{1/2}\frac{\left(2\omega_{c}-\omega_{n}-\omega_{0}\right)}{t_{n}^{3/2}}e^{\alpha\left(\omega_{n}-\omega_{0}\right)-\frac{1}{2}\alpha^{2}t_{n}}e^{-\frac{\left(2\omega_{c}-\omega_{n}-\omega_{0}\right)^{2}}{2t_{n}}}
\]

\begin{equation}
=\left(\frac{2}{\pi}\right)^{1/2}\frac{\left(2\omega_{c}-\omega_{n}-\omega_{0}\right)}{t_{n}^{3/2}}e^{2\alpha\left(\omega_{n}-\omega_{c}\right)}e^{-\frac{\left(2\omega_{c}-\omega_{n}-\omega_{0}-\alpha t_{n}\right)^{2}}{2t_{n}}}.\label{eq:135}
\end{equation}

\medskip{}

\selectlanguage{brazil}%
The RHS (\ref{eq:134}) is given by (\ref{eq:128}). We notice that

\selectlanguage{english}%
\medskip{}

\selectlanguage{brazil}%
\begin{equation}
\Pi_{\epsilon\rightarrow0}^{gm}\left(\omega_{c},\omega_{n};t_{n}\right)=\Pi_{\epsilon\rightarrow0}^{gm}\left(\omega_{n},\omega_{c};t_{n}\right).\label{eq:136}
\end{equation}

\selectlanguage{english}%
\medskip{}

This can be seen by (\ref{eq:densProbConfinamento}) and (\ref{eq:30}):

\medskip{}

\[
\Pi_{\epsilon\rightarrow0}^{gm}\left(\omega_{c},\omega_{n};t_{n}\right)\Rightarrow
\]

\begin{equation}
W^{gm}\left(\omega_{c},\omega_{1},...,\omega_{n};t_{n}\right)=\frac{1}{\left(2\pi\epsilon\right)^{n/2}}e^{-\frac{\left(\omega_{1}-\omega_{c}\right)^{2}}{2\epsilon}-\stackrel[i=1]{n-2}{\sum}\left[\frac{\left(\omega_{i+1}-\omega_{i}\right)^{2}}{2\epsilon}\right]-\frac{\left(\omega_{n}-\omega_{n-1}\right)^{2}}{2\epsilon}}
\end{equation}

\[
\Pi_{\epsilon\rightarrow0}^{gm}\left(\omega_{n},\omega_{c};t_{n}\right)\Rightarrow
\]

\begin{equation}
W^{gm}\left(\omega_{n},\omega_{1},...,\omega_{c};t_{n}\right)=\frac{1}{\left(2\pi\epsilon\right)^{n/2}}e^{-\frac{\left(\omega_{1}-\omega_{n}\right)^{2}}{2\epsilon}-\stackrel[i=1]{n-2}{\sum}\left[\frac{\left(\omega_{i+1}-\omega_{i}\right)^{2}}{2\epsilon}\right]-\frac{\left(\omega_{c}-\omega_{n-1}\right)^{2}}{2\epsilon}}.
\end{equation}

\medskip{}

\selectlanguage{brazil}%
This happens because $\omega_{0}\left(=\omega_{c}\right)$ and $\omega_{n}$
are not integration variables in \foreignlanguage{english}{(\ref{eq:densProbConfinamento}),
that is, $\omega_{c}$ to be with $\omega_{n-1}$ or $\omega_{1}$
doen not matter, because they are integrated in the same integration
limits. Therefore, (\ref{eq:136}) is valid. Equation (\ref{eq:128}),
with $\omega_{0}\neq0$ is written as:}

\selectlanguage{english}%
\medskip{}

\begin{equation}
\Pi_{\epsilon\rightarrow0}^{gm}\left(\omega_{0},\omega_{c};t_{n}\right)=\sqrt{\epsilon}\gamma\frac{\omega_{c}-\omega_{0}}{t_{n}^{3/2}}e^{-\frac{1}{2}\alpha^{2}t_{n}}e^{\alpha\left(\omega_{c}-\omega_{0}\right)}e^{-\frac{1}{2t_{n}}\left(\omega_{c}-\omega_{0}\right)^{2}}.\label{eq:139}
\end{equation}

\medskip{}

\selectlanguage{brazil}%
Then,

\selectlanguage{english}%
\medskip{}

\[
\Pi_{\epsilon\rightarrow0}^{gm}\left(\omega_{c},\omega_{n};t_{n}\right)=\Pi_{\epsilon\rightarrow0}^{gm}\left(\omega_{n},\omega_{c};t_{n}\right)
\]

\begin{equation}
=\sqrt{\epsilon}\gamma\frac{\omega_{c}-\omega_{n}}{t_{n}^{3/2}}e^{-\frac{1}{2}\alpha^{2}t_{n}}e^{\alpha\left(\omega_{c}-\omega_{n}\right)}e^{-\frac{1}{2t_{n}}\left(\omega_{c}-\omega_{n}\right)^{2}}.\label{eq:140}
\end{equation}
\medskip{}

Returning to (\ref{eq:134}), in $\Pi_{\epsilon}^{gm}\left(\omega_{0},\omega_{c};t_{i}\right)$
we use (\ref{eq:128}), and, in $\Pi_{\epsilon}^{gm}\left(\omega_{c},\omega_{n};t_{n}-t_{i}\right)$,
(\ref{eq:140}). We also use

\medskip{}

\begin{equation}
\int_{0}^{c}dx\frac{1}{x^{3/2}\left(c-x\right)^{3/2}}e^{-\frac{a^{2}}{2x}-\frac{b^{2}}{2\left(c-x\right)}}=\sqrt{2\pi}\frac{a+b}{ab}\frac{1}{c^{3/2}}e^{-\frac{\left(a+b\right)^{2}}{2c}},\label{eq:141}
\end{equation}

\medskip{}

\noindent so,

\medskip{}

\[
\int_{0}^{t_{n}}dt_{i}\underset{\epsilon\rightarrow0}{lim}\frac{1}{\epsilon}\Pi_{\epsilon}^{gm}\left(\omega_{0},\omega_{c};t_{i}\right)\Pi_{\epsilon}^{gm}\left(\omega_{c},\omega_{n};t_{n}-t_{i}\right)
\]

\[
=\int_{0}^{t_{n}}dt_{i}\underset{\epsilon\rightarrow0}{lim}\frac{1}{\epsilon}\left(\sqrt{\epsilon}\gamma\frac{\omega_{c}-\omega_{0}}{t_{i}^{3/2}}e^{-\frac{1}{2}\alpha^{2}t_{i}}e^{\alpha\left(\omega_{c}-\omega_{0}\right)}e^{-\frac{1}{2t_{i}}\left(\omega_{c}-\omega_{0}\right)^{2}}\right)\cdot
\]

\[
\cdot\left(\sqrt{\epsilon}\gamma\frac{\omega_{c}-\omega_{n}}{\left(t_{n}-t_{i}\right)^{3/2}}e^{-\frac{1}{2}\alpha^{2}\left(t_{n}-t_{i}\right)}e^{\alpha\left(\omega_{c}-\omega_{n}\right)}e^{-\frac{1}{2\left(t_{n}-t_{i}\right)}\left(\omega_{c}-\omega_{n}\right)^{2}}\right)
\]

\[
=\int_{0}^{t_{n}}dt_{i}\gamma^{2}\frac{\left(\omega_{c}-\omega_{0}\right)\left(\omega_{c}-\omega_{n}\right)}{\left(t_{n}-t_{i}\right)^{3/2}t_{i}^{3/2}}e^{\alpha\left(\omega_{c}-\omega_{0}\right)}e^{\alpha\left(\omega_{c}-\omega_{n}\right)}e^{-\frac{1}{2t_{i}}\left(\omega_{c}-\omega_{0}\right)^{2}}e^{-\frac{1}{2\left(t_{n}-t_{i}\right)}\left(\omega_{c}-\omega_{n}\right)^{2}}e^{-\frac{1}{2}\alpha^{2}t_{n}}
\]

\selectlanguage{brazil}%
\begin{equation}
=\sqrt{2\pi}\frac{\left(2\omega_{c}-\omega_{n}-\omega_{0}\right)}{t_{n}^{3/2}}\gamma^{2}e^{-\frac{\left[\left(2\omega_{c}-\omega_{n}-\omega_{0}\right)-\alpha t_{n}\right]^{2}}{2t_{n}}}.
\end{equation}

\selectlanguage{english}%
\medskip{}

\selectlanguage{brazil}%
Equating with (\ref{eq:135}),

\selectlanguage{english}%
\medskip{}

\begin{equation}
\gamma=\frac{1}{\sqrt{\pi}}e^{\alpha\left(\omega_{n}-\omega_{c}\right)}.\label{eq:gamma1}
\end{equation}

\medskip{}

Back to (\ref{eq:139}), with $\omega_{0}\neq0$,

\medskip{}

\[
\Pi_{\epsilon}^{gm}\left(\omega_{0},\omega_{c},t_{n}=T\right)=\sqrt{\epsilon}\frac{1}{\sqrt{\pi}}e^{\alpha\left(\omega_{n}-\omega_{c}\right)}\frac{\omega_{c}-\omega_{0}}{t_{n}^{3/2}}e^{-\frac{1}{2}\alpha^{2}t_{n}}e^{\alpha\left(\omega_{c}-\omega_{0}\right)}e^{-\frac{1}{2t_{n}}\left(\omega_{c}-\omega_{0}\right)^{2}}
\]

\[
=\sqrt{\epsilon}\frac{1}{\sqrt{\pi}}e^{\alpha\left(\omega_{n}-\omega_{0}\right)}\frac{\omega_{c}-\omega_{0}}{t_{n}^{3/2}}e^{-\frac{1}{2}\alpha^{2}t_{n}}e^{-\frac{1}{2t_{n}}\left(\omega_{c}-\omega_{0}\right)^{2}}.
\]

\selectlanguage{brazil}%
\begin{equation}
\therefore\Pi_{\epsilon}^{gm}\left(\omega_{0},\omega_{c},t_{n}=T\right)=\sqrt{\epsilon}\frac{1}{\sqrt{\pi}}e^{\alpha\left(\omega_{n}-\omega_{c}\right)}\frac{\omega_{c}-\omega_{0}}{t_{n}^{3/2}}e^{-\frac{\left[\left(\omega_{c}-\omega_{0}\right)-\alpha t_{n}\right]^{2}}{2t_{n}}}.\label{eq:143}
\end{equation}

\selectlanguage{english}%
\medskip{}

In the case of $\omega_{n}<\omega_{c}$, we use (\ref{eq:136}) and
(\ref{eq:143}) to analyze the situation where the process starts
near the barrier and finishes at $\omega_{n}$:

\medskip{}

\begin{equation}
\Pi_{\epsilon}^{gm}\left(\omega_{c},\omega_{n},t_{n}=T\right)=\sqrt{\epsilon}\frac{1}{\sqrt{\pi}}e^{\alpha\left(\omega_{n}-\omega_{c}\right)}\frac{\omega_{c}-\omega_{n}}{t_{n}^{3/2}}e^{-\frac{\left[\left(\omega_{c}-\omega_{n}\right)-\alpha t_{n}\right]^{2}}{2t_{n}}};\;\omega_{n}<\omega_{c}.\label{eq:144}
\end{equation}
\medskip{}

\selectlanguage{brazil}%
In the case of $\Pi_{\epsilon}^{gm}\left(\omega_{c},\omega_{c},t_{n}=T\right)$,
we must have translation invariance, and $\Pi_{\epsilon}^{gm}\left(\omega_{0},\omega_{c},t_{n}=T\right)$
can only depend on $\omega_{0}$ and $\omega_{c}$ with $\omega_{c}-\omega_{0}$.
The expansion (\ref{eq:128}), already developed in the case of (\ref{eq:139}),
with $\omega_{n}\rightarrow\omega_{c}$ (or $\eta\rightarrow0$),
tends to zero, as in (\ref{eq:143}). The next term of the expansion
(\ref{eq:127}) must be proportional to $\epsilon/t_{n}^{3/2}$ and
to the drift term:\foreignlanguage{english}{\medskip{}
 }

\selectlanguage{english}%
\begin{equation}
\Pi_{\epsilon}^{gm}\left(\omega_{c},\omega_{c},t_{n}=T\right)=c\frac{\epsilon}{t_{n}^{3/2}}e^{-\frac{\alpha^{2}t_{n}}{2}}.
\end{equation}

\medskip{}

In \cite{MagI}, the identification of the constant is done by (\ref{eq:pi gm}),
and it is sufficient to study the case of $n=2$ variables.

\medskip{}

\selectlanguage{brazil}%
\begin{equation}
\Pi_{\epsilon}^{gm}\left(\omega_{0},\omega_{2},t_{2}\right)=\int_{-\infty}^{\omega_{c}}d\omega_{1}\frac{1}{2\pi\epsilon}e^{-\frac{1}{2\epsilon}\left[\left(\omega_{1}-\omega_{0}-\epsilon\alpha\right)^{2}+\left(\omega_{2}-\omega_{1}-\epsilon\alpha\right)^{2}\right]},
\end{equation}

\selectlanguage{english}%
\medskip{}

We can solve this integral in Mathematica, taking into account that
in two steps $t_{n}=2\epsilon$:

\medskip{}

\begin{equation}
\Pi_{\epsilon}^{gm}\left(\omega_{0},\omega_{2};t_{2}\right)=\frac{1}{2\pi\epsilon}\frac{1}{2}e^{-\frac{\left(2\epsilon\alpha+\omega_{0}-\omega_{2}\right)^{2}}{2\left(2\epsilon\right)}}\sqrt{\pi}\sqrt{\epsilon}\left(1+Erf\left[\frac{\omega_{0}-\omega_{2}}{2\sqrt{\epsilon}}\right]\right).
\end{equation}

\[
e^{-\frac{\left(2\epsilon\alpha+\omega_{0}-\omega_{2}\right)^{2}}{2\left(2\epsilon\right)}}=e^{-\frac{4\epsilon^{2}\alpha^{2}}{2\left(2\epsilon\right)}-2\epsilon\alpha\left(\omega_{0}-\omega_{2}\right)\frac{1}{2\epsilon}-\frac{\left(\omega_{0}-\omega_{2}\right)^{2}}{2\left(2\epsilon\right)}}
\]

We make $\omega_{2}-\omega_{0}\sim0$, because we are in the case
$\omega_{0}\rightarrow\omega_{c}$ and $\omega_{n}\rightarrow\omega_{c}$:

\[
e^{-\frac{\left(\omega_{0}-\omega_{2}\right)^{2}}{2\left(2\epsilon\right)}}\sim1+\left[\frac{2\left(\omega_{0}-\omega_{2}\right)}{2\left(2\epsilon\right)}e^{-\frac{\left(\omega_{0}-\omega_{2}\right)^{2}}{2\left(2\epsilon\right)}}\right]_{\omega_{2}-\omega_{0}=0}\left(\omega_{2}-\omega_{0}\right)
\]

\[
+\frac{1}{2}\left[\frac{2}{2\left(2\epsilon\right)}e^{-\frac{\left(\omega_{0}-\omega_{2}\right)^{2}}{2\left(2\epsilon\right)}}-\left(\frac{2\left(\omega_{0}-\omega_{2}\right)}{2\left(2\epsilon\right)}\right)^{2}e^{-\frac{\left(\omega_{0}-\omega_{2}\right)^{2}}{2\left(2\epsilon\right)}}\right]_{\omega_{2}-\omega_{0}=0}\left(\omega_{2}-\omega_{0}\right)^{2}
\]

\[
=1+\frac{1}{2t_{n}}\left(\omega_{2}-\omega_{0}\right)^{2}\rightarrow1
\]

\medskip{}

Keeping the term in $\epsilon$ and, given that $Erf(\frac{\omega_{0}-\omega_{2}}{2\sqrt{\epsilon}}\sim0)=0$,

\medskip{}

\begin{equation}
\Pi_{\epsilon}^{gm}\left(\omega_{0}=\omega_{c},\omega_{2}=\omega_{c};t_{2}\right)=\frac{1}{\sqrt{2\pi}\sqrt{2\epsilon}}\frac{1}{2}.\frac{2\epsilon}{2\epsilon}e^{-\frac{t_{n}\alpha^{2}}{2}}=\frac{1}{\sqrt{2\pi}}\frac{\epsilon}{t_{n}^{3/2}}e^{-\frac{\alpha^{2}t_{n}}{2}}.
\end{equation}

\medskip{}

Hence,

\medskip{}

\begin{equation}
c=\frac{1}{\sqrt{2\pi}}
\end{equation}

\medskip{}

\begin{equation}
\Pi_{\epsilon}^{gm}\left(\omega_{c},\omega_{c},t_{n}=T\right)=\frac{1}{\sqrt{2\pi}}\frac{\epsilon}{t_{n}^{3/2}}e^{-\frac{\alpha^{2}t_{n}}{2}}.\label{eq:148}
\end{equation}

\medskip{}

In a nutshell, we analyzed the behaviour near the barrier: $\Pi_{\epsilon\rightarrow0}^{gm}\left(\omega_{0},\omega_{c};t_{n}=T\right)$,
$\Pi_{\epsilon}^{gm}\left(\omega_{c},\omega_{n},t_{n}=T\right)$ and
$\Pi_{\epsilon}^{gm}\left(\omega_{c},\omega_{c},t_{n}=T\right)$,
described by (\ref{eq:143}), (\ref{eq:144}) and (\ref{eq:148}),
respectively.

\medskip{}

\pagebreak{}

\subsection{Analysis of divergent and finite terms in $\stackrel[i,j=1]{n-1}{\sum}\partial_{i}\partial_{j}$\label{sec:An=00003D00003D0000E1lise-de-termos}}

\medskip{}

\hspace*{0.25in}In (\ref{eq:57}) there is the term $\stackrel[i,j=1]{n-1}{\sum}\partial_{i}\partial_{j}$,
which was used in (\ref{eq:107}) in the following way:

\medskip{}

\begin{equation}
\stackrel[i,j=1]{n-1}{\sum}\partial_{i}\partial_{j}=2\underset{i<j}{\sum}\partial_{i}\partial_{j}+\stackrel[i=1]{n-1}{\sum}\partial_{i}^{2},\label{eq:217}
\end{equation}

\medskip{}

\noindent where the first term of the RHS can be expressed as

\medskip{}

\begin{equation}
\underset{i<j}{\sum}\partial_{i}\partial_{j}=\stackrel[i=1]{n-2}{\sum}\stackrel[j=i+1]{n-1}{\sum}\partial_{i}\partial_{j}\label{eq:218}
\end{equation}
\medskip{}

\noindent or\medskip{}

\begin{equation}
\stackrel[j=2]{n-1}{\sum}\stackrel[i=1]{j-1}{\sum}\partial_{i}\partial_{j}.\label{eq:219}
\end{equation}
\medskip{}

By equation (\ref{eq:57}), when $W^{gm}\left(\omega_{0},\omega_{1},...,\omega_{n};t_{n}\right)$
receives the application of (i) the (\ref{eq:218}) summation operator,
according to (\ref{eq:103}), and also (ii) the second term of RHS
(\ref{eq:217}), according to (\ref{eq:104}), we have

\medskip{}

\[
2\stackrel[i=1]{n-2}{\sum}\stackrel[j=i+1]{n-1}{\sum}\Pi_{\epsilon}^{gm}\left(\omega_{0},\omega_{c};t_{i}\right)\Pi_{\epsilon}^{gm}\left(\omega_{c},\omega_{c};t_{j}-t_{i}\right)\Pi_{\epsilon}^{gm}\left(\omega_{c},\omega_{n};t_{n}-t_{j}\right)
\]

\begin{equation}
+\stackrel[i=1]{n-1}{\sum}\partial_{i}\left[\Pi_{\epsilon}^{gm}\left(\omega_{0},\omega_{i};t_{i}\right)\Pi_{\epsilon}^{gm}\left(\omega_{i},\omega_{n};t_{n}-t_{i}\right)\right]_{\omega_{i}=\omega_{c}}.\label{eq:220}
\end{equation}

\medskip{}

Given the form of (\ref{eq:30}),\medskip{}

\begin{equation}
\partial_{i}\left[\Pi_{\epsilon}^{gm}\left(\omega_{0},\omega_{i};t_{i}\right)\right]=\partial_{i}\left[\Pi_{\epsilon}^{gm}\left(\omega_{i},\omega_{n};t_{n}-t_{i}\right)\right].
\end{equation}
\medskip{}

Thus, the second term of (\ref{eq:220}) becomes\medskip{}

\begin{equation}
I_{1}\equiv\stackrel[i=1]{n-1}{\sum}2\partial_{i}\left[\Pi_{\epsilon}^{gm}\left(\omega_{0},\omega_{i};t_{i}\right)\right]_{\omega_{i}=\omega_{c}}\Pi_{\epsilon}^{gm}\left(\omega_{c},\omega_{n};t_{n}-t_{i}\right).\label{eq:222}
\end{equation}

\medskip{}

Let us compute this term. Regarding $\Pi_{\epsilon}^{gm}\left(\omega_{0},\omega_{i};t_{i}\right)$,
we saw in (\ref{eq:133-1}) that in the limit $\eta\rightarrow0$,
the expression $v\left(\eta\right)$ of (\ref{eq:C*v}) leads to $\gamma$,
given by (\ref{eq:gamma1}):

\medskip{}

\[
\gamma=\frac{1}{\sqrt{\pi}}e^{\alpha\left(\omega_{n}-\omega_{c}\right)}
\]

\medskip{}

More broadly, in (\ref{eq:133-1}):

\medskip{}

\[
v(\eta)\varpropto\frac{\gamma\sqrt{\pi}}{2\eta}=\frac{1}{2\eta}e^{\alpha\left(\omega_{n}-\omega_{c}\right)}
\]

\medskip{}

\noindent we can propose that in this limit, still respecting $\gamma$,
$v\left(\eta\right)$ is given by

\medskip{}

\begin{equation}
v\left(\eta\right)=e^{\alpha\left(\omega_{n}-\omega_{c}\right)}\left(\frac{1}{2\eta}+v_{0}+v_{1}\eta+\mathcal{O}\left(\eta^{2}\right)\right).
\end{equation}

\medskip{}

In (\ref{eq:139}),

\medskip{}

\[
\Pi_{\epsilon\rightarrow0}^{gm}\left(\omega_{0},\omega_{c};t_{n}\right)=\sqrt{\epsilon}\gamma\frac{\omega_{c}-\omega_{0}}{t_{n}^{3/2}}e^{-\frac{1}{2}\alpha^{2}t_{n}}e^{\alpha\left(\omega_{c}-\omega_{0}\right)}e^{-\frac{1}{2t_{n}}\left(\omega_{c}-\omega_{0}\right)^{2}}
\]

\[
=\sqrt{\epsilon}\frac{2\eta}{\sqrt{\pi}}v\left(\eta\right)\frac{\omega_{c}-\omega_{0}}{t_{n}^{3/2}}e^{-\frac{1}{2}\alpha^{2}t_{n}}e^{\alpha\left(\omega_{c}-\omega_{0}\right)}e^{-\frac{1}{2t_{n}}\left(\omega_{c}-\omega_{0}\right)^{2}}
\]

\[
=\sqrt{\epsilon}\frac{2\eta}{\sqrt{\pi}}e^{\alpha\left(\omega_{n}-\omega_{c}\right)}\left(\frac{1}{2\eta}+v_{0}+v_{1}\eta+...\right)\frac{\omega_{c}-\omega_{0}}{t_{n}^{3/2}}e^{-\frac{1}{2}\alpha^{2}t_{n}}e^{\alpha\left(\omega_{c}-\omega_{0}\right)}e^{-\frac{1}{2t_{n}}\left(\omega_{c}-\omega_{0}\right)^{2}}
\]

\[
=\sqrt{\epsilon}\frac{1}{\sqrt{\pi}}e^{\alpha\left(\omega_{n}-\omega_{c}\right)}\frac{\omega_{c}-\omega_{0}}{t_{n}^{3/2}}e^{-\frac{1}{2}\alpha^{2}t_{n}}e^{\alpha\left(\omega_{c}-\omega_{0}\right)}e^{-\frac{1}{2t_{n}}\left(\omega_{c}-\omega_{0}\right)^{2}}
\]

\begin{equation}
+\sqrt{\epsilon}\frac{2}{\sqrt{\pi}}e^{\alpha\left(\omega_{n}-\omega_{c}\right)}\left(v_{0}\eta+v_{1}\eta^{2}+...\right)\frac{\omega_{c}-\omega_{0}}{t_{n}^{3/2}}e^{-\frac{1}{2}\alpha^{2}t_{n}}e^{\alpha\left(\omega_{c}-\omega_{0}\right)}e^{-\frac{1}{2t_{n}}\left(\omega_{c}-\omega_{0}\right)^{2}}.
\end{equation}
\medskip{}

By the definition of (\ref{eq:def eta}),

\medskip{}

\begin{equation}
\partial_{i}\equiv\partial/\partial\omega_{i}=\frac{d\eta}{d\omega_{i}}\frac{\partial}{\partial\eta}
\end{equation}

\begin{equation}
\frac{d\eta}{d\omega_{i}}=-\frac{1}{\sqrt{2\epsilon}}
\end{equation}

\medskip{}

\selectlanguage{brazil}%
then

\selectlanguage{english}%
\noindent \medskip{}

\[
\underset{\epsilon\rightarrow0}{lim}\underset{\omega\rightarrow\omega_{c}^{-}}{lim}\frac{\partial}{\partial\omega}\Pi_{\epsilon}^{gm}\left(\omega_{0},\omega;t_{n}\right)
\]

\begin{equation}
=-v_{0}\left(\frac{2}{\pi}\right)^{1/2}\frac{\omega_{c}-\omega_{0}}{t_{n}^{3/2}}e^{-\frac{1}{2}\alpha^{2}t_{n}}e^{\alpha\left(\omega_{c}-\omega_{0}\right)}e^{-\frac{1}{2t_{n}}\left(\omega_{c}-\omega_{0}\right)^{2}}e^{\alpha\left(\omega_{n}-\omega_{c}\right)}.
\end{equation}
\medskip{}

There is not the term $u_{1}\eta+...$ after the derivative because
we have taken the limit $\omega\rightarrow\omega_{c}^{-}$ and, in
this case, $\eta\rightarrow0$, annulling this term. Back to (\ref{eq:222}),
also using (\ref{eq:144}), with $t_{n}$ replaced by $t_{n}-t_{i}$,
and (\ref{eq:133}):\medskip{}

\[
I_{1}=\stackrel[i=1]{n-1}{\sum}\left(-v_{0}\left(\frac{2}{\pi}\right)^{1/2}\frac{\omega_{c}-\omega_{0}}{t_{i}^{3/2}}e^{-\frac{1}{2}\alpha^{2}t_{n}}e^{\alpha\left(\omega_{c}-\omega_{0}\right)}e^{-\frac{1}{2t_{n}}\left(\omega_{c}-\omega_{0}\right)^{2}}e^{\alpha\left(\omega_{n}-\omega_{c}\right)}\right)\cdot
\]

\[
\left(\sqrt{\epsilon}\frac{1}{\sqrt{\pi}}e^{\alpha\left(\omega_{n}-\omega_{c}\right)}\frac{\omega_{c}-\omega_{n}}{\left(t_{n}-t_{i}\right)^{3/2}}e^{-\frac{\left[\left(\omega_{c}-\omega_{n}\right)-\alpha\left(t_{n}-t_{i}\right)\right]^{2}}{2\left(t_{n}-t_{i}\right)}}\right)
\]

\[
=-\stackrel[i=1]{n-1}{\sum}\left(v_{0}\frac{\sqrt{2}}{\pi\sqrt{\epsilon}}\epsilon\frac{\omega_{c}-\omega_{0}}{t_{i}^{3/2}}\frac{\omega_{c}-\omega_{n}}{\left(t_{n}-t_{i}\right)^{3/2}}e^{-\frac{\left[\left(\omega_{c}-\omega_{0}\right)-\alpha t_{i}\right]^{2}}{2t_{i}}}e^{-\frac{\left[\left(\omega_{c}-\omega_{n}\right)-\alpha\left(t_{n}-t_{i}\right)\right]^{2}}{2\left(t_{n}-t_{i}\right)}}e^{2\alpha\left(\omega_{n}-\omega_{c}\right)}\right)
\]

\begin{equation}
\Rightarrow-\frac{v_{0}\sqrt{2}}{\pi\sqrt{\epsilon}}\left(\omega_{c}-\omega_{0}\right)\left(\omega_{c}-\omega_{n}\right)e^{2\alpha\left(\omega_{n}-\omega_{c}\right)}\intop_{0}^{t_{n}}dt_{i}\frac{e^{-\frac{\left[\left(\omega_{c}-\omega_{0}\right)-\alpha t_{i}\right]^{2}}{2t_{i}}}e^{-\frac{\left[\left(\omega_{c}-\omega_{n}\right)-\alpha\left(t_{n}-t_{i}\right)\right]^{2}}{2\left(t_{n}-t_{i}\right)}}}{t_{i}^{3/2}\left(t_{n}-t_{i}\right)^{3/2}}.
\end{equation}

\medskip{}

\selectlanguage{brazil}%
\begin{equation}
\therefore I_{1}=-\frac{2v_{0}}{\sqrt{\pi}\sqrt{\epsilon}}\left(2\omega_{c}-\omega_{n}-\omega_{0}\right)\frac{1}{t_{n}^{3/2}}e^{-\frac{\left[\left(2\omega_{c}-\omega_{n}-\omega_{0}\right)-\alpha t_{n}\right]^{2}}{2t_{n}}}e^{2\alpha\left(\omega_{n}-\omega_{c}\right)}.
\end{equation}

\selectlanguage{english}%
\medskip{}

Therefore, $I_{1}$ diverges with $1/\sqrt{\epsilon}$. Now we analyze
the first term of \ref{eq:220},

\medskip{}

\begin{equation}
I_{2}\equiv\stackrel[i=1]{n-2}{\sum}\stackrel[j=i+1]{n-1}{\sum}\Pi_{\epsilon}^{gm}\left(\omega_{0},\omega_{c};t_{i}\right)\Pi_{\epsilon}^{gm}\left(\omega_{c},\omega_{c};t_{j}-t_{i}\right)\Pi_{\epsilon}^{gm}\left(\omega_{c},\omega_{n};t_{n}-t_{j}\right).\label{eq:232}
\end{equation}
\medskip{}

By (\ref{eq:143}), (\ref{eq:148}) and (\ref{eq:144}) we write

\[
I_{2}=\stackrel[i=1]{n-2}{\sum}\Pi_{\epsilon}^{gm}\left(\omega_{0},\omega_{c};t_{i}\right)\stackrel[j=i+1]{n-1}{\sum}\Pi_{\epsilon}^{gm}\left(\omega_{c},\omega_{c};t_{j}-t_{i}\right)\Pi_{\epsilon}^{gm}\left(\omega_{c},\omega_{n};t_{n}-t_{j}\right)
\]

\[
=\stackrel[i=1]{n-2}{\sum}\frac{\sqrt{\epsilon}}{\sqrt{\pi}}e^{\alpha\left(\omega_{n}-\omega_{c}\right)}\frac{\omega_{c}-\omega_{0}}{t_{i}^{3/2}}e^{-\frac{\left[\left(\omega_{c}-\omega_{0}\right)-\alpha t_{i}\right]^{2}}{2t_{i}}}\cdot
\]

\begin{equation}
\stackrel[j=i+1]{n-1}{\sum}\frac{\epsilon}{\sqrt{2\pi}}\frac{1}{\left(t_{j}-t_{i}\right)^{3/2}}e^{-\frac{\alpha^{2}\left(t_{j}-t_{i}\right)}{2}}\frac{\sqrt{\epsilon}}{\sqrt{\pi}}e^{\alpha\left(\omega_{n}-\omega_{c}\right)}\frac{\omega_{c}-\omega_{n}}{\left(t_{n}-t_{j}\right)^{3/2}}e^{-\frac{\left[\left(\omega_{c}-\omega_{n}\right)-\alpha\left(t_{n}-t_{j}\right)\right]^{2}}{2\left(t_{n}-t_{j}\right)}}.\label{eq:233}
\end{equation}

\medskip{}

In (\ref{eq:233}) it is not possible to use $\epsilon\stackrel[j=i+1]{n-1}{\sum}=\intop_{t_{i}}^{t_{n}}dt_{j}$,
since it is only valid when the sum and the integral are finite when
$\epsilon\rightarrow0$. Here this does not happen, because the integrand
in

\medskip{}

\begin{equation}
\intop_{t_{i}}^{t_{n}}dt_{j}\frac{1}{\left(t_{j}-t_{i}\right)^{3/2}\left(t_{n}-t_{j}\right)^{3/2}}e^{-\frac{\left[\left(\omega_{c}-\omega_{n}\right)-\alpha\left(t_{n}-t_{j}\right)\right]^{2}}{2\left(t_{n}-t_{j}\right)}}e^{-\frac{\alpha^{2}\left(t_{j}-t_{i}\right)}{2}}\label{eq:234}
\end{equation}

\medskip{}

\noindent diverges due to the annulment of $\left(t_{j}-t_{i}\right)^{3/2}$
when the sum initiates, that is, when $t_{j}=t_{i}$. We add the exponential
term in $\beta\epsilon$, which cuts the expression in $t_{j}=t_{i}$:\medskip{}

\[
\stackrel[j=i+1]{n-1}{\sum}\frac{\epsilon}{\left(t_{j}-t_{i}\right)^{3/2}\left(t_{n}-t_{j}\right)^{3/2}}e^{-\frac{\left[\left(\omega_{c}-\omega_{n}\right)-\alpha\left(t_{n}-t_{j}\right)\right]^{2}}{2\left(t_{n}-t_{j}\right)}}e^{-\frac{\alpha^{2}\left(t_{j}-t_{i}\right)}{2}}=
\]

\begin{equation}
=\intop_{t_{i}}^{t_{n}}dt_{j}\frac{e^{-\frac{\beta\epsilon}{2\left(t_{j}-t_{i}\right)}}}{\left(t_{j}-t_{i}\right)^{3/2}\left(t_{n}-t_{j}\right)^{3/2}}e^{-\frac{\left[\left(\omega_{c}-\omega_{n}\right)-\alpha\left(t_{n}-t_{j}\right)\right]^{2}}{2\left(t_{n}-t_{j}\right)}}e^{-\frac{\alpha^{2}\left(t_{j}-t_{i}\right)}{2}}.\label{eq:235}
\end{equation}
\medskip{}

This factor still keeps the singular part of $1/\sqrt{\beta\epsilon}$
and the finite part. To see that there is a singular part in $1/\sqrt{\epsilon}$,
we can open the integral (\ref{eq:234}) in the divergent term, integrating
from $t_{i}$ to $t_{i}+\epsilon$, and another from $t_{i}+2\epsilon$
to $t_{n}$ (finite term). To make it easier, we notice that the integral
is dominated by $t_{j}=t_{i}$, which can be taken out of the expression.

\medskip{}

\[
I_{3}=\frac{1}{\left(t_{n}-t_{i}\right)^{3/2}}e^{-\frac{\left[\left(\omega_{c}-\omega_{n}\right)-\alpha\left(t_{n}-t_{i}\right)\right]^{2}}{2\left(t_{n}-t_{i}\right)}}\intop_{t_{i}}^{t_{i}+\epsilon}dt_{j}\frac{e^{-\frac{\alpha^{2}\left(t_{j}-t_{i}\right)}{2}}}{\left(t_{j}-t_{i}\right)^{3/2}}
\]

\begin{equation}
+\intop_{t_{i}+2\epsilon}^{t_{n}}dt_{j}\frac{e^{-\frac{\left[\left(\omega_{c}-\omega_{n}\right)-\alpha\left(t_{n}-t_{j}\right)\right]^{2}}{2\left(t_{n}-t_{j}\right)}}e^{-\frac{\alpha^{2}\left(t_{j}-t_{i}\right)}{2}}}{\left(t_{j}-t_{i}\right)^{3/2}\left(t_{n}-t_{j}\right)^{3/2}}.
\end{equation}
\medskip{}

\selectlanguage{brazil}%
The term

\selectlanguage{english}%
\medskip{}

\[
e^{-\frac{\alpha^{2}\left(t_{j}-t_{i}\right)}{2}}
\]

\medskip{}

\noindent in the first line of the integral converges to 1. Continuing,

\medskip{}

\begin{equation}
I_{3}=\frac{2}{\sqrt{\epsilon}}\frac{1}{\left(t_{n}-t_{i}\right)^{3/2}}e^{-\frac{\left[\left(\omega_{c}-\omega_{n}\right)-\alpha\left(t_{n}-t_{i}\right)\right]^{2}}{2\left(t_{n}-t_{i}\right)}}+finite\;term.\label{eq:237}
\end{equation}
\medskip{}

\selectlanguage{brazil}%
Back to (\ref{eq:235}) and using (\ref{eq:141}),

\selectlanguage{english}%
\medskip{}

\[
-\frac{\left[\left(\omega_{c}-\omega_{n}\right)-\alpha\left(t_{n}-t_{j}\right)\right]^{2}}{2\left(t_{n}-t_{j}\right)}-\frac{\alpha^{2}\left(t_{j}-t_{i}\right)}{2}
\]

\[
=-\frac{\left(\omega_{c}-\omega_{n}\right)^{2}}{2\left(t_{n}-t_{j}\right)}+\left(\omega_{c}-\omega_{n}\right)\alpha-\frac{\alpha^{2}}{2}\left(t_{n}-t_{i}\right).
\]

\medskip{}

\[
I_{3}=\intop_{t_{i}}^{t_{n}}dt_{j}\frac{e^{-\frac{\beta\epsilon}{2\left(t_{j}-t_{i}\right)}}}{\left(t_{j}-t_{i}\right)^{3/2}\left(t_{n}-t_{j}\right)^{3/2}}e^{-\frac{\left[\left(\omega_{c}-\omega_{n}\right)-\alpha\left(t_{n}-t_{j}\right)\right]^{2}}{2\left(t_{n}-t_{j}\right)}}e^{-\frac{\alpha^{2}\left(t_{j}-t_{i}\right)}{2}}
\]

\[
=\intop_{t_{i}}^{t_{n}}dt_{j}\frac{e^{-\frac{\beta\epsilon}{2\left(t_{j}-t_{i}\right)}}}{\left(t_{j}-t_{i}\right)^{3/2}\left(t_{n}-t_{j}\right)^{3/2}}e^{-\frac{\left(\omega_{c}-\omega_{n}\right)^{2}}{2\left(t_{n}-t_{j}\right)}+\left(\omega_{c}-\omega_{n}\right)\alpha-\frac{\alpha^{2}}{2}\left(t_{n}-t_{i}\right)}
\]

\begin{equation}
=\sqrt{2\pi}\left(\frac{1}{\sqrt{\beta\epsilon}}+\frac{1}{\omega_{c}-\omega_{n}}\right)\frac{1}{\left(t_{n}-t_{i}\right)^{3/2}}e^{-\frac{\left[\left(\omega_{c}-\omega_{n}\right)+\sqrt{\beta\epsilon}\right]^{2}}{2\left(t_{n}-t_{i}\right)}}e^{\left(\omega_{c}-\omega_{n}\right)\alpha}e^{-\frac{\alpha^{2}}{2}\left(t_{n}-t_{i}\right)}.\label{eq:238}
\end{equation}
\medskip{}

We take $I_{3}$ to (\ref{eq:233}) to evaluate $I_{2}$:

\medskip{}

\[
I_{2}=\frac{\epsilon}{\pi}\left(\omega_{c}-\omega_{0}\right)\left(\omega_{c}-\omega_{n}\right)\stackrel[i=1]{n-2}{\sum}\frac{1}{t_{i}^{3/2}}e^{-\frac{\left[\left(\omega_{c}-\omega_{0}\right)-\alpha t_{i}\right]^{2}}{2t_{i}}}e^{2\alpha\left(\omega_{n}-\omega_{c}\right)}\cdot
\]

\begin{equation}
\left(\frac{1}{\sqrt{\beta\epsilon}}+\frac{1}{\omega_{c}-\omega_{n}}\right)\frac{1}{\left(t_{n}-t_{i}\right)^{3/2}}e^{-\frac{\left[\left(\omega_{c}-\omega_{n}\right)+\sqrt{\beta\epsilon}\right]^{2}}{2\left(t_{n}-t_{i}\right)}}e^{\left(\omega_{c}-\omega_{n}\right)\alpha}e^{-\frac{\alpha^{2}}{2}\left(t_{n}-t_{i}\right)}.
\end{equation}

\medskip{}

\selectlanguage{brazil}%
Now we use $\stackrel[i=1]{n-2}{\sum}\rightarrow\frac{1}{\epsilon}\int_{0}^{t_{n}}dt_{i}$
and (\ref{eq:141}), regruping terms:

\selectlanguage{english}%
\medskip{}

\[
I_{2}=\frac{1}{\pi}\left(\omega_{c}-\omega_{0}\right)\left(\omega_{c}-\omega_{n}\right)\left(\frac{1}{\sqrt{\beta\epsilon}}+\frac{1}{\omega_{c}-\omega_{n}}\right)e^{\alpha\left(\omega_{n}-\omega_{c}\right)}\cdot
\]

\begin{equation}
\intop_{0}^{t_{n}}dt_{i}\frac{1}{t_{i}^{3/2}}\frac{1}{\left(t_{n}-t_{i}\right)^{3/2}}e^{-\frac{\left[\left(\omega_{c}-\omega_{n}\right)+\sqrt{\beta\epsilon}\right]^{2}}{2\left(t_{n}-t_{i}\right)}}e^{-\frac{\left[\left(\omega_{c}-\omega_{0}\right)-\alpha t_{i}\right]^{2}}{2t_{i}}}e^{-\frac{\alpha^{2}}{2}\left(t_{n}-t_{i}\right)}.
\end{equation}

\[
=\frac{1}{\pi}\left(\omega_{c}-\omega_{0}\right)\left(\omega_{c}-\omega_{n}\right)\left(\frac{1}{\sqrt{\beta\epsilon}}+\frac{1}{\omega_{c}-\omega_{n}}\right)e^{\alpha\left(\omega_{n}-\omega_{c}\right)}e^{\alpha\left(\omega_{c}-\omega_{0}\right)}e^{-\frac{\alpha^{2}}{2}t_{n}}\cdot
\]

\begin{equation}
\intop_{0}^{t_{n}}dt_{i}\frac{1}{t_{i}^{3/2}}\frac{1}{\left(t_{n}-t_{i}\right)^{3/2}}e^{-\frac{\left[\left(\omega_{c}-\omega_{n}\right)+\sqrt{\beta\epsilon}\right]^{2}}{2\left(t_{n}-t_{i}\right)}}e^{-\frac{\left(\omega_{c}-\omega_{0}\right)^{2}}{2t_{i}}}.
\end{equation}

\medskip{}

\[
\therefore I_{2}=\sqrt{\frac{2}{\pi}}\left(\omega_{c}-\omega_{0}\right)\left(\omega_{c}-\omega_{n}\right)\left(\frac{1}{\sqrt{\beta\epsilon}}+\frac{1}{\omega_{c}-\omega_{n}}\right)e^{\alpha\left(\omega_{n}-\omega_{c}\right)}e^{\alpha\left(\omega_{c}-\omega_{0}\right)}e^{-\frac{\alpha^{2}}{2}t_{n}}\cdot
\]

\begin{equation}
e^{-\frac{\left[\left(2\omega_{c}-\omega_{0}-\omega_{n}\right)+\sqrt{\beta\epsilon}\right]^{2}}{2t_{n}}}\frac{1}{t_{n}^{3/2}}\frac{\left[\left(2\omega_{c}-\omega_{0}-\omega_{n}\right)+\sqrt{\beta\epsilon}\right]}{\left(\omega_{c}-\omega_{0}\right)\left(2\omega_{c}-\omega_{n}+\sqrt{\beta\epsilon}\right)}.
\end{equation}
\medskip{}

Expanding the exponential of $\sqrt{\beta\epsilon}$ around zero:

\medskip{}

\begin{equation}
e^{-\frac{\left[\left(2\omega_{c}-\omega_{0}-\omega_{n}\right)+\sqrt{\beta\epsilon}\right]^{2}}{2t_{n}}}\simeq e^{-\frac{\left(2\omega_{c}-\omega_{0}-\omega_{n}\right)^{2}}{2t_{n}}}-\frac{1}{t_{n}}\left(2\omega_{c}-\omega_{0}-\omega_{n}\right)\sqrt{\beta\epsilon}e^{-\frac{\left(2\omega_{c}-\omega_{0}-\omega_{n}\right)^{2}}{2t_{n}}}+...
\end{equation}
\medskip{}

\[
I_{2}=\sqrt{\frac{2}{\pi}}\frac{\left[\left(2\omega_{c}-\omega_{0}-\omega_{n}\right)+\sqrt{\beta\epsilon}\right]}{t_{n}^{3/2}}\left(\omega_{c}-\omega_{n}\right)e^{\alpha\left(\omega_{n}-\omega_{c}\right)}e^{\alpha\left(\omega_{c}-\omega_{0}\right)}e^{-\frac{\alpha^{2}}{2}t_{n}}\cdot
\]

\[
e^{-\frac{\left(2\omega_{c}-\omega_{0}-\omega_{n}\right)^{2}}{2t_{n}}}\left(1-\frac{\sqrt{\beta\epsilon}}{t_{n}}\left(2\omega_{c}-\omega_{0}-\omega_{n}\right)\right)\frac{1}{\omega_{c}-\omega_{n}+\sqrt{\beta\epsilon}}\frac{\omega_{c}-\omega_{n}+\sqrt{\beta\epsilon}}{\left(\omega_{c}-\omega_{n}\right)\sqrt{\beta\epsilon}}
\]

\[
=\sqrt{\frac{2}{\pi}}\left[\frac{\left(2\omega_{c}-\omega_{0}-\omega_{n}\right)}{\sqrt{\beta\epsilon}}+1\right]\frac{1}{t_{n}^{3/2}}e^{-\frac{\left[\left(2\omega_{c}-\omega_{0}-\omega_{n}\right)-\alpha t_{n}\right]^{2}}{2t_{n}}}e^{2\alpha\left(\omega_{n}-\omega_{c}\right)}\cdot
\]

\[
\left(1-\frac{\sqrt{\beta\epsilon}}{t_{n}}\left(2\omega_{c}-\omega_{0}-\omega_{n}\right)\right)
\]

\[
=\sqrt{\frac{2}{\pi}}\left[\frac{\left(2\omega_{c}-\omega_{0}-\omega_{n}\right)}{\sqrt{\beta\epsilon}}-\frac{\left(2\omega_{c}-\omega_{0}-\omega_{n}\right)^{2}}{t_{n}}+1-\frac{\sqrt{\beta\epsilon}}{t_{n}}\left(2\omega_{c}-\omega_{0}-\omega_{n}\right)\right]\cdot
\]

\begin{equation}
\frac{1}{t_{n}^{3/2}}e^{-\frac{\left[\left(2\omega_{c}-\omega_{0}-\omega_{n}\right)-\alpha t_{n}\right]^{2}}{2t_{n}}}e^{2\alpha\left(\omega_{n}-\omega_{c}\right)}.
\end{equation}

\medskip{}

The last term in the brackets becomes zero as $\epsilon\rightarrow0$.

\medskip{}

\[
\therefore I_{2}=\sqrt{\frac{2}{\pi}}\left[\frac{\left(2\omega_{c}-\omega_{0}-\omega_{n}\right)}{\sqrt{\beta\epsilon}}+\left(1-\frac{\left(2\omega_{c}-\omega_{0}-\omega_{n}\right)^{2}}{t_{n}}\right)\right]\cdot
\]

\begin{equation}
\frac{1}{t_{n}^{3/2}}e^{-\frac{\left[\left(2\omega_{c}-\omega_{0}-\omega_{n}\right)-\alpha t_{n}\right]^{2}}{2t_{n}}}e^{2\alpha\left(\omega_{n}-\omega_{c}\right)}.\label{eq:245}
\end{equation}
\medskip{}

Therefore, in (\ref{eq:217}):\medskip{}

\[
\stackrel[i,j=1]{n-1}{\sum}\partial_{i}\partial_{j}=
\]

\[
2\left\{ \frac{1}{\sqrt{\epsilon}}\left(\frac{1}{\sqrt{\beta}}-v_{0}\sqrt{2}\right)\sqrt{\frac{2}{\pi}}\left(2\omega_{c}-\omega_{n}-\omega_{0}\right)\frac{1}{t_{n}^{3/2}}e^{-\frac{\left[\left(2\omega_{c}-\omega_{n}-\omega_{0}\right)-\alpha t_{n}\right]^{2}}{2t_{n}}}e^{2\alpha\left(\omega_{n}-\omega_{c}\right)}\right.
\]

\begin{equation}
\left.+\sqrt{\frac{2}{\pi}}\left[1-\frac{\left(2\omega_{c}-\omega_{0}-\omega_{n}\right)^{2}}{t_{n}}\right]\frac{1}{t_{n}^{3/2}}e^{-\frac{\left[\left(2\omega_{c}-\omega_{n}-\omega_{0}\right)-\alpha t_{n}\right]^{2}}{2t_{n}}}e^{2\alpha\left(\omega_{n}-\omega_{c}\right)}\right\} .\label{eq:246}
\end{equation}
\medskip{}

The LHS (\ref{eq:217}) is applied to $W^{gm}\left(\omega_{0},\omega_{1},...,\omega_{n};t_{n}\right)$,
according to (\ref{eq:57}). In (\ref{eq:106}), we saw that

\medskip{}

\[
\frac{\partial\Pi_{\epsilon}^{gm}}{\partial\omega_{c}}\left(\omega_{0}=0,\omega_{n};t_{n}\right)
\]

\begin{equation}
=\stackrel[i=1]{n-1}{\sum}\int_{-\infty}^{\omega_{c}}d\omega_{1}...d\hat{\omega}_{i}...d\omega_{n-1}W^{gm}\left(\omega_{0},\omega_{1},...,\omega_{i}=\omega_{c},...,\omega_{n-1},\omega_{n};t_{n}\right).
\end{equation}

\medskip{}

Now,

\medskip{}

\[
\frac{\partial^{2}\Pi_{\epsilon}^{gm}}{\partial\omega_{c}^{2}}\left(\omega_{0}=0,\omega_{n};t_{n}\right)
\]

\[
=\stackrel[i,j=1]{n-1}{\sum}\int_{-\infty}^{\omega_{c}}d\omega_{1}...d\hat{\omega}_{i}...d\hat{\omega}_{j}...d\omega_{n-1}W^{gm}\left(\omega_{0},\omega_{1},...,\omega_{i}=\omega_{c},...,\omega_{j}=\omega_{c},...\omega_{n-1},\omega_{n};t_{n}\right)
\]

\medskip{}

Using (\ref{eq:217}), (\ref{eq:101}) and (\ref{eq:104}):

\medskip{}

\[
\frac{\partial^{2}\Pi_{\epsilon}^{gm}}{\partial\omega_{c}^{2}}\left(\omega_{0}=0,\omega_{n};t_{n}\right)
\]

\[
=2\stackrel[i=1]{n-2}{\sum}\stackrel[j=i+1]{n-1}{\sum}\int_{-\infty}^{\omega_{c}}d\omega_{1}...d\hat{\omega}_{i}...d\hat{\omega}_{j}...d\omega_{n-1}
\]

\[
W^{gm}\left(\omega_{0},\omega_{1},...,\omega_{i}=\omega_{c},...,\omega_{j}=\omega_{c},...\omega_{n-1},\omega_{n};t_{n}\right)
\]

\[
+\stackrel[i=1]{n-1}{\sum}\int_{-\infty}^{\omega_{c}}d\omega_{1}...d\hat{\omega}_{i}...d\omega_{n-1}
\]

\[
\frac{\partial}{\partial\omega_{c}}W^{gm}\left(\omega_{0},\omega_{1},...,\omega_{i}=\omega_{c},...\omega_{n-1},\omega_{n};t_{n}\right)
\]

\[
=2\stackrel[i=1]{n-2}{\sum}\stackrel[j=i+1]{n-1}{\sum}\int_{-\infty}^{\omega_{c}}d\omega_{1}...d\omega_{n-1}\partial_{i}\partial_{j}W^{gm}\left(\omega_{0},\omega_{1},...\omega_{n-1},\omega_{n};t_{n}\right)
\]

\[
+\stackrel[i=1]{n-1}{\sum}\int_{-\infty}^{\omega_{c}}d\omega_{1}...d\omega_{n-1}\partial_{i}^{2}W^{gm}\left(\omega_{0},\omega_{1},...\omega_{n-1},\omega_{n};t_{n}\right)
\]

\begin{equation}
=\stackrel[i,j=1]{n-1}{\sum}\int_{-\infty}^{\omega_{c}}d\omega_{1}...d\omega_{n-1}\partial_{i}\partial_{j}W^{gm}\left(\omega_{0},\omega_{1},...\omega_{n-1},\omega_{n};t_{n}\right).\label{eq:248}
\end{equation}

\medskip{}

Then, the RHS (\ref{eq:248}), which corresponds to the LHS (\ref{eq:217}),
can be evaluated by the LHS of (\ref{eq:248}). In the continuum,
$\epsilon\rightarrow0$, with $\omega_{0}\neq0$, we compute the second
derivative of $\Pi_{\epsilon\rightarrow0}^{gm}\left(\omega_{0},\omega_{n},t_{n}\right)$
with respect to $\omega_{c}$, by (\ref{eq:PiGaussianaBarreira}).
The first derivative was computed in (\ref{eq:135}). Thus,

\medskip{}

\[
\frac{\partial^{2}\Pi_{\epsilon\rightarrow0}^{gm}\left(\omega_{0},\omega_{n},t_{n}\right)}{\partial\omega_{c}^{2}}=\frac{\partial}{\partial\omega_{c}}\left(\frac{\partial\Pi_{\epsilon\rightarrow0}^{gm}}{\partial\omega_{c}}\left(\omega_{0},\omega_{n};t_{n}\right)\right)
\]

\[
=\frac{\partial}{\partial\omega_{c}}\left(\left(\frac{2}{\pi}\right)^{1/2}\frac{\left(2\omega_{c}-\omega_{n}-\omega_{0}\right)}{t_{n}^{3/2}}e^{2\alpha\left(\omega_{n}-\omega_{0}-\omega_{c}\right)}e^{-\frac{\left(2\omega_{c}-\omega_{n}-\omega_{0}-\alpha t_{n}\right)^{2}}{2t_{n}}}\right)
\]

\[
=\left(\frac{2}{\pi}\right)^{1/2}\left[\frac{2}{t_{n}^{3/2}}e^{2\alpha\left(\omega_{n}-\omega_{0}-\omega_{c}\right)}e^{-\frac{\left(2\omega_{c}-\omega_{n}-\omega_{0}-\alpha t_{n}\right)^{2}}{2t_{n}}}\right.
\]

\[
-\frac{2\left(2\omega_{c}-\omega_{n}-\omega_{0}\right)}{t_{n}^{5/2}}\left[\left(2\omega_{c}-\omega_{n}-\omega_{0}\right)-\alpha t_{n}\right]e^{2\alpha\left(\omega_{n}-\omega_{0}-\omega_{c}\right)}e^{-\frac{\left(2\omega_{c}-\omega_{n}-\omega_{0}-\alpha t_{n}\right)^{2}}{2t_{n}}}
\]

\[
\left.-\frac{2\alpha t_{n}\left(2\omega_{c}-\omega_{n}-\omega_{0}\right)}{t_{n}^{5/2}}e^{2\alpha\left(\omega_{n}-\omega_{0}-\omega_{c}\right)}e^{-\frac{\left(2\omega_{c}-\omega_{n}-\omega_{0}-\alpha t_{n}\right)^{2}}{2t_{n}}}\right].
\]

\[
\therefore\frac{\partial^{2}\Pi_{\epsilon\rightarrow0}^{gm}\left(\omega_{0},\omega_{n},t_{n}\right)}{\partial\omega_{c}^{2}}=
\]

\begin{equation}
2\left(\frac{2}{\pi}\right)^{1/2}\left[1-\frac{\left(2\omega_{c}-\omega_{n}-\omega_{0}\right)^{2}}{t_{n}}\right]\frac{1}{t_{n}^{3/2}}e^{-\frac{\left(2\omega_{c}-\omega_{n}-\omega_{0}-\alpha t_{n}\right)^{2}}{2t_{n}}}e^{2\alpha\left(\omega_{n}-\omega_{0}-\omega_{c}\right)}.\label{eq:249}
\end{equation}
\medskip{}

Now we equate (\ref{eq:249}) to (\ref{eq:246}). The second term
of this equation, that is, of (\ref{eq:246}), is just equal to (\ref{eq:249}).
Therefore, canceling both sides, we conclude that the first term of
the RHS (\ref{eq:246}) must be null. In order to nullify it, we must
have\medskip{}

\begin{equation}
\left(\frac{1}{\sqrt{\beta}}-v_{0}\sqrt{2}\right)=0.\label{eq:250}
\end{equation}

\medskip{}

This means that $I_{1}$ (term in $\nu_{0}$), which diverges (with
$1/\sqrt{\epsilon}$, cancels with the divergent terms of $I_{2}$
(term in $1/\sqrt{\beta}$), which were separated when the exponential
term in $\beta\epsilon$ was introduced. It still remains the non-divergent
part of the first term of RHS (\ref{eq:217}). Concluding, the summation

\medskip{}

\[
\stackrel[i,j=1]{n-1}{\sum}\partial_{i}\partial_{j}
\]
\medskip{}

\noindent has two divergent terms, whose effects cancel mutually.
Thus, non-divergent terms remain in the first summation of the RHS
(\ref{eq:217}), and we can transform summations into integrals: \medskip{}

\begin{equation}
2\stackrel[i=1]{n-2}{\sum}\stackrel[j=i+1]{n-1}{\sum}\rightarrow2\frac{1}{\epsilon^{2}}\intop_{0}^{t_{n}}dt_{i}\intop_{t_{i}}^{t_{n}}dt_{j}\label{eq:251}
\end{equation}

\medskip{}

\medskip{}

\section{\emph{Drift} of the non-gaussian distribution: the specific case
of $\kappa_{15}$\label{sec:DriftDistrNG}}

\medskip{}

\hspace*{0.25in}In the case of the density of probability (\ref{eq:240})
be expanded up to the $15th$ order in derivatives, we evaluate the
integral (\ref{eq:242}).

\medskip{}

\[
\alpha=\frac{1}{t_{n}\sigma}\left\{ \left[\left(r-r_{f}\right)-\frac{1}{2}t_{n}\sigma^{2}\right]+\right.
\]

\[
\left\{ 1,307,674,368,000\cdot\left[1,307,674,368,000\right.\right.
\]

\[
+\sigma^{3}\left[217,945,728,000\kappa_{3}+1,816,214,400\cdot\left(10\kappa_{3}^{2}+\kappa_{6}\right)\sigma^{3}\right.
\]

\[
+259,459,200\cdot\left(35\kappa_{3}\kappa_{4}+\kappa_{7}\right)\sigma^{4}
\]

\[
+32,432,400\cdot\left(35\kappa_{4}^{2}+56\kappa_{3}\kappa_{5}+\kappa_{8}\right)\sigma^{5}+3,603,600\cdot\left(126\kappa_{4}\kappa_{5}+84\kappa_{3}\kappa_{6}+\kappa_{9}\right)\sigma^{6}
\]

\[
+360,360\cdot\left(\kappa_{10}+6\cdot\left(21\kappa_{5}^{2}+35\kappa_{4}\kappa_{6}+20\kappa_{3}\kappa_{7}\right)\right)\sigma^{7}
\]

\[
+32.760\cdot\left(\kappa_{11}+33\cdot\left(14\kappa_{5}\kappa_{6}+10\kappa_{4}\kappa_{7}+5\kappa_{3}\kappa_{8}\right)\right)\sigma^{8}
\]

\[
+2,730\cdot\left(\kappa_{12}+11\cdot\left(42\kappa_{6}^{2}+72\kappa_{5}\kappa_{7}+45\kappa_{4}\kappa_{8}+20\kappa_{3}\kappa_{9}\right)\right)\sigma^{9}
\]

\[
+210\cdot\left(\kappa_{13}+143\cdot\left(2\kappa_{10}\kappa_{3}+12\kappa_{6}\kappa_{7}+9\kappa_{5}\kappa_{8}+5\kappa_{4}\kappa_{9}\right)\right)\sigma^{10}
\]

\[
+15\cdot\left(\kappa_{14}+13\cdot\left(28\kappa_{11}\kappa_{3}+11\cdot\left(7\kappa_{10}\kappa_{4}+12\kappa_{7}^{2}+21\kappa_{8}\kappa_{6}+14\kappa_{5}\kappa_{9}\right)\right)\right)\sigma^{11}
\]

\[
+\left(\kappa_{15}+13\cdot\left(35\kappa_{12}\kappa_{3}+105\kappa_{11}\kappa_{4}+231\kappa_{10}\kappa_{5}+495\kappa_{7}\kappa_{8}+385\kappa_{6}\kappa_{9}\right)\right)\sigma^{12}
\]

\begin{equation}
+\left.\left.\left.\left.10,897,286,400\cdot\sigma\cdot\left(5\kappa_{4}+\kappa_{5}\sigma\right)\right]\right]^{-1}\right\} \right\} .\label{eq:315}
\end{equation}

\medskip{}

\end{document}